\documentclass[fleqn,10pt]{wlscirep}
\usepackage[utf8]{inputenc}
\usepackage[T1]{fontenc}
\usepackage{multirow}
\RequirePackage{graphicx}
\RequirePackage{fancyhdr}
\RequirePackage{bookmark}
\RequirePackage{caption} 
\usepackage{arydshln}
\usepackage{lineno}

\usepackage{hyperref}
\hypersetup{colorlinks = true, allcolors = blue}

\usepackage[nameinlink]{cleveref}

\crefname{figure}{Fig.}{Figs.}
\crefformat{equation}{Eq.~#2(#1)#3}
\crefformat{section}{Section~#2#1#3}
\AtBeginDocument{%
	\let\citet\cite
}

\usepackage[labelfont=bf]{caption}
\usepackage{babel}

\title{Towards Unified AI-Driven Fracture Mechanics: The Extended Deep Energy Method (XDEM)}

\author[1,2,*]{Yizheng Wang}

\author[1]{Yuzhou Lin}

\author[3]{Somdatta Goswami}

\author[5]{Luyang Zhao}

\author[1]{Huadong Zhang}
\author[1]{Jinshuai Bai}

\author[2]{Cosmin Anitescu}

\author[4]{Mohammad Sadegh Eshaghi}

\author[4]{Xiaoying Zhuang}

\author[2]{Timon Rabczuk}

\author[1,*]{Yinghua Liu}

\affil[1]{Department of Engineering Mechanics, Tsinghua University, Beijing 100084, China}

\affil[2]{Institute of Structural Mechanics, Bauhaus-Universit\"{a}t Weimar, Marienstr. 15, D-99423 Weimar, Germany}

\affil[3]{Department of Civil and Systems Engineering, Johns Hopkins University, Baltimore, MD, 21218, United States of America}

\affil[4]{ Institute of Photonics, Department of Mathematics and Physics, Leibniz University Hannover, Germany}

\affil[5]{Department of Mechanical and Materials Engineering, Western University, London N6A 5B9, Ontario, Canada}

\affil[*]{Corresponding.wang-yz19@tsinghua.org.cn, yhliu@mail.tsinghua.edu.cn}

\begin{abstract}
Physics-Informed Neural Networks (PINNs) have recently emerged as powerful tools for solving partial differential equations (PDEs), with the Deep Energy Method (DEM) proving especially effective in fracture mechanics due to its energy-based formulation. Despite these advances, existing DEM approaches require dense collocation near cracks, face stability challenges, and typically treat discrete and continuous fracture models separately. To overcome these limitations, we introduce the Extended Deep Energy Method (XDEM), a unified deep learning framework that incorporates both displacement discontinuities and crack-tip asymptotics in the discrete setting, while flexibly coupling displacement and phase fields in the continuous setting. This integration enables accurate fracture predictions using uniformly distributed, relatively sparse collocation points. Validation across benchmark problems including stress intensity factor evaluation, straight and kinked crack growth, and complex crack initiation demonstrates that XDEM consistently outperforms standard DEM in accuracy and efficiency. By bridging discrete and phase-field models within a single framework, XDEM establishes a robust foundation for applying AI to fracture mechanics and opens new avenues for predictive modeling in engineering and materials science.
\end{abstract}

\begin{document}

\flushbottom
\maketitle 
\section*{Introduction}

The accurate modeling of the failure of materials and thereby structures has long been recognized as a grand challenge in mechanics, shaping decades of research in engineering and materials science. Fracture is a primary driver of material failure, and the accurate simulation of crack initiation and propagation remains a central problem. Classical computational approaches to fracture fall broadly into two categories: discrete fracture models and continuous damage models. Discrete approaches including the Virtual Crack Closure Technique (VCCT) \cite{krueger2004virtual}, the Cohesive Zone Method (CZM) \cite{dugdale1960yielding,barenblatt1962mathematical}, and extended finite element methods (XFEM \cite{moes1999finite}/XIGA \cite{ghorashi2012extended,ghorashi2015t}) offer computational efficiency but require explicit crack representation and fracture criteria, making them difficult to apply in problems with complex crack networks or three-dimensional geometries. Continuous approaches, such as the phase-field method \cite{miehe2010phase,miehe2010thermodynamically,francfort1998revisiting,goswami2019adaptive,goswami2020adaptiveCMAME} and peridynamics \cite{ren2016dual}, provide a more flexible description by allowing cracks to nucleate and evolve without any tracking. However, this flexibility comes at the cost of significantly higher computational expense, primarily due to the need for fine spatial discretization to resolve crack features accurately.


\begin{figure}[!t]
	\begin{centering}
		\includegraphics[scale=0.53]{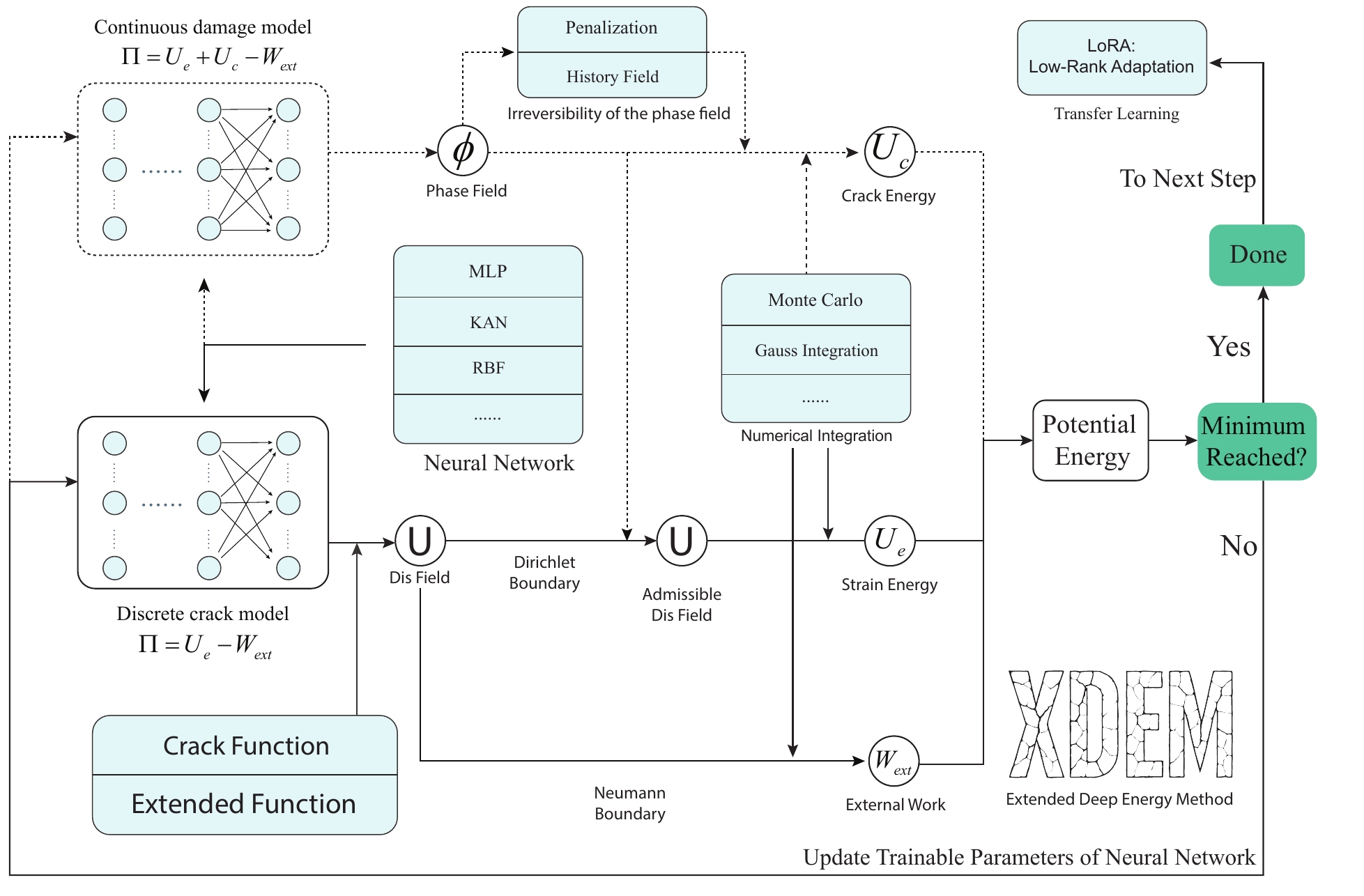}
		\par\end{centering}
	\caption{Schematic illustration of the Extended Deep Energy Method (XDEM), comprising unified discrete and continuous formulations. The continuous model is indicated by dashed lines.} \label{fig:Extended-Deep-Energy-Method}
\end{figure}

It is important to note that these modeling frameworks—whether discrete or continuous—ultimately lead to governing partial differential equations (PDEs) that must be solved numerically. Traditional numerical schemes such as the finite element method (FEM), finite difference method (FDM), or meshfree methods have long been the workhorses for these PDEs. Recently, a new class of solvers based on machine learning, particularly Physics-Informed Neural Networks (PINNs) \cite{PINN_original_paper,loss_is_minimum_potential_energy}, has emerged as an alternative paradigm for solving such equations. PINNs and their variational counterpart, the Deep Energy Method (DEM), embed the underlying physical laws directly into the neural network training objective, enabling the solution of PDEs without requiring explicit meshing or labeled data. The DEM formulation is particularly well suited for fracture mechanics, where the governing equations arise naturally from variational energy principles \cite{goswami2020transfer}. DEM has been successfully applied to both discrete and continuous fracture models, yielding several representative advances. In discrete formulations, Zhao et al. employed DEM to simulate crack propagation \cite{zhao2025denns}, though their approach required additional collocation points along the evolving crack path. Chen et al. \cite{chen2024crack} combined strong-form PINNs with asymptotic fracture solutions to model fatigue crack growth, yet did not exploit the variational DEM framework. Chen et al. \cite{chen2024crack} applied the PINNs (employing the PDE in its strong form) combined with asymptotic fracture solutions to simulate fatigue crack growth. For continuous models, Goswami et al. pioneered the use of DEM for phase-field fracture, demonstrating its effectiveness for both second-order \cite{goswami2020transfer} and fourth-order formulations \cite{goswami2020adaptive}. However, accurate modeling of shear-dominated (mode-II) failure remained challenging. 
Zheng \cite{zheng2022physics} proposed a FEM-inspired approach, where the displacement field and the phase field at the nodes of the mesh were predicted by neural networks and subsequently interpolated with shape functions to construct displacement and phase fields. However, this method required mesh refinement along the crack path, similar to FEM. Building on this idea, Manav et al. \cite{manav2024phase} conducted a more systematic study, applying DEM to crack nucleation, propagation, kinking, branching, and coalescence. Compared with traditional numerical solvers like FEM and isogeometric analysis (IGA), a key advantage of DEM is that it allows larger load increments \cite{goswami2020transfer}, enabling accelerated simulation of crack propagation.  
Despite these advances, existing DEM approaches to fracture mechanics have treated discrete and continuous models independently. Since each model has its own strengths, this separation limits their overall effectiveness: discrete models offer higher computational efficiency and are advantageous for relatively simple crack growth problems, while continuous phase-field models do not require explicit fracture criteria and are better suited for complex crack patterns and 3D systems. Moreover, existing DEM frameworks typically require refined collocation near crack tips to accurately capture the highly localized fields \cite{manav2024phase,goswami2020transfer,goswami2020adaptive,zhao2025denns}, which in turn demands a priori knowledge of the crack path or the development of efficient adaptive refinement schemes \cite{goswami2020adaptive}.

To address these limitations, we propose the Extended Deep Energy Method (XDEM) (Fig. \ref{fig:Extended-Deep-Energy-Method}), a unified AI framework that bridges discrete and continuous formulations of fracture mechanics. In the discrete regime, XDEM introduces a crack function to represent displacement discontinuities and an extended function to capture near-tip asymptotic fields. In the continuous phase-field regime, XDEM decouples displacement and damage evolution through flexible neural architectures, enforcing irreversibility via penalization or history-field strategies. Together, these developments eliminate the need for problem-specific collocation refinement, while substantially enhancing accuracy, robustness, and computational efficiency. The main contributions of this work are summarized as follows:
\begin{itemize}[leftmargin=*,nosep]
	\item	We establish a unified DEM-based framework applicable to both discrete and continuous fracture models.
	\item	We introduce crack and extended functions that enable accurate stress intensity factor identification with sparse, uniformly distributed collocation points.
	\item	We demonstrate the versatility of XDEM through benchmark studies on stress intensity factors, crack propagation (straight, kinked, and branching), and crack initiation.
\end{itemize}

\section*{Results}

This section presents the performance of the proposed Extended Deep Energy Method (XDEM) in modeling fracture evolution under varying loading and boundary conditions. 
XDEM is presented in two forms: discrete (XDEM-D) and continuous (XDEM-C). The results discussed in the main text pertain to XDEM-D, while results for XDEM-C are provided in the Supplementary Section.
The results are organized into two parts: (i) prediction of stress intensity factors (SIFs), which quantify the near-tip stress fields, and (ii) simulation of crack path propagation under different fracture modes (modes I, II and III).

\bigbreak
\noindent
\textbf{Stress intensity factor}

\noindent The stress intensity factor is a fundamental quantity in linear elastic fracture mechanics, characterizing the intensity of the stress field near the crack tip. We begin by evaluating the ability of XDEM to accurately predict SIFs. Three representative cases: mode I (opening), mode II (sliding), and mode III (tearing) are considered, along with mixed-mode loading conditions.
Since the individual mode I, II, and III cracks can be regarded as special cases of mixed-mode fracture, we focus here on the mixed-mode configuration, while detailed results for the pure modes are provided in Supplementary Section.

\begin{figure}
	\begin{centering}
	\includegraphics[scale=0.33]{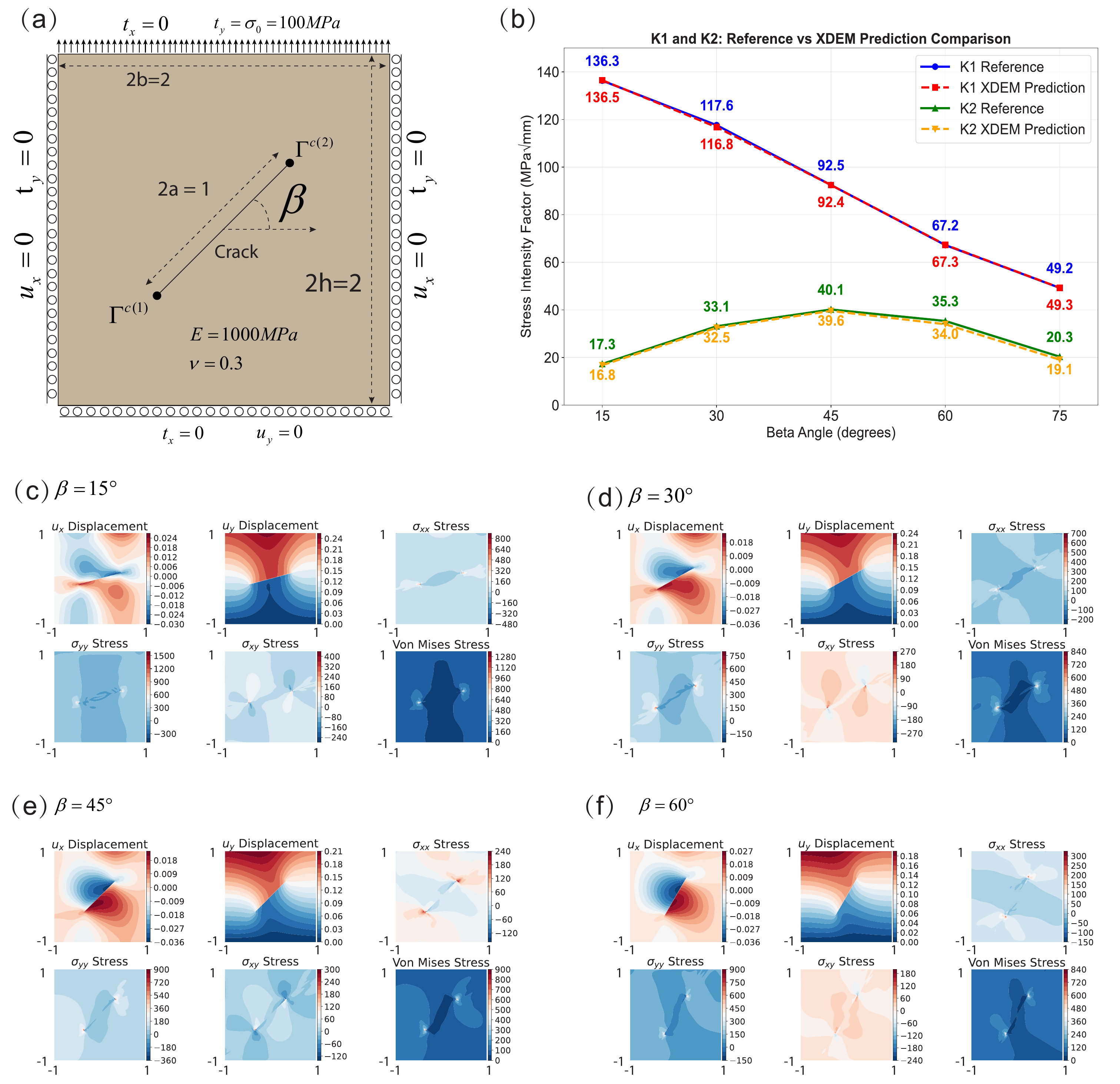}
	\par\end{centering}
	\caption{Mixed-mode crack problem: (a) geometry, material properties, and boundary conditions; (b) comparison of XDEM predicted stress intensity factors ($K_{I}$ and $K_{II}$) with FEM reference solutions for $a=0.5$ and $\sigma_{0}=100,\text{MPa}$ at different crack inclination angles $\beta$; (c–f) displacement and stress contours predicted by XDEM for different crack angles $\beta$ in the mixed mode configuration.}
    \label{fig:Mode-mixed-crack_schematic}
\end{figure}

The mixed-mode problem involves a rectangular plate of length $2b=2$ and height $2h=2$, containing a central crack of length $2a=1$ inclined at an angle $\beta$ (see \Cref{fig:Mode-mixed-crack_schematic} a). The material parameters are $E=1000,\text{MPa}$ and Poisson’s ratio $\nu=0.3$, under plane strain conditions. A uniform tensile stress $\sigma_{0}=100,\text{MPa}$ is applied on the top boundary, while the left and right edges are constrained in the $x$-direction ($u_x=0$) and the bottom boundary in the $y$-direction ($u_y=0$). The displacement fields used in the XDEM formulation are defined in Supplementary Section.

\Cref{fig:Mode-mixed-crack_schematic}b compares the XDEM-predicted mode-I and mode-II stress intensity factors ($K_{I}$ and $K_{II}$) with finite element method (FEM) reference solutions for crack inclination angles $\beta \in {15^{\circ}, 30^{\circ}, 45^{\circ}, 60^{\circ}, 75^{\circ}}$. XDEM shows excellent agreement across all orientations, accurately capturing both components of the stress intensity factor. The SIFs are computed using the interaction integral method, as described in Supplementary Section 4.8.

These results demonstrate that XDEM reliably captures near-tip stress fields—a key prerequisite for accurate crack propagation modeling, which is explored further in Supplementary Section 2.2. Training parameters are consistent with the mode-I case, using $100\times100$ uniformly distributed collocation points. The displacement and stress contours for various crack angles (\Cref{fig:Mode-mixed-crack_schematic}c–f) clearly exhibit the expected discontinuities across the crack surface, underscoring the effectiveness of the crack function in representing displacement jumps within the neural architecture.

To examine the sensitivity of XDEM to collocation density, \Cref{tab:XDEM_points_num} summarizes its accuracy and computational efficiency across different point distributions. The results indicate that, for most crack angles, a moderate resolution of $30\times30$ points suffices to achieve high accuracy. At very low point densities, however, integration inaccuracies during training can occasionally lead to spurious stress fields or non-physical fracture patterns (see Supplementary Section 2). Such effects can be mitigated by employing early stopping criteria or by modestly increasing the collocation resolution.

\begin{figure}[t]
\centering
\includegraphics[scale=0.23]{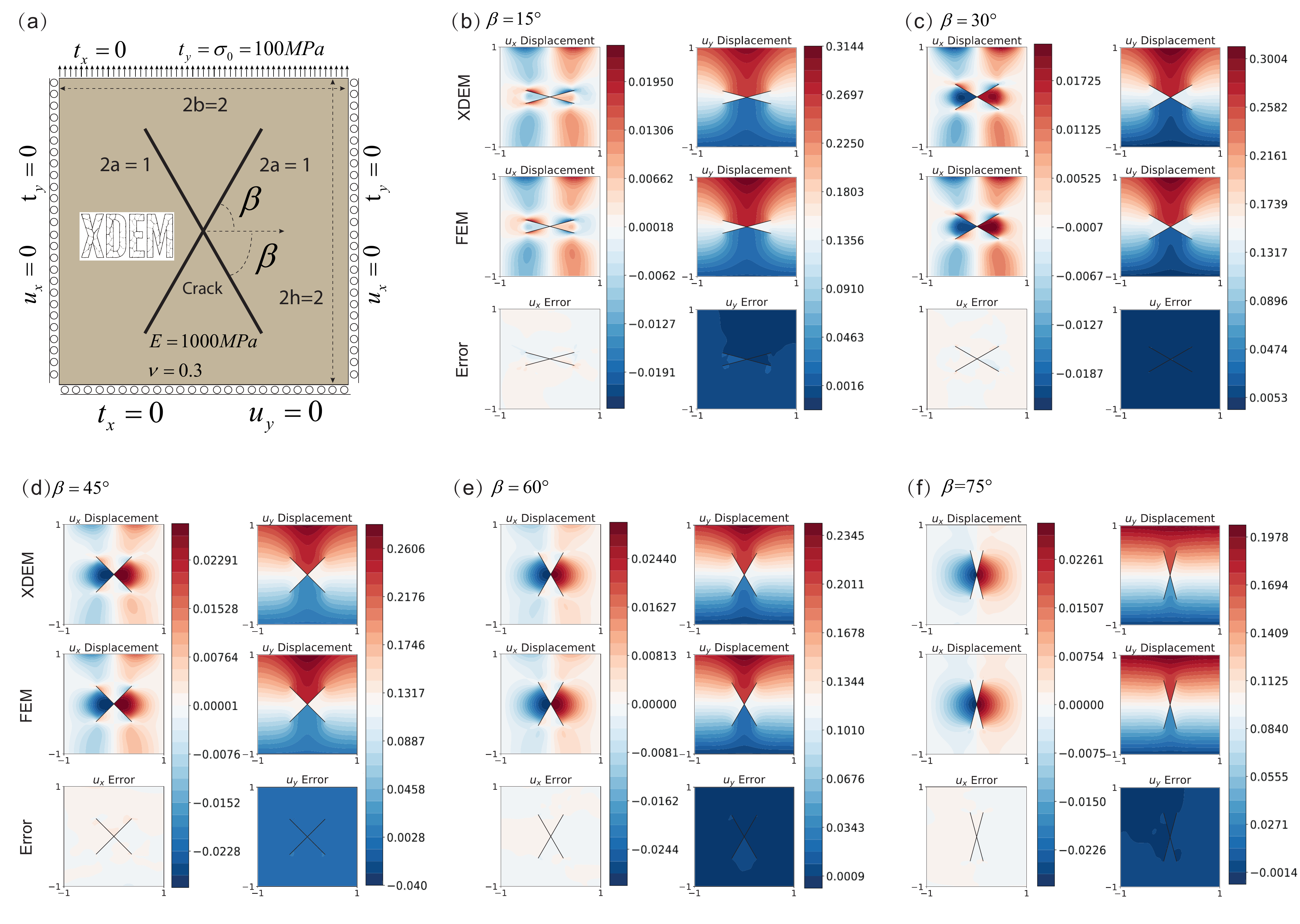}
\caption{Displacement contours predicted by XDEM and FEM (reference solution) for intersecting cracks, demonstrating the accurate capture of displacement discontinuities by XDEM. The error map is computed with FEM taken as the reference solution. \label{fig:crack_intersecting_contourf}}
\end{figure}

Finally, to further illustrate the generality of XDEM, \Cref{fig:crack_intersecting_contourf} presents results for intersecting cracks. XDEM accurately reproduces displacement discontinuities along multiple crack interfaces in close agreement with the reference solution \cite{zhao2025denns}. These results highlight the robustness of XDEM in representing complex crack topologies without requiring adaptive meshing or problem-specific refinement.

\bigbreak
\noindent
\textbf{Crack propagation}

\noindent
The preceding section established that XDEM can accurately predict SIFs, forming the basis for modeling dynamic crack evolution. Here, we further validate XDEM’s performance in simulating crack propagation under mixed-mode loading, focusing on three representative scenarios: straight crack growth, crack kinking, and crack initiation. Detailed results for the additional examples are provided in Supplementary Section 2.2.

\begin{figure}[t]
\centering
\includegraphics[scale=0.26]{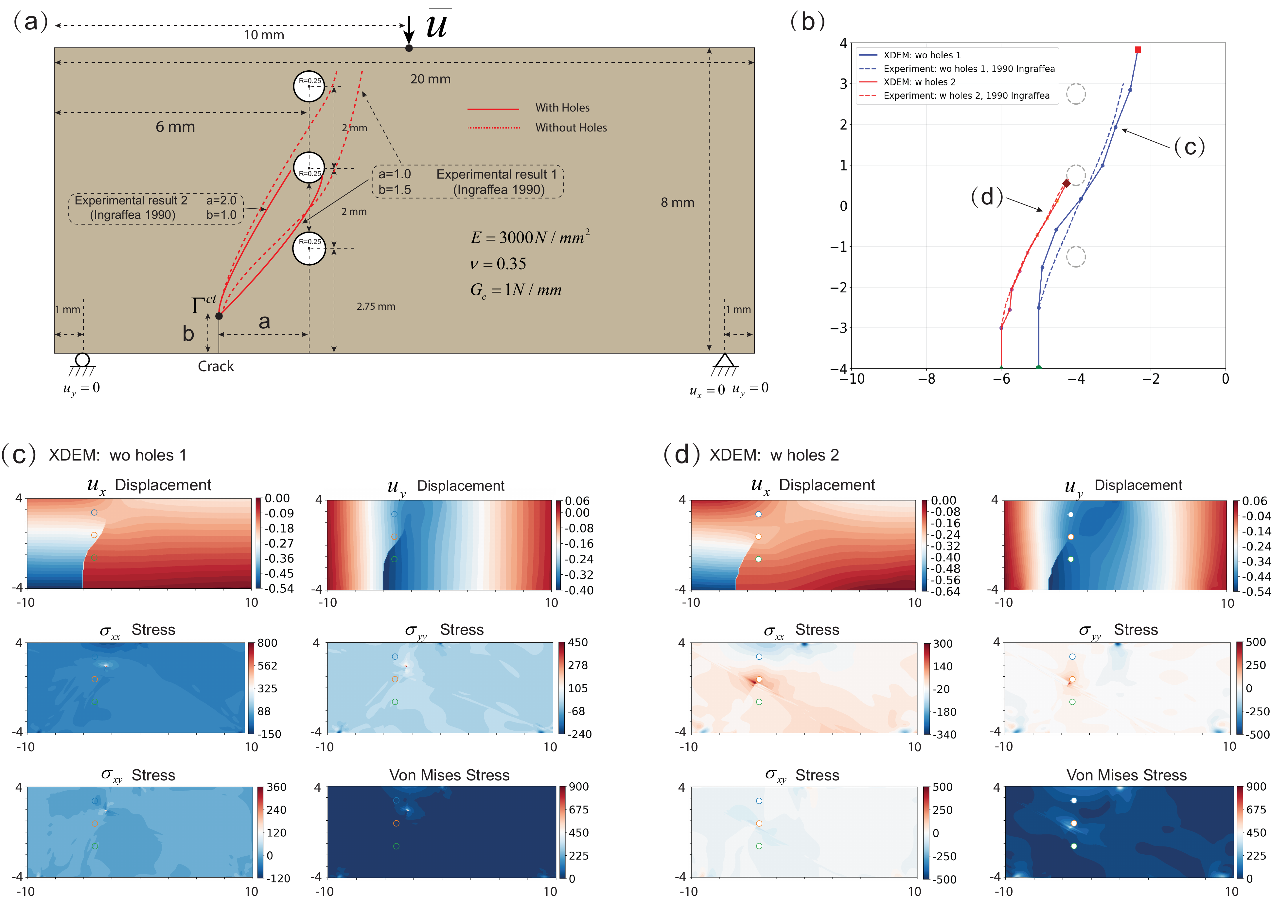}
\caption{Performance of XDEM on the Bittencourt problem: (a) Geometry of the benchmark. (b) Predicted crack propagation paths compared with experimental observations \cite{ingraffea1990probabilistic} for two cases: without a hole (Experiment 1) and with a hole (Experiment 2). (c) Displacement and stress contours for Experiment 1. (d) Displacement and stress contours for Experiment 2. \label{fig:XDEM_Bittencourt}}
\end{figure}

We first examine the classical Bittencourt problem \cite{bittencourt1996quasi}, a widely used benchmark in computational fracture mechanics. Experimental observations for this setup are available in \cite{ingraffea1990probabilistic}. As shown in \Cref{fig:XDEM_Bittencourt}a, this configuration is particularly interesting because the crack trajectory varies depending on the initial crack position, leading the crack to deflect toward different holes. The displacement field formulation used in XDEM for this problem is given in Supplementary Section 2.2.

In the XDEM simulations, crack propagation initiates at a normalized displacement $\bar{u}\approx0.4$, consistent with previous experimental and numerical findings \cite{areias2013element}. The predicted crack paths (Fig.~\ref{fig:XDEM_Bittencourt} b) show excellent agreement with the experimental observations \cite{ingraffea1990probabilistic}, accurately reproducing the deflection behavior observed in both scenarios—with and without a hole. The corresponding displacement and stress contours (\Cref{fig:XDEM_Bittencourt} c,d) clearly illustrate XDEM’s ability to capture displacement discontinuities and stress concentration near the crack tip during propagation.

To further assess the robustness of XDEM, we next consider crack propagation in materials with stiffness inclusions, where the primary challenge arises from the interaction between materials of differing elastic properties. The benchmark setup follows \citet{bouchard2003numerical,hirshikesh2019fenics} and is illustrated in \Cref{fig:XDEM_inclusion}a. The displacement field used in XDEM is detailed in Supplementary Section 2.2.3.

\Cref{fig:XDEM_inclusion}b compares the predicted crack paths for soft and hard inclusions in linear elastic materials with corresponding reference solutions \cite{bouchard2003numerical,hirshikesh2019fenics}. XDEM reproduces the reference trajectories with high fidelity for both cases, confirming its robustness across strong material heterogeneities. The load–displacement response predicted by XDEM (\Cref{fig:XDEM_inclusion}c) exhibits excellent agreement with established results, while the displacement and stress contours (\Cref{fig:XDEM_inclusion}d,e) demonstrate accurate resolution of both the stress concentration near the crack tip and the stress discontinuity across the inclusion interface.

\begin{figure}[t]
\centering
\includegraphics[scale=0.18]{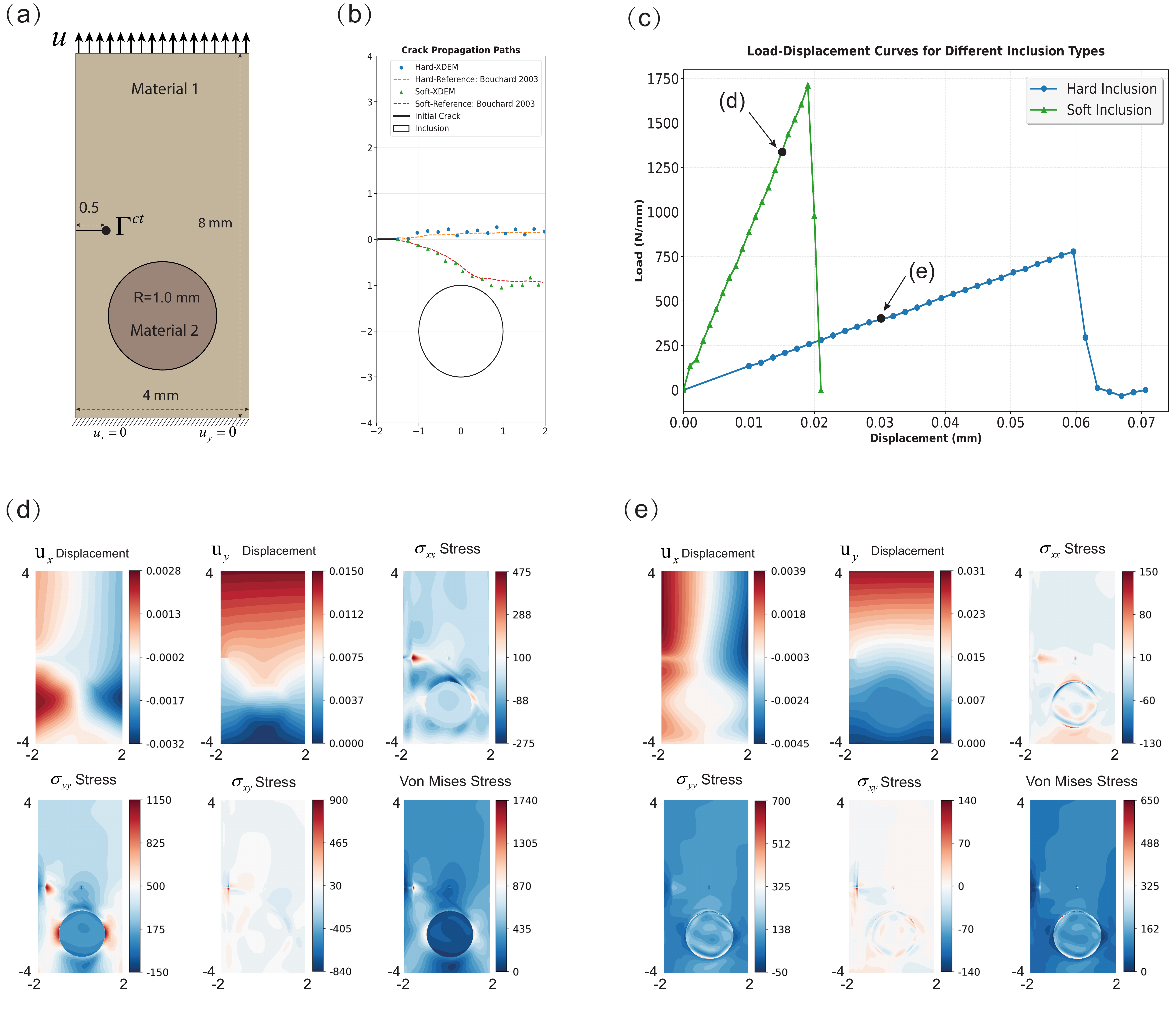}
\caption{Performance of XDEM on inclusion-driven crack propagation: (a) Schematic of the inclusion setup for two cases—soft inclusion ($E_{1}=210{,}000~\mathrm{N/mm^{2}}$, $E_{2}=21{,}000~\mathrm{N/mm^{2}}$) and hard inclusion ($E_{1}=21{,}000~\mathrm{N/mm^{2}}$, $E_{2}=210{,}000~\mathrm{N/mm^{2}}$), with $\nu=0.3$ and $G_{c}=2.7~\mathrm{N/m}$ for all materials. (b) Comparison of predicted crack paths with reference solutions \cite{bouchard2003numerical,hirshikesh2019fenics}. (c) Load–displacement response obtained by XDEM. (d–e) Displacement and stress contours for the soft and hard inclusion cases, respectively. \label{fig:XDEM_inclusion}}
\end{figure}

\section*{Discussion}\label{sec12}

In this work, we introduced the Extended Deep Energy Method (XDEM): a unified, physics-informed AI framework for fracture mechanics that seamlessly integrates discrete and continuous damage representations within a single variational formulation. By embedding explicit crack discontinuities and near-tip enrichments in the discrete setting, and coupling displacement and phase fields in the continuous setting, XDEM effectively overcomes several long-standing limitations of conventional DEM frameworks. In particular, XDEM eliminates the need for refined collocation near crack tips, maintains high accuracy with uniformly distributed sampling points, and enhances numerical stability across a wide spectrum of fracture scenarios.
Comprehensive validation across canonical benchmarks—including stress intensity factor prediction, straight and kinked crack growth, crack initiation, inclusion-induced fracture, and three-dimensional crack propagation—demonstrates that XDEM achieves accuracy and efficiency on par with, and in many cases surpassing, traditional numerical approaches and standard DEM formulations. These results highlight XDEM’s ability to accurately capture both localized crack-tip fields and complex crack trajectories, thereby bridging the methodological gap between discrete and phase-field models.
Beyond addressing key challenges in fracture simulation, XDEM establishes a scalable foundation for AI-driven computational mechanics. Its unified structure, variational grounding, and compatibility with transfer learning techniques position it as a powerful and data-efficient framework for large-scale simulations of complex materials and structures. Future extensions of XDEM to dynamic fracture, multiphysics coupling (e.g., thermo-mechanical or fluid-driven cracking), and neural operator–based generalization could further broaden its applicability and impact.
In summary, XDEM provides a robust, accurate, and versatile framework that advances the state of the art in fracture mechanics while exemplifying how physics-informed AI can transform the predictive modeling of complex failure phenomena in engineering and materials science.

\section*{Method}\label{sec:Method}

We now provide an overview of the proposed Extended Deep Energy Method (XDEM). Full technical details and derivations are presented in Supplementary Section 3.

\bigbreak
\noindent
\textbf{Extended Deep Energy Method Framework}

\noindent XDEM unifies both discrete and continuous formulations of the Deep Energy Method (DEM) within a single framework, as illustrated schematically in \Cref{fig:Extended-Deep-Energy-Method}. The discrete formulation captures sharp displacement discontinuities across cracks, while the continuous formulation models diffusive damage evolution through a phase-field representation.

\bigbreak
\noindent
\textbf{Discrete Crack Formulation}

\noindent The loss functional of the discrete XDEM is defined as:
\begin{equation} 
\begin{aligned} 
\boldsymbol{u}^{n+1} & = \arg\min_{\boldsymbol{u}} \Pi,\\ \Pi & = U_{e} - W_{ext},\\ U_{e} & = \int_{\Omega} \tfrac{1}{2}\,\boldsymbol{\varepsilon}(\boldsymbol{x};\boldsymbol{\theta}_{\boldsymbol{u}}):\boldsymbol{C}:\boldsymbol{\varepsilon}(\boldsymbol{x};\boldsymbol{\theta}_{\boldsymbol{u}})\,dV,\\ W_{ext} & = \int_{\Omega}\boldsymbol{f}\cdot\boldsymbol{u}(\boldsymbol{x};\boldsymbol{\theta}_{\boldsymbol{u}})\,dV + \int_{\Gamma^{\boldsymbol{t}}}\bar{\boldsymbol{t}}\cdot\boldsymbol{u}(\boldsymbol{x};\boldsymbol{\theta}_{\boldsymbol{u}})\,dS,\\ \text{s.t. } & \; u_{i}(\boldsymbol{x};\boldsymbol{\theta}_{\boldsymbol{u}})=\bar{u}_{i}(\boldsymbol{x},t^{n+1}),\;\boldsymbol{x}\in\Gamma^{\boldsymbol{u}};\quad u_{i}^{+}\not\equiv u_{i}^{-},\;\boldsymbol{x}\in\Gamma^{c}. \end{aligned} 
\label{eq:dis_DEM} 
\end{equation}

Here, $\boldsymbol{\sigma}$, $\boldsymbol{C}$, $\boldsymbol{\varepsilon}$, and $\boldsymbol{u}$ represent the stress tensor, stiffness tensor, strain tensor, and displacement vector, respectively. $\Omega$ denotes the computational domain, while $\Gamma^{\boldsymbol{u}}$, $\Gamma^{\boldsymbol{t}}$, and $\Gamma^{c}$ correspond to the Dirichlet, Neumann, and crack boundaries. The external load contributions $\boldsymbol{f}$, $\bar{\boldsymbol{t}}$, and $\bar{\boldsymbol{u}}$ represent body forces, prescribed traction, and imposed displacements. Traction-free conditions are imposed along the crack surface $\Gamma^{c}$, and mixed (Robin) boundaries are not considered here. In this formulation, the proposed crack function automatically enforces displacement discontinuity across $\Gamma^{c}$, while the extended function enhances the representation of near-tip singular fields, significantly improving accuracy and convergence. Detailed definitions and schematics are provided in Supplementary Section 3.1.

\bigbreak
\noindent \textbf{Continuous Damage Formulation}

\noindent For the continuous (phase-field) setting, the loss functional of XDEM is expressed as:
\begin{equation} 
\begin{aligned} 
& \{\boldsymbol{u}^{n+1},\phi^{n+1}\} = \arg\min_{\boldsymbol{\theta}_{\boldsymbol{u}},\boldsymbol{\theta}_{\phi}} \Pi\big(\boldsymbol{u}(\boldsymbol{x};\boldsymbol{\theta}_{\boldsymbol{u}}),\phi(\boldsymbol{x};\boldsymbol{\theta}_{\phi})\big),\\ & \Pi = U_{e} + U_{c} - W_{ext},\\ & U_{e}(\boldsymbol{u},\phi) = \int_{\Omega}\!\Big[w\!\big(\phi(\boldsymbol{x};\boldsymbol{\theta}_{\phi})\big)\,\varPsi^{+}\!\big(\boldsymbol{u}(\boldsymbol{x};\boldsymbol{\theta}_{\boldsymbol{u}})\big) + \varPsi^{-}\!\big(\boldsymbol{u}(\boldsymbol{x};\boldsymbol{\theta}_{\boldsymbol{u}})\big)\Big]\,dV,\\ 
& U_{c}(\boldsymbol{u},\phi) = \frac{G_{c}}{c_{w}}\int_{\Omega}\!\frac{g\!\big(\phi(\boldsymbol{x};\boldsymbol{\theta}_{\phi})\big)}{l_{0}} + l_{0}\,\nabla\phi(\boldsymbol{x};\boldsymbol{\theta}_{\phi})\cdot\nabla\phi(\boldsymbol{x};\boldsymbol{\theta}_{\phi})\,dV,\\ 
& W_{ext} = \int_{\Omega}\boldsymbol{f}\cdot\boldsymbol{u}(\boldsymbol{x};\boldsymbol{\theta}_{\boldsymbol{u}})\,dV + \int_{\Gamma^{\boldsymbol{t}}}\bar{\boldsymbol{t}}\cdot\boldsymbol{u}(\boldsymbol{x};\boldsymbol{\theta}_{\boldsymbol{u}})\,dS,\\ \text{s.t. } 
& \; u_{i}(\boldsymbol{x};\boldsymbol{\theta}_{\boldsymbol{u}})=\bar{u}_{i}(\boldsymbol{x},t^{n+1}),\;\boldsymbol{x}\in\Gamma^{\boldsymbol{u}};\quad \phi^{n+1}\geq\phi^{n}. \end{aligned} \label{eq:variational_principle_DEM} 
\end{equation}
Here, $\boldsymbol{\theta}_{\boldsymbol{u}}$ and $\boldsymbol{\theta}_{\phi}$ denote the trainable parameters of the displacement and phase-field neural networks, respectively. Unlike the discrete formulation, the phase-field approach naturally captures crack nucleation and evolution without requiring an explicit crack-propagation criterion. The irreversibility constraint $\phi^{n+1}\geq\phi^{n}$ is enforced to prevent crack healing, implemented via penalization or history-field mechanisms (see Supplementary Section 3.2).

In both discrete and continuous settings, the fracture process is simulated incrementally across loading steps. To enhance computational efficiency and reduce retraining costs, XDEM employs LoRA-based transfer learning between load increments—details of which are provided in Supplementary Section 3.3.

\subsection*{Code availability}
The code of this work is available at \url{https://github.com/yizheng-wang/XDEM}.

\section*{Acknowledgement}
The study was supported by the Key Project of the National Natural Science Foundation of China (12332005) and scholarship from Bauhaus University in Weimar. We would like to thank the Fracture Mechanics course at Tsinghua University for the insightful discussions, especially Professor Bin Liu and Teaching Assistant Xinyang Xia for their excellent lectures and dedicated guidance. We would like to thank Kan Lin from the Academy of Arts and Design, Tsinghua University, for helping with the graphical enhancement of the figures.

\subsection*{Author contributions}
\textbf{Yizheng Wang}: Conceptualization, Methodology, Formal analysis, Investigation, Data curation, Validation, Visualization, Writing, original draft, Writing, review \& editing. \\ 
\textbf{Yuzhou Lin}: Implementation of the fracture phase-field model using FEM with UEL.  \\
\textbf{Somdatta Goswami}: Supervision, Writing, review \& editing.  \\
\textbf{Luyang Zhao}: Technical discussions, including the interaction integral and crack function.\\
\textbf{Huadong Zhang}: Exploration of 3D fracture problems.  \\
\textbf{Cosmin Anitescu}: Supervision, Writing, review \& editing. \\
\textbf{Mohammad Sadegh Eshaghi}: Investigation.  \\
\textbf{Xiaoying Zhuang}: Supervision, Writing, review \& editing.\\  
\textbf{Timon Rabczuk}: Supervision, Writing, review \& editing.\\ 
\textbf{Yinghua Liu}: Supervision, Funding acquisition.\\

\subsection*{Competing interests}
The authors declare that they have no known competing financial interests or personal relationships that could
have appeared to influence the work reported in this paper.

\setcounter{figure}{0}
\setcounter{table}{0}
\setcounter{section}{0}
\setcounter{page}{1}

\newpage
\section*{Figures}\label{figures}

\begin{figure}[ht!]
\begin{center}
\caption{Schematic illustration of the Extended Deep Energy Method (XDEM), which consists of discrete and continuous models. The continuous formulation is indicated by dashed lines.}
\end{center}
\captionsetup{justification=centering}
\end{figure}

\begin{figure}[ht!]
\begin{center}
\caption{Mixed-mode crack problem: (a) geometry, material properties, and boundary conditions; (b) comparison of XDEM-predicted SIFs ($K_{I}$ and $K_{II}$) with FEM reference solutions for $a=0.5$ and $\sigma_{0}=100\,\text{MPa}$ at different crack inclination angles $\beta$. (c-f) Displacement and stress contours predicted by XDEM for different crack angles $\beta$ in the mixed-mode crack problem.}
\end{center}
\captionsetup{justification=centering}
\end{figure}

\begin{figure}[ht!]
\begin{center}
\caption{Displacement and stress contours predicted by XDEM for intersecting cracks, demonstrating accurate capture of displacement discontinuities and stress concentration at crack tips.}
\end{center}
\captionsetup{justification=centering}
\end{figure}

\begin{figure}[ht!]
\begin{center}
\caption{XDEM results for the Bittencourt problem: (a) Geometry of the Bittencourt setup. (b) Predicted crack propagation paths by XDEM compared against experimental observations \cite{ingraffea1990probabilistic} for two scenarios: without a hole (Experiment 1) and with a hole (Experiment 2). (c) Displacement and stress contours for Experiment 1 (without hole). (d) Displacement and stress contours for Experiment 2 (with hole).}
\end{center}
\captionsetup{justification=centering}
\end{figure}

\begin{figure}[ht!]
\begin{center}
\caption{Performance of XDEM on crack inclusion problems: (a) schematic of the inclusion setup with two scenarios. 
		Case 1: linear elasticity with soft inclusion, where the inclusion is a softer material with $E_{1}=210{,}000~\mathrm{N/mm^{2}}$ and $E_{2}=21{,}000~\mathrm{N/mm^{2}}$. 
		Case 2: linear elasticity with hard inclusion, where the inclusion is stiffer, $E_{1}=21{,}000~\mathrm{N/mm^{2}}$ and $E_{2}=210{,}000~\mathrm{N/mm^{2}}$. 
		The Poisson's ratio and fracture energy are $\nu=0.3$ and $G_{c}=2.7~\mathrm{N/m}$ for all materials. 
		(b) Comparison of crack propagation paths between XDEM and the reference solutions \cite{bouchard2003numerical,hirshikesh2019fenics}. 
		(c) Load-displacement curve predicted by XDEM. 
		(d) Displacement and stress contour at $\bar{u}=0.015$ for the soft inclusion case. 
		(e) Displacement and stress contour at $\bar{u}=0.0302$ for the hard inclusion case. }
\end{center}
\captionsetup{justification=centering}
\end{figure}

\newpage
\section*{Tables}\label{tables}

\begin{table}
	\caption{Accuracy and efficiency of XDEM for different crack angles $\beta$ and collocation point densities in the mixed-mode crack problem. FEM solutions are used as the reference.  \label{tab:XDEM_points_num}}
	
	\centering{}%
	\begin{tabular}{ccccccc}
		\toprule 
		Angle $\beta$ & Points & $K_{I}$ (Ref., XDEM) & Rel. error & $K_{II}$ (Ref., XDEM) & Rel. error & Epochs, Time (s)  \tabularnewline
		\midrule 
		\multirow{3}{*}{$15^{\circ}$} & 30$\times$30 & 136.3, 132.9 & 2.52\% & 17.26, 17.63 & 2.19\% & 4000, 117  \tabularnewline
		& 50$\times$50 & 136.3, 138.2 & 1.39\% & 17.26, 18.73 & 8.51\% & 2900, 83.9  \tabularnewline
		& 80$\times$80 & 136.3, 135.9 & \textbf{0.28\%} & 17.26, 17.35 & \textbf{0.52\%} & 2600, 74.8 \tabularnewline
		\midrule 
		\multirow{3}{*}{$30^{\circ}$} & 30$\times$30 & 117.6, 112.9 & 3.96\% & 33.10, 32.28 & 2.47\% & 4400, 127  \tabularnewline
		& 50$\times$50 & 117.6, 115.0 & 2.18\% & 33.10, 32.44 & 2.00\% & 2900, 83.9  \tabularnewline
		& 80$\times$80 & 117.6, 118.3 & \textbf{0.62\%} & 33.10, 33.22 & \textbf{0.37\%} & 5500, 158 
        \tabularnewline
		\midrule 
		\multirow{3}{*}{$45^{\circ}$} & 30$\times$30 & 92.45, 93.61 & 1.25\% & 40.15, 41.45 & 3.25\% & 4000, 116  \tabularnewline
		& 50$\times$50 & 92.45, 93.86 & 1.52\% & 40.15, 41.19 & 2.60\% & 3600, 104  \tabularnewline
		& 80$\times$80 & 92.45, 93.23 & \textbf{0.85\%} & 40.15, 39.58 & \textbf{1.43\%} & 3300, 95.2  \tabularnewline
		\midrule 
		\multirow{3}{*}{$60^{\circ}$} & 30$\times$30 & 67.20, 69.80 & 3.87\% & 35.32, 37.09 & 5.02\% & 3500, 101  \tabularnewline
		& 50$\times$50 & 67.20, 69.02 & 2.71\% & 35.32, 35.50 & 5.15\% & 4800, 139  \tabularnewline
		& 80$\times$80 & 67.20, 66.63 & \textbf{0.84\%} & 35.32, 34.94 & \textbf{1.08\%} & 8300, 239  \tabularnewline
		\midrule
		\multirow{3}{*}{$75^{\circ}$} & 30$\times$30 & 49.23, 52.03 & 5.70\% & 20.30, 19.19 & 5.48\% & 4400, 127  \tabularnewline
		& 50$\times$50 & 49.23, 51.09 & 3.76\% & 20.30, 19.48 & 4.03\% & 6500, 188  \tabularnewline
		& 80$\times$80 & 49.23, 49.63 & \textbf{0.81\%} & 20.30, 19.95 & \textbf{1.72\%} & 9700, 279 \tabularnewline
		\bottomrule
	\end{tabular}
\end{table}

\clearpage

\newpage
\vspace*{\baselineskip}
\centerline{\Large{\textbf{Supplementary Information}}}
\vspace*{\baselineskip}
\centerline{\Large{\textbf{Towards Unified AI-Driven Fracture Mechanics: }}}
\vspace{4pt}
\centerline{\Large{\textbf{The Extended Deep Energy Method (XDEM) }}}

\makeatletter
\renewcommand \thesection{S\@arabic\c@section}
\renewcommand\thetable{S\@arabic\c@table}
\renewcommand \thefigure{S\@arabic\c@figure}
\makeatother
\setcounter{figure}{0}
\setcounter{table}{0}
\setcounter{section}{0}
\setcounter{page}{1}

\normalsize

\section*{Nomenclature}
\label{sec:Nomenclature} 

\begin{table}[h!]
	\caption{Summary of the main symbols and notation used in this work.}
	\centering
	\begin{tabular}{c l}
		\toprule
		Notation & Description  \\
		\toprule
		$\boldsymbol{u}$ & Displacement field \\
		$\boldsymbol{\varepsilon}$ & Strain field \\
		$\boldsymbol{\sigma}$ & Stress field \\
		$\boldsymbol{C}$ & Elastic stiffness tensor \\
		$\boldsymbol{f}$ & Body force \\
		$\phi$ & Phase field variable \\
		$\Gamma^{c}$ & Crack surface \\
		$\lambda$ & Lamé constant \\
		$E$ & Young's modulus \\
		$\nu$ & Poisson's ratio \\
		$G$ & Shear modulus \\
		$K$ & Bulk modulus \\
		$G_{c}$ & Fracture energy release rate \\
		$l$ & Regularization (length scale) parameter \\
		$K_{I}$, $K_{II}$, $K_{III}$ & Mode I, II, and III stress intensity factors \\
		$J$ & $J$-integral \\
		$\Psi^{+}(\boldsymbol{\varepsilon})$ & Tensile part of the strain energy density \\
		$\Psi^{-}(\boldsymbol{\varepsilon})$ & Compressive part of the strain energy density \\
		$w(\phi)$ & Degradation function that represents the reduction of material stiffness \\
		$g(\phi)$ & Local dissipation function \\
		$\boldsymbol{X}$ & Learnable Williams series expansion in the extend function \\
		$D(\boldsymbol{x})$ & Distance function \\
		$T$ & Decay function \\
		FEM & Finite Element Method \\
		XDEM-D & Extended Deep Energy Method in discrete crack models \\
		XDEM-C & Extended Deep Energy Method in continuous damage models \\
		\bottomrule
	\end{tabular}
	\label{table:nomenclature}
\end{table}

\section{Prerequisite knowledge}
\label{sec:History-of-fracture}

This section introduces two fundamental frameworks in fracture mechanics, discrete and continuous crack models, under the assumptions of linear elasticity, small deformations, and quasi-static loading.

\subsection{Discrete crack model in fracture mechanics\label{subsec:Discrete-crack-model}}

\begin{figure}
	\begin{centering}
		\includegraphics[scale=0.8]{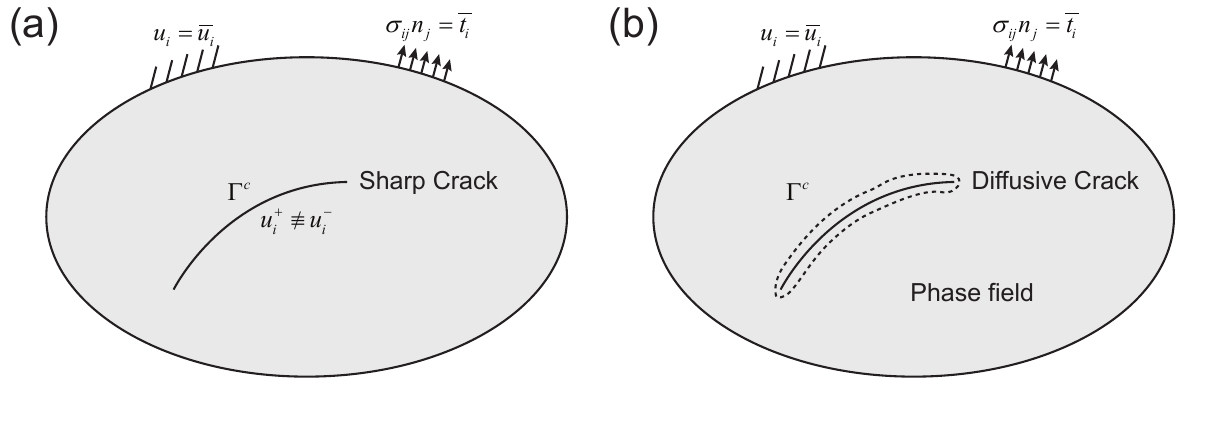}
		\par\end{centering}
	\caption{Schematic of discrete and continuous damage models in fracture mechanics: (a) Discrete crack model with a sharp crack; (b) Continuous damage model where the sharp crack is regularized into a diffused zone.\label{fig:Discrete_contin_schematic}}
\end{figure}

The discrete crack model is essentially governed by the PDEs of linear elasticity:  
\begin{equation}
	\begin{cases}
		\sigma_{ij,j} (\boldsymbol x) +f_{i} (\boldsymbol x)=0 & \boldsymbol{x}\in\Omega,\\
		\sigma_{ij} (\boldsymbol x) =C_{ijkl}\varepsilon_{kl} (\boldsymbol x) & \boldsymbol{x}\in\Omega,\\
		\varepsilon_{ij} (\boldsymbol x) =\tfrac{1}{2}(u_{i,j} (\boldsymbol x)+u_{j,i} (\boldsymbol x)) & \boldsymbol{x}\in\Omega\setminus\Gamma^{c},\\
		u_{i}^{+}(\boldsymbol x)\not\equiv u_{i}^{-}(\boldsymbol x) & \boldsymbol{x}\in\Gamma^{c},\\
		\sigma_{ij}^{\pm}n_{j}^{\pm}=\bar{q}_{i}^{\pm} & \boldsymbol{x}\in\Gamma^{c},\\
		\sigma_{ij}(\boldsymbol x) n_{j}=\bar{t}_{i}(\boldsymbol x) & \boldsymbol{x}\in\Gamma^{\boldsymbol{t}},\\
		u_{i} (\boldsymbol x) =\bar{u}_{i}(\boldsymbol x) & \boldsymbol{x}\in\Gamma^{\boldsymbol{u}}.
	\end{cases}\label{eq:dis_PDEs}
\end{equation}
Here, $\boldsymbol{\sigma}$, $\boldsymbol{C}$, $\boldsymbol{\varepsilon}$, and $\boldsymbol{u}$ denote the stress tensor, stiffness tensor, strain tensor, and displacement vector, respectively. The domain is $\Omega$, with $\Gamma^{\boldsymbol{u}}$, $\Gamma^{\boldsymbol{t}}$, and $\Gamma^{c}$ representing the Dirichlet boundary, Neumann boundary, and crack surface, respectively. $\boldsymbol{f}$, $\bar{\boldsymbol{t}}$, and $\bar{\boldsymbol{u}}$ denote the body force, prescribed traction, and prescribed displacement.
$\bar{\boldsymbol{q}}$ denotes the external traction vector applied on the crack surface. The superscripts $\pm$ indicate the two opposing sides of the crack.
The symbol $\not\equiv$ indicates a displacement discontinuity across $\Gamma^{c}$.
In this work, only Dirichlet and Neumann boundary conditions are considered, with traction-free conditions $\bar{\boldsymbol{q}}=0$ assumed on $\Gamma^{c}$. The boundaries satisfy $\Gamma^{\boldsymbol{t}}\cup\Gamma^{\boldsymbol{u}}=\Gamma$ and $\Gamma^{\boldsymbol{t}}\cap\Gamma^{\boldsymbol{u}}=\emptyset$.
A schematic of the discrete crack model is shown in \Cref{fig:Discrete_contin_schematic}a. 

\noindent The system of equations \Cref{eq:dis_PDEs} can often be reformulated by the principle of minimum potential energy:  
\begin{equation}
	\begin{aligned}
		& \boldsymbol{u} = \arg\min_{\boldsymbol{u}} \Pi, \\
		& \Pi = U_{e}-W_{ext}, \\
		& U_{e} = \int_{\Omega}\tfrac{1}{2}\,\boldsymbol{\varepsilon}(\boldsymbol x):\boldsymbol{C}:\boldsymbol{\varepsilon}(\boldsymbol x)\,dV, \\
		& W_{ext} = \int_{\Omega}\boldsymbol{f}\cdot\boldsymbol{u}(\boldsymbol x)\,dV + \int_{\Gamma^{\boldsymbol{t}}}\bar{\boldsymbol{t}}\cdot\boldsymbol{u}\,dS + \int_{\Gamma^{\boldsymbol{c}}}\bar{\boldsymbol{q}}^{+}\cdot\boldsymbol{u}^{+}dS+\int_{\Gamma^{\boldsymbol{c}}}\bar{\boldsymbol{q}}^{-}\cdot\boldsymbol{u}^{-}dS, \\
		\text{s.t. } & u_{i}(\boldsymbol x)=\bar{u}_{i}(\boldsymbol x), \ \boldsymbol{x}\in\Gamma^{\boldsymbol{u}}; \quad \forall \;\; u_{i}^{+}(\boldsymbol x)\not\equiv u_{i}^{-}(\boldsymbol x), \ \boldsymbol{x}\in\Gamma^{c}.
	\end{aligned}
	\label{eq:dis_energy}
\end{equation}
It is straightforward to show that $\delta\Pi=0$ is equivalent to \Cref{eq:dis_PDEs}. By solving \Cref{eq:dis_energy}, the displacement field can be obtained, from which the stress field follows via the constitutive and geometric relations. To further simulate crack propagation, a fracture criterion must be specified. Common criteria include the maximum circumferential stress criterion \citet{erdogan1963crack}, the maximum energy release rate criterion \citet{hussain1974strain}, and the minimum strain energy density criterion \citet{sih1974strain}. In this work, we adopt the maximum circumferential stress criterion, whereby the crack propagates in the direction of the maximum hoop tensile stress.

\subsection{Continuous damage model in fracture mechanics\label{subsec:Continuous-damage-model}}

Here we introduce the classical phase-field fracture model as a representative of continuous damage models. The phase-field method for fracture was first proposed by Bourdin et al. in 2000 \citet{bourdin2000numerical}, based on the variational principle of Francfort and Marigo \citet{francfort1998revisiting}. Unlike discrete models, the phase-field approach regularizes sharp cracks into diffused regions by introducing an additional damage-related phase-field variable, as illustrated in \Cref{fig:Discrete_contin_schematic}b.  

\noindent Starting from the energy functional in \Cref{eq:dis_energy}, the fracture energy is incorporated as follows:  
\begin{equation}
	\begin{aligned}
		& \boldsymbol{u},\phi = \arg\min_{\boldsymbol{u},\phi}\Pi, \\
		& \Pi = U_{e}+U_{c}-W_{ext}, \\
		& U_{e}(\boldsymbol{u},\phi) = \int_{\Omega} w(\phi)\,\varPsi^{+}(\boldsymbol{\varepsilon}) + \varPsi^{-}(\boldsymbol{\varepsilon})\,dV, \\
		& U_{c}(\boldsymbol{u},\phi) = \frac{G_{c}}{c_{w}}\int_{\Omega}\frac{g(\phi)}{l_{0}}+l_{0}\,(\nabla\phi)\cdot(\nabla\phi)\,dV, \\
		& W_{ext} = \int_{\Omega}\boldsymbol{f}\cdot\boldsymbol{u}\,dV+\int_{\Gamma^{\boldsymbol{t}}}\bar{\boldsymbol{t}}\cdot\boldsymbol{u}\,dS, \\
		\text{s.t. } & u_{i}=\bar{u}_{i}, \ \boldsymbol{x}\in\Gamma^{\boldsymbol{u}};\quad \phi^{n+1}\geq\phi^{n}.
	\end{aligned}
	\label{eq:con_PFM}
\end{equation}

Here, $w(\phi)$ is the degradation function that represents the reduction of material stiffness. It must satisfy the following conditions: $w(0)=1$, $w(1)=0$, $w^{\prime}(1)=0$, and $w^{\prime}(\phi)<0$. A common choice is $w(\phi)=(1-\phi)^{2}$. $\varPsi^{+}(\boldsymbol{\varepsilon})$ and $\varPsi^{-}(\boldsymbol{\varepsilon})$ denote the tensile and compressive contributions of the strain energy, respectively. Typical decompositions include the formulations of Miehe \citet{miehe2010phase} and Amor \citet{amor2009regularized}:  
\begin{equation}
	\begin{aligned}
		\text{Miehe:} \quad & \varPsi^{+}(\boldsymbol{\varepsilon})=\tfrac{1}{2}\lambda\langle\varepsilon_{ii}\rangle_{+}^{2}
		+ G \sum_{i=1}^{3}\langle\lambda_{i}\rangle_{+}^{2}, \\
		& \varPsi^{-}(\boldsymbol{\varepsilon})=\tfrac{1}{2}\lambda\langle\varepsilon_{ii}\rangle_{-}^{2}
		+ G \sum_{i=1}^{3}\langle\lambda_{i}\rangle_{-}^{2}, \\
		\text{Amor:} \quad & \varPsi^{+}(\boldsymbol{\varepsilon})=\tfrac{1}{2}K\langle\varepsilon_{ii}\rangle_{+}^{2}
		+ G\,\varepsilon_{ij}^{\prime}\varepsilon_{ij}^{\prime}, \\
		& \varPsi^{-}(\boldsymbol{\varepsilon})=\tfrac{1}{2}K\langle\varepsilon_{ii}\rangle_{-}^{2},
	\end{aligned}
\end{equation}
where $\lambda$ and $G$ are the Lame constants, $K=\lambda+2G/3$ is the bulk modulus, $\lambda_{i}$ are the eigenvalues of the strain tensor $\boldsymbol{\varepsilon}$, and $\varepsilon_{ij}^{\prime}=\varepsilon_{ij}-\varepsilon_{kk}\delta_{ij}/3$ is the deviatoric strain tensor. The positive and negative parts of a scalar are defined as $\langle x\rangle_{+}=(x+|x|)/2$ and $\langle x\rangle_{-}=(x-|x|)/2$.   In the manuscript, we use Miehe as the form of the energy decomposition.

In \Cref{eq:con_PFM}, $G_{c}$ is the critical energy release rate, and $c_{w}=4\int_{0}^{1}\sqrt{g(\phi)}\,d\phi$ is a normalization constant. The function $g(\phi)$ denotes the local dissipation function, commonly chosen as in the AT1 ($g=\phi$, $c_{w}=8/3$) or AT2 ($g=\phi^{2}$, $c_{w}=2$) models.  

In practical phase-field simulations, the choices of $w(\phi)$ and $g(\phi)$, the type of energy decomposition, and the length-scale parameter $l_{0}$ must be specified in advance. By minimizing $\Pi$ in \Cref{eq:con_PFM}, both the displacement field $\boldsymbol{u}$ and the phase-field variable $\phi$ can be obtained. Compared with discrete crack models, the phase-field model has the advantage of allowing spontaneous crack nucleation without prescribing a fracture criterion. However, its major drawback is the significantly higher computational cost. It is also important to note that the irreversibility condition $\phi^{n+1}\geq\phi^{n}$ is typically enforced by either a history field approach \citet{miehe2010phase} or a penalization technique \citet{gerasimov2019penalization}, as further discussed in \Cref{subsec:irreversibility_phase_field}.

\subsection{Deep Energy Method}

The core idea of the Deep Energy Method (DEM) is to employ neural networks as trial functions to approximate the solution fields by directly minimizing the energy functional \cite{loss_is_minimum_potential_energy}. In fracture mechanics, the energy functional can be selected either from the discrete formulation \citet{zhao2025denns} in \Cref{eq:dis_energy}  or from the continuous phase-field formulation  \citet{goswami2020transfer} in \Cref{eq:con_PFM}.  

For the discrete crack model, the DEM optimization process is formulated as:  
\begin{equation}
	\begin{aligned}\boldsymbol{u}^{n+1} & =\arg\min_{\boldsymbol{u}}\Pi\\
		\Pi & =U_{e}-W_{ext}\\
		U_{e} & =\int_{\Omega}\frac{1}{2}\boldsymbol{\varepsilon}(\boldsymbol{x};\boldsymbol{\theta}_{\boldsymbol{u}}):\boldsymbol{C}:\boldsymbol{\varepsilon}(\boldsymbol{x};\boldsymbol{\theta}_{\boldsymbol{u}})dV\\
		W_{ext} & =\int_{\Omega}\boldsymbol{f}\cdot\boldsymbol{u}(\boldsymbol{x};\boldsymbol{\theta}_{\boldsymbol{u}})dV+\int_{\Gamma^{\boldsymbol{t}}}\bar{\boldsymbol{t}}\cdot\boldsymbol{u}(\boldsymbol{x};\boldsymbol{\theta}_{\boldsymbol{u}})dS.\\
		\text{s.t. } & u_{i}(\boldsymbol{x};\boldsymbol{\theta}_{\boldsymbol{u}})=\bar{u}_{i}(\boldsymbol{x},t^{n+1}),\boldsymbol{x}\in\Gamma^{\boldsymbol{u}};u_{i}^{+}\not\equiv u_{i}^{-},\boldsymbol{x}\in\Gamma^{c}
	\end{aligned}
	\label{eq:dis_DEM}
\end{equation}
Here, $\boldsymbol{\theta}_{\boldsymbol{u}}$ denotes the trainable parameters of the displacement neural network $NN(\boldsymbol{x};\boldsymbol{\theta}_{\boldsymbol{u}})$. The discontinuity of the displacement field across the crack surface, $u_{i}^{+}\not\equiv u_{i}^{-}$, can be incorporated using subdomain DEM (CENN) \citet{wang2022cenn} or discontinuity-embedded neural networks \citet{zhao2025denns}.  

For the continuous phase-field fracture model, the DEM optimization problem is given by:  
\begin{equation}
	\begin{aligned}\{\boldsymbol{u}^{n+1},\phi^{n+1}\} & =\arg\min_{\boldsymbol{\theta}_{\boldsymbol{u}},\boldsymbol{\theta}_{\phi}}\Pi(\boldsymbol{u}(\boldsymbol{x};\boldsymbol{\theta}_{\boldsymbol{u}}),\phi(\boldsymbol{x};\boldsymbol{\theta}_{\phi}))\\
		\Pi & =U_{e}+U_{c}-W_{ext}\\
		U_{e}(\boldsymbol{u},\phi) & =\int_{\Omega}[w(\phi(\boldsymbol{x};\boldsymbol{\theta}_{\phi}))\varPsi^{+}(\boldsymbol{u}(\boldsymbol{x};\boldsymbol{\theta}_{\boldsymbol{u}}))+\varPsi^{-}(\boldsymbol{u}(\boldsymbol{x};\boldsymbol{\theta}_{\boldsymbol{u}}))]dV\\
		U_{c}(\boldsymbol{u},\phi) & =\frac{G_{c}}{c_{w}}\int_{\Omega}\frac{g(\phi(\boldsymbol{x};\boldsymbol{\theta}_{\phi}))}{l_{0}}+l_{0}(\nabla\phi(\boldsymbol{x};\boldsymbol{\theta}_{\phi}))\cdot(\nabla\phi(\boldsymbol{x};\boldsymbol{\theta}_{\phi}))dV\\
		W_{ext} & =\int_{\Omega}\boldsymbol{f}\cdot\boldsymbol{u}(\boldsymbol{x};\boldsymbol{\theta}_{\boldsymbol{u}})dV+\int_{\Gamma^{\boldsymbol{t}}}\bar{\boldsymbol{t}}\cdot\boldsymbol{u}(\boldsymbol{x};\boldsymbol{\theta}_{\boldsymbol{u}})dS.\\
		\text{s.t. } & u_{i}(\boldsymbol{x};\boldsymbol{\theta}_{\boldsymbol{u}})=\bar{u}_{i}(\boldsymbol{x},t^{n+1}),\boldsymbol{x}\in\Gamma^{\boldsymbol{u}};\phi^{n+1}\geq\phi^{n}
	\end{aligned}
	,\label{eq:variational_principle_DEM}
\end{equation}
where $\boldsymbol{\theta}_{\boldsymbol{u}}$ and $\boldsymbol{\theta}_{\phi}$ denote the trainable parameters of the displacement neural network $NN(\boldsymbol{x};\boldsymbol{\theta}_{\boldsymbol{u}})$ and the phase-field neural network $NN(\boldsymbol{x};\boldsymbol{\theta}_{\phi})$, respectively. Unlike discrete models, the phase-field formulation does not require a predefined crack propagation criterion, as cracks can nucleate and evolve naturally. However, the irreversibility condition $\phi^{n+1}\geq \phi^{n}$ must be satisfied, ensuring that cracks cannot heal once formed.

\section{Additional numerical examples} \label{sec: SI_results}

\subsection{Stress intensity factor}  \label{subsec:SI_Stress-intensity-factor}

\subsubsection{Mode I crack} \label{subsec:SI_Mode-I-crack}

We first consider a standard mode I crack problem. The geometry consists of a plate with crack length $2a=1$, centered at the middle of the structure. The overall dimensions are $2b=2$ in length and $2h=6$ in height. The material properties are $E=1000\,\text{MPa}$ and Poisson's ratio $\nu=0.3$ under plane strain conditions. A uniform tensile stress $\sigma_{0}=100\,\text{MPa}$ is applied on the top boundary, while the bottom boundary is constrained only in the $y$-direction ($u_{y}=0$). The left and right boundaries are traction-free. To suppress rigid body motions, we additionally fix $u_{x}=0$ at the crack center, which coincides with the origin of the coordinate system.  

\begin{figure}
	\begin{centering}
		\includegraphics[scale=0.35]{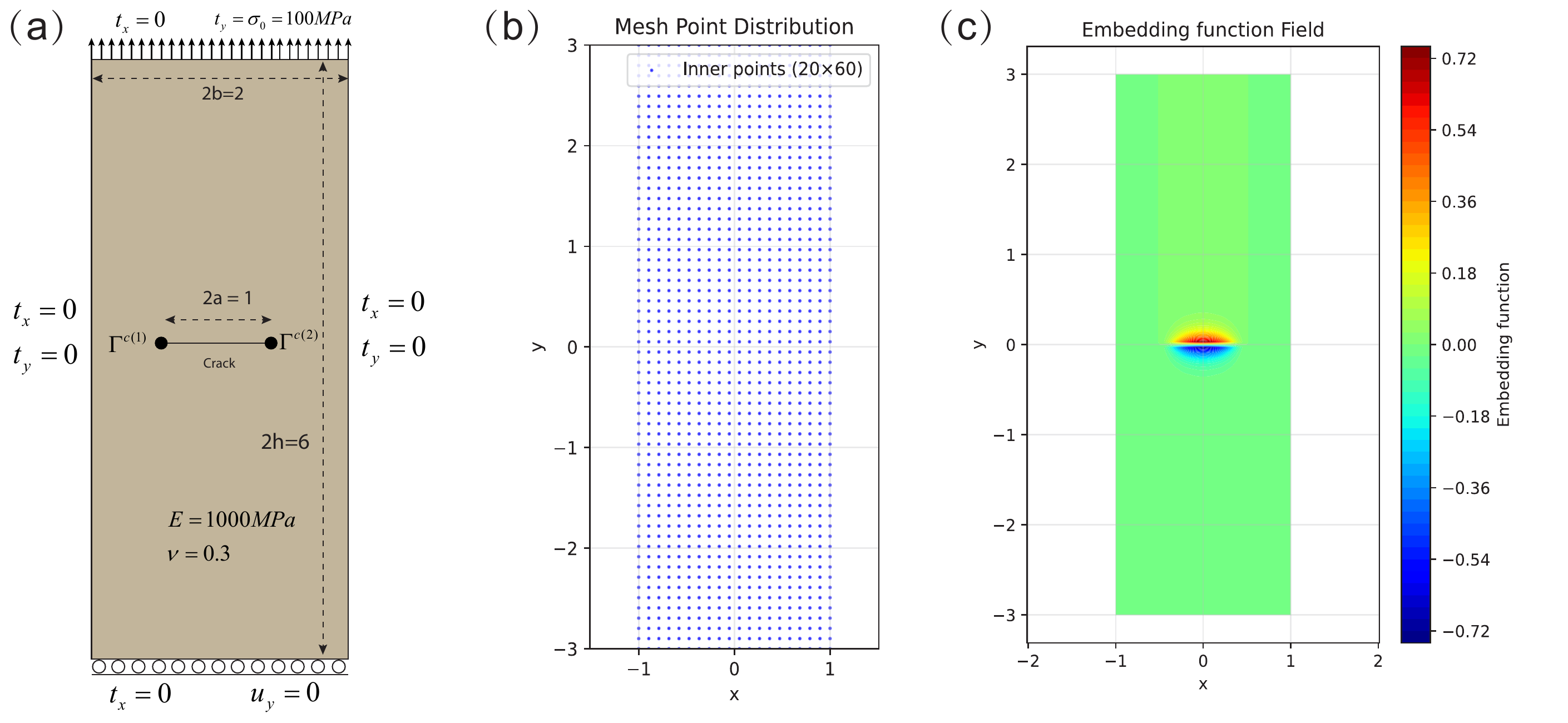}
		\par\end{centering}
	\caption{Schematic of the mode I crack problem: (a) geometry, material properties, and boundary conditions; (b) collocation points for XDEM, uniformly distributed with $20\times 60$ points; (c) contour of the embedding function. \label{fig:Mode-=002160-crack_schematic}}
\end{figure}

It is worth noting that, theoretically, only half of the structure (either left or right) needs to be modeled due to symmetry in the $x$-direction. However, in order to better illustrate the role of the embedding function and to make the problem more accessible for non-mechanics readers, we simulate the entire domain, as shown in \Cref{fig:Mode-=002160-crack_schematic}. The displacement fields in XDEM are expressed as:
\begin{equation}
	\begin{aligned}
		u_{1}(\boldsymbol{x},\varrho;\boldsymbol{\theta}_{\boldsymbol{u}}) &= \frac{x}{b}\Big[NN_{x}(\boldsymbol{x},\varrho;\boldsymbol{\theta}_{\boldsymbol{u}}) + \sum_{i=2} T(\boldsymbol{x};\Gamma^{c(i)}) \cdot X_{1}(\boldsymbol{x};\Gamma^{c(i)})\Big], \\
		u_{2}(\boldsymbol{x},\varrho;\boldsymbol{\theta}_{\boldsymbol{u}}) &= \frac{y+h}{2h}\Big[NN_{y}(\boldsymbol{x},\varrho;\boldsymbol{\theta}_{\boldsymbol{u}}) + \sum_{i=2} T(\boldsymbol{x};\Gamma^{c(i)}) \cdot X_{2}(\boldsymbol{x};\Gamma^{c(i)})\Big],
	\end{aligned}
	\label{eq:crack1_dis}
\end{equation}

\begin{figure}
	\begin{centering}
		\includegraphics[scale=0.65]{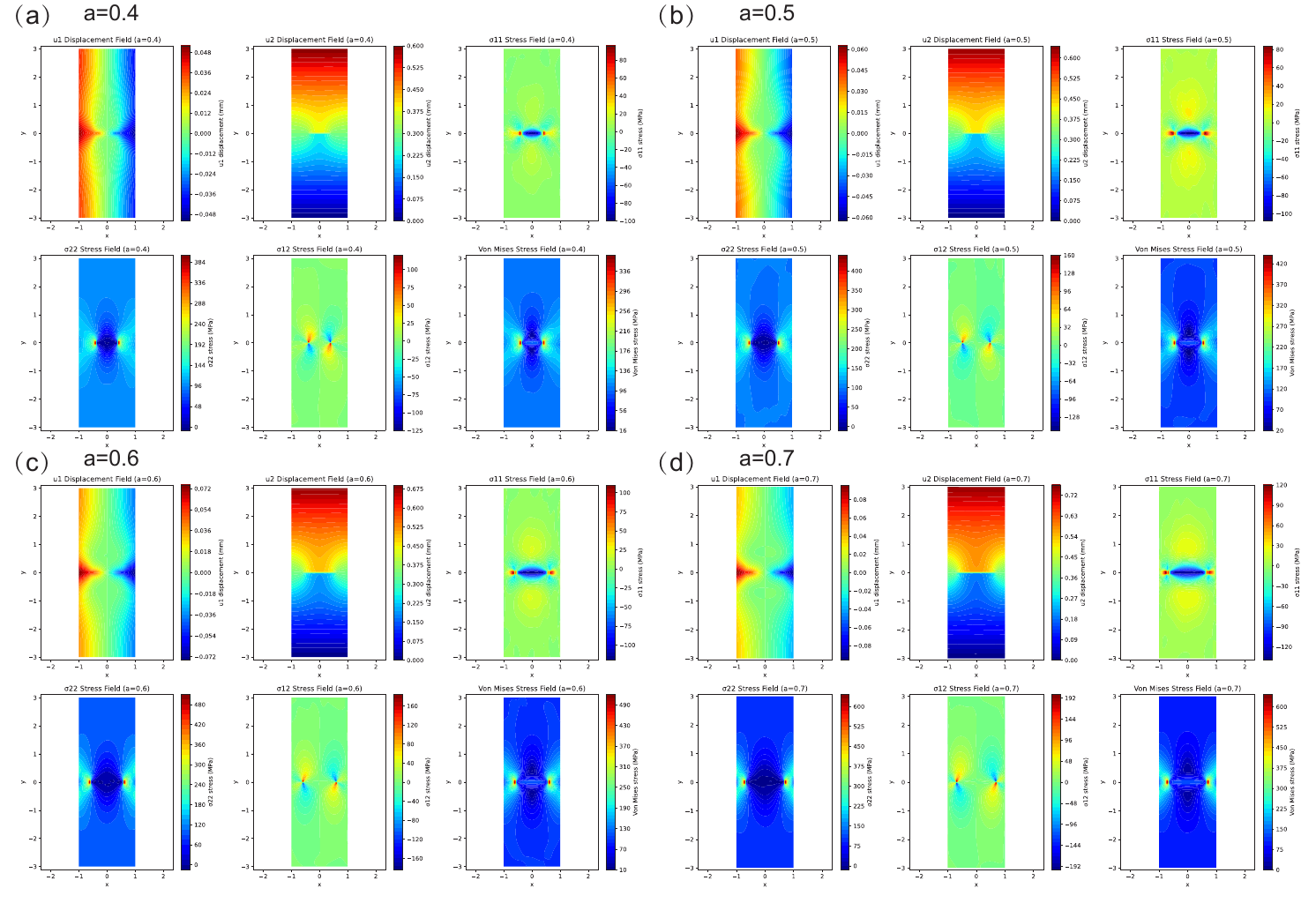}
		\par\end{centering}
	\caption{XDEM-predicted displacement and stress contours for different crack lengths in the mode I crack problem. \label{fig:crack1_contourf}}
\end{figure}

\Cref{fig:crack1_contourf} shows the displacement and stress contours predicted by XDEM for different crack lengths $a$. The discontinuity in $u_{y}$ and the stress concentration near the crack tips can be clearly observed. The training setup employs a multilayer perceptron (MLP) with 4 hidden layers and 30 neurons per layer, using the $\tanh$ activation function. The network input consists of the spatial coordinates $(x,y)$ together with the embedded crack function, and the outputs are the displacement components $(u,v)$. The optimizer is Adam with an initial learning rate of $0.001$. Training is performed for $15{,}000$ epochs with early stopping (patience = 10) to prevent overfitting. A step-wise learning rate scheduler reduces the learning rate by a factor of $0.5$ every $5{,}000$ epochs.  

For validation, we compare the predicted stress intensity factors against the reference solution from \citet{tada1973stress}:
\begin{equation}
	K_{I} = \sigma_{0}\sqrt{\pi a}\left[1-0.025\left(\frac{a}{b}\right)^{2}+0.06\left(\frac{a}{b}\right)^{4}\right]\sqrt{\sec\left(\frac{\pi a}{2b}\right)}.
	\label{eq:exact_SIF_crack1}
\end{equation}

\begin{figure}
	\begin{centering}
		\includegraphics[scale=0.70]{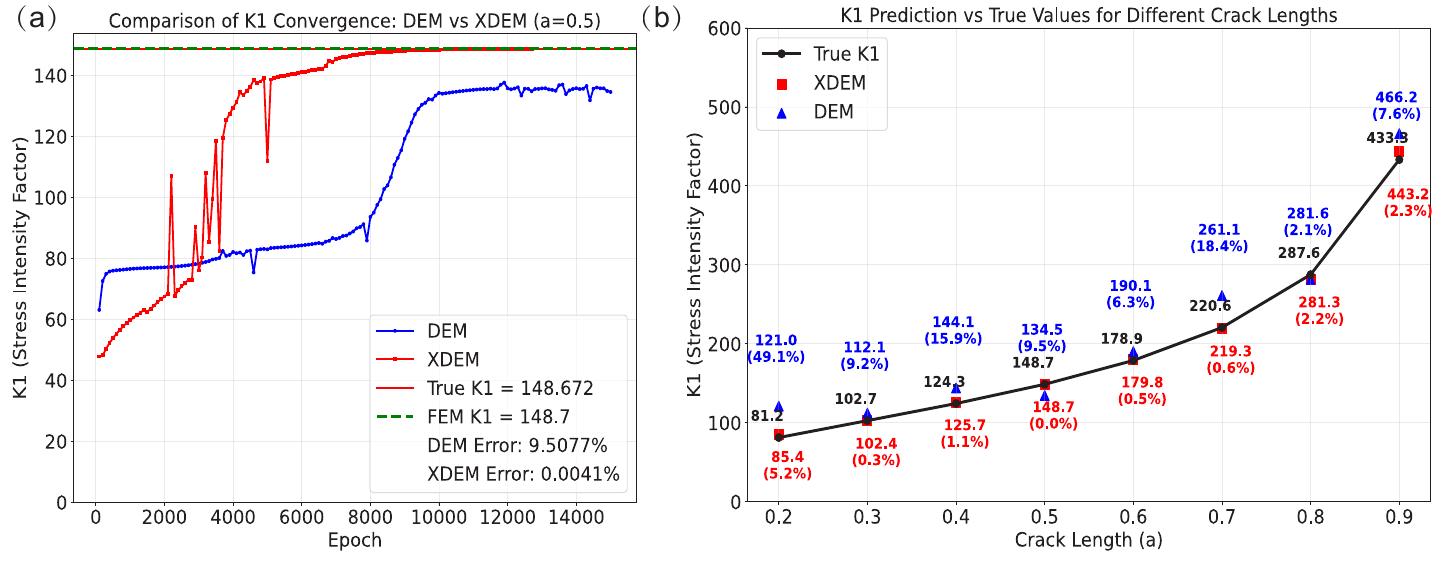}
		\par\end{centering}
	\caption{Comparison of stress intensity factors in the mode I crack problem: (a) convergence of SIFs with uniformly distributed $20\times 60$ points for XDEM and DEM without enrichment, with the reference value $K_{I}=148.672$; (b) XDEM and DEM predicted SIFs for different crack lengths with uniformly distributed $100\times 600$ points. The relative errors are in brackets. \label{fig:crack1_SIF}}
\end{figure}

\Cref{fig:crack1_SIF}a compares XDEM with the standard DEM. It is evident that, under the same collocation scheme, XDEM captures the SIF more accurately than DEM. Importantly, XDEM and DEM exhibit comparable efficiency since the extended function in XDEM is analytical and only requires learning a few additional coefficients. \Cref{fig:crack1_SIF}b shows the predictions of XDEM for different crack lengths, which agree very well with the analytical solution in \Cref{eq:exact_SIF_crack1}, confirming the accuracy of the displacement and stress fields in \Cref{fig:crack1_contourf}. The SIFs are computed using the interaction integral method, as described in \Cref{subsec:Calculation-of-sif}, with the $J$-integral taken along a circular contour centered at the crack tip with radius $\min\left(\frac{a}{2},\frac{b-a}{2}\right)$:
\begin{equation}
	K_{I} = \sqrt{J\frac{E}{1-\nu^{2}}}.
\end{equation}

For comparison, FEM simulations were also performed. With $46{,}014$ CPE8 elements and $138{,}878$ nodes, the FEM-predicted SIF was $148.7$, requiring $35$ seconds of computation. By contrast, XDEM required only $20\times 60 = 1200$ collocation points and $1000$ training epochs, taking $14$ seconds to achieve comparable accuracy. Thus, XDEM, similar in spirit to XFEM, effectively captures displacement discontinuities through the crack function and enhances the representation of the crack-tip fields with the extended function.  

\subsubsection{Mode II crack}

\begin{figure}
	\begin{centering}
		\includegraphics[scale=0.35]{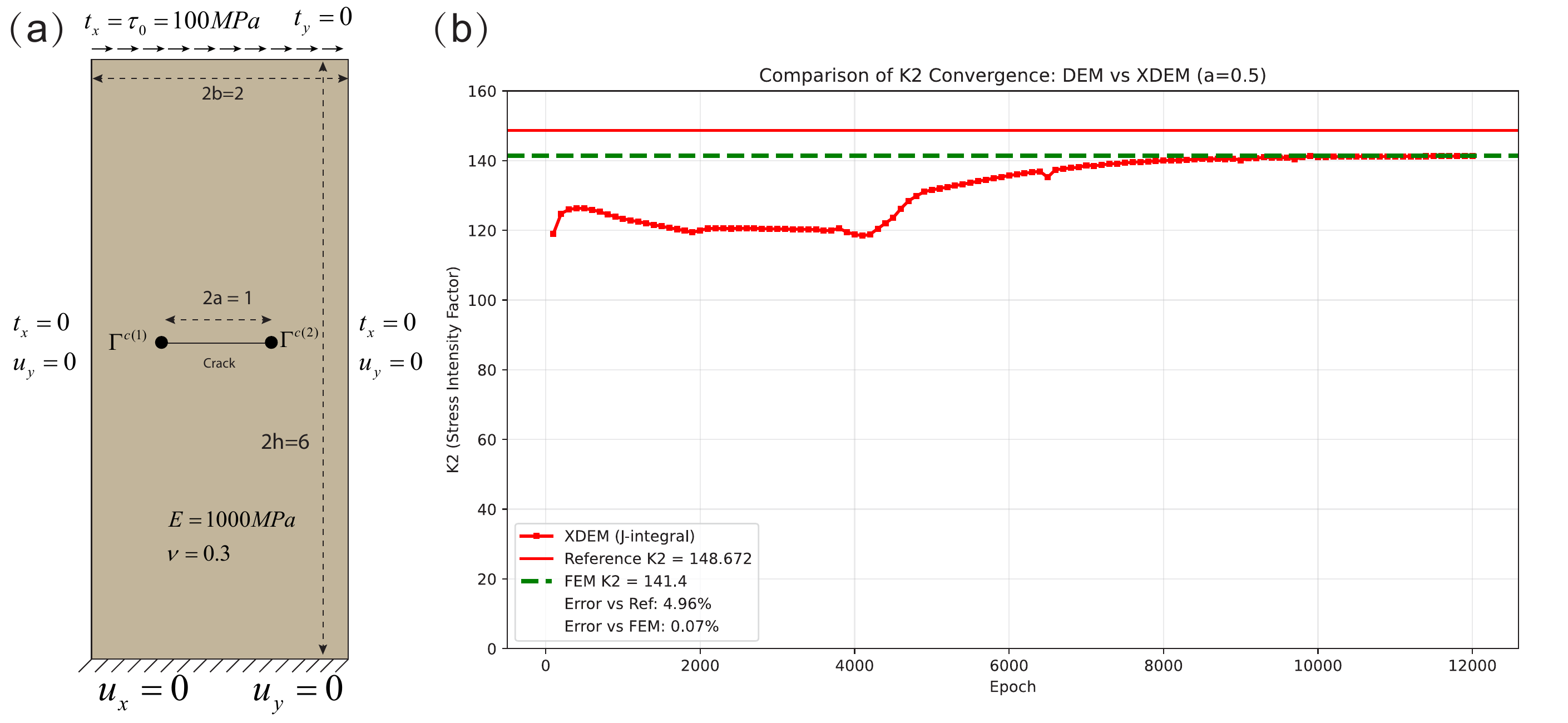}
		\par\end{centering}
	\caption{Mode II crack problem: (a) geometry, material properties, and boundary conditions; (b) comparison of stress intensity factors predicted by XDEM, FEM, and the reference solution for $a=0.5$ and $\tau_{0}=100\,\text{MPa}$, where the SIF is computed using the $J$-integral method. \label{fig:Mode-II-crack_schematic}}
\end{figure}

We next consider the standard mode II crack problem. The geometry consists of a plate with crack length $2a=1$, centered at the middle of the structure. The overall dimensions are $2b=2$ in length and $2h=6$ in height. The material properties are $E=1000\,\text{MPa}$ and Poisson's ratio $\nu=0.3$ under plane strain conditions. A uniform shear stress $\tau_{0}=100\,\text{MPa}$ is applied on the top boundary. All boundaries are constrained in the $y$-direction ($u_{y}=0$), and the bottom boundary is further constrained in the $x$-direction ($u_{x}=0$). The schematic of the mode II crack problem is shown in \Cref{fig:Mode-II-crack_schematic}a. The displacement fields in XDEM are expressed as:
\begin{equation}
	\begin{aligned}
		u_{1}(\boldsymbol{x},\varrho;\boldsymbol{\theta}_{\boldsymbol{u}}) &= \frac{y+h}{2h}\Big[NN_{x}(\boldsymbol{x},\varrho;\boldsymbol{\theta}_{\boldsymbol{u}})+\sum_{i=2}T(\boldsymbol{x};\Gamma^{c(i)}) \cdot X_{1}(\boldsymbol{x};\Gamma^{c(i)})\Big], \\
		u_{2}(\boldsymbol{x},\varrho;\boldsymbol{\theta}_{\boldsymbol{u}}) &= \frac{h+y}{2h}\cdot\frac{h-y}{2h}\cdot\frac{b+x}{2b}\cdot\frac{b-x}{2b}\Big[NN_{y}(\boldsymbol{x},\varrho;\boldsymbol{\theta}_{\boldsymbol{u}})+\sum_{i=2}T(\boldsymbol{x};\Gamma^{c(i)}) \cdot X_{2}(\boldsymbol{x};\Gamma^{c(i)})\Big].
	\end{aligned}
	\label{eq:crack2_dis}
\end{equation}

The training setup is consistent with that of the mode I crack problem. In this example, we use $100\times 600$ uniformly distributed collocation points. \Cref{fig:Mode-II-crack_schematic}b demonstrates that XDEM provides accurate predictions of the stress intensity factor, highlighting its effectiveness. The reference SIF is given by:
\begin{equation}
	K_{II}=\tau_{0}\sqrt{\pi a}\left[1-0.025\left(\frac{a}{b}\right)^{2}+0.06\left(\frac{a}{b}\right)^{4}\right]\sqrt{\sec\left(\frac{\pi a}{2b}\right)}.
	\label{eq:exact_SIF_crack2}
\end{equation}
It is worth noting that, due to the boundary conditions not being purely shear loading, the analytical expression in \Cref{eq:exact_SIF_crack2} yields a small deviation. Therefore, the SIF predicted by XDEM is evaluated using the $J$-integral:
\begin{equation}
	K_{II} = \sqrt{J\frac{E}{1-\nu^{2}}}.
\end{equation}

\begin{figure}
	\begin{centering}
		\includegraphics[scale=0.35]{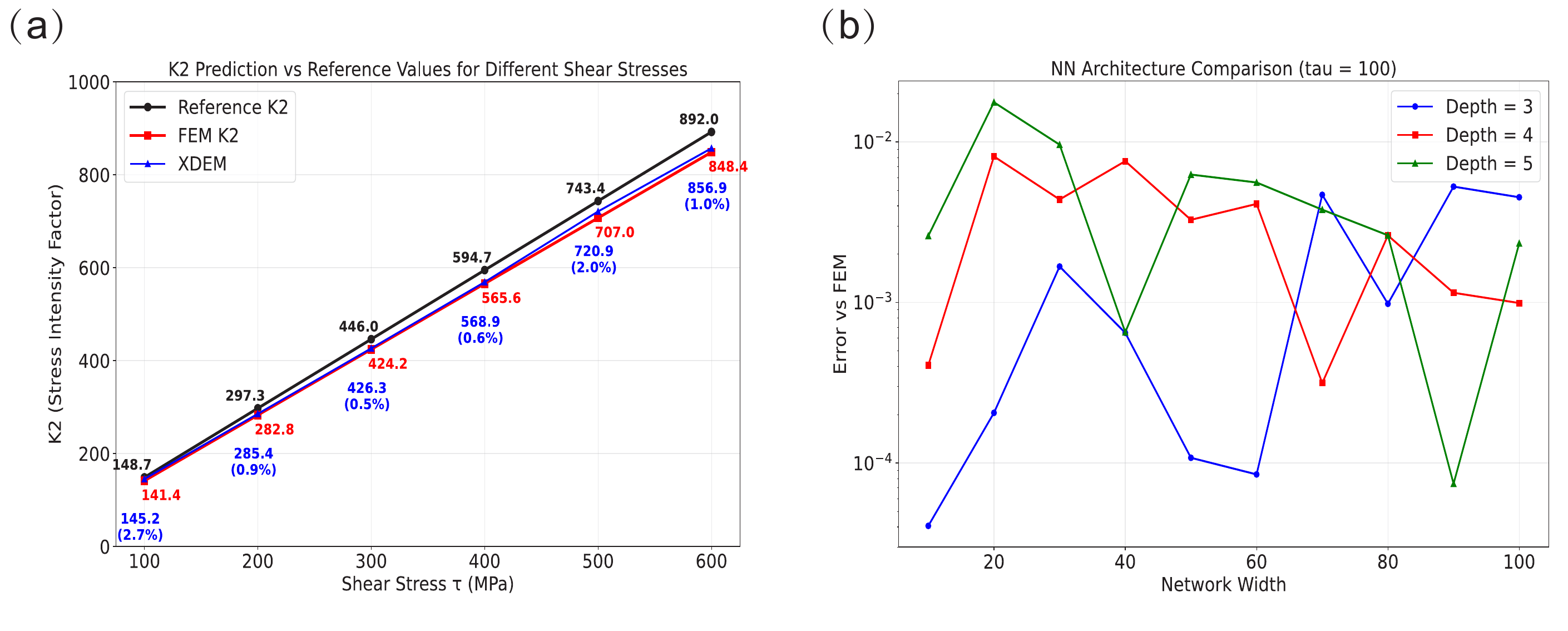}
		\par\end{centering}
	\caption{Performance of XDEM for the mode II crack problem: (a) predicted SIFs for different load levels with $a=0.5$ (trained for 10,000 steps); (b) effect of neural network architecture, with errors evaluated against the FEM reference solution. \label{fig:crack2_tau_nn}}
\end{figure}

\begin{figure}
	\begin{centering}
		\includegraphics[scale=0.65]{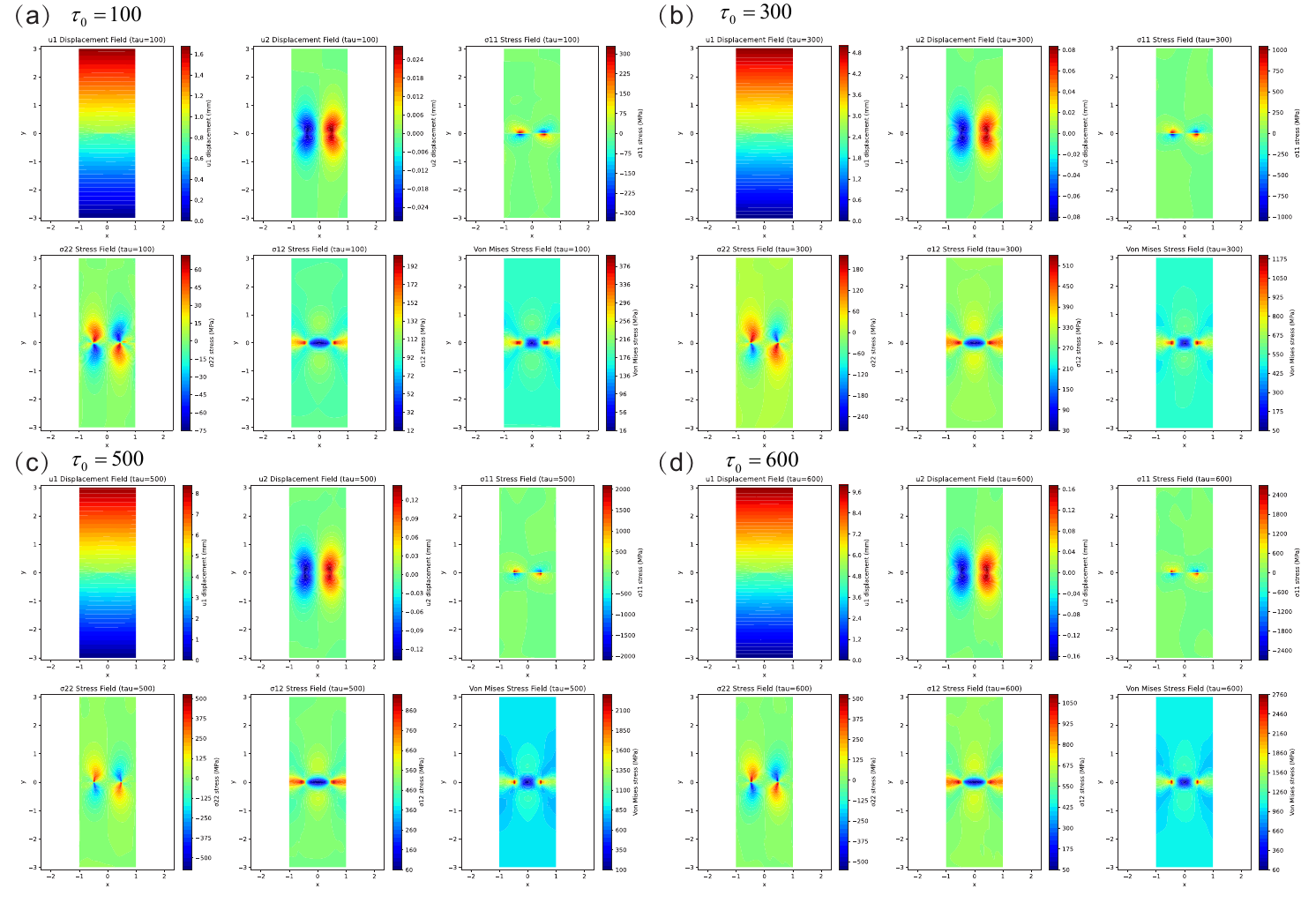}
		\par\end{centering}
	\caption{Displacement and stress contours predicted by XDEM under different load levels for the mode II crack problem. \label{fig:crack2_contourf}}
\end{figure}

For reference, we also compute the SIF using FEM, which serves as a reliable benchmark after convergence analysis. The FEM model consists of $174{,}633$ CPE8 elements and $525{,}787$ nodes, with a total runtime of $61$ seconds. \Cref{fig:crack2_tau_nn}a shows the XDEM predictions of $K_{II}$ under different loading conditions, while \Cref{fig:crack2_contourf} presents the corresponding displacement and stress contours, clearly exhibiting the discontinuity in $u_{y}$ and stress concentration near the crack tips. \Cref{fig:crack2_tau_nn}b further shows the effect of network architecture: as long as the neural network exceeds a certain threshold in complexity, XDEM is able to accurately predict the SIF for mode II cracks. These results confirm that XDEM delivers robust and accurate performance for mode II fracture problems across different loads and network configurations.  

\subsubsection{Mode III crack}
Next, we analyze the performance of XDEM for a Mode III crack. 
Mode III is an anti-plane shear problem in which the displacement field is out of the plane. The governing PDE simplifies to the Poisson equation: $\nabla^{2} u(\boldsymbol{x}) = 0$.

In the previous problems, the results were verified using reference solutions such as FEM rather than analytical solutions. Therefore, here we employ the analytical solution:
\begin{equation}
	w=u_{3}=\frac{K_{3}}{G}\sqrt{\frac{2r}{\pi}}\sin\left(\frac{\theta}{2}\right),
	\label{eq:exact_mode_3}
\end{equation}
where $K_{3}=1~\mathrm{MPa}\sqrt{\mathrm{mm}}$ is the Mode III stress intensity factor, and $G=1~\mathrm{MPa}$ is the shear modulus. The displacement boundary conditions defined by \Cref{eq:exact_mode_3} are imposed on $x=\pm1$ and $y=\pm1$, while traction-free conditions are applied along the crack faces. The admissible displacement field in XDEM is expressed as
\begin{equation}
	\begin{aligned}
		w(\boldsymbol{x},\varrho;\boldsymbol{\theta}_{\boldsymbol{u}}) &=
		u_{p}(\boldsymbol{x})+
		\left(\frac{1+y}{2}\right)
		\left(\frac{1-y}{2}\right)
		\left(\frac{1+x}{2}\right)
		\left(\frac{1-x}{2}\right)
		\left[ NN(\boldsymbol{x},\varrho;\boldsymbol{\theta}_{\boldsymbol{u}}) + 
		T(\boldsymbol{x};\Gamma^{ct})X_{3}(\boldsymbol{x};\Gamma^{ct}) \right],
	\end{aligned}
	\label{eq:crack3_dis}
\end{equation}
where $X_{3}$ denotes the extended function, and $u_{p}(\boldsymbol{x})$ is the particular solution satisfying $u_{p}(\boldsymbol{x})=u_{3}$ on the essential boundaries.  
The total loss function of XDEM is formulated as
\begin{equation}
	\mathcal{L}_{\mathrm{XDEM}}=\int_{\Omega}\frac{1}{2}(\nabla w)\cdot(\nabla w)\,d\Omega.
\end{equation}
The neural network architecture and training setup are identical to those used in the Mode I and II crack problems.

\begin{figure}
	\begin{centering}
		\includegraphics[scale=0.7]{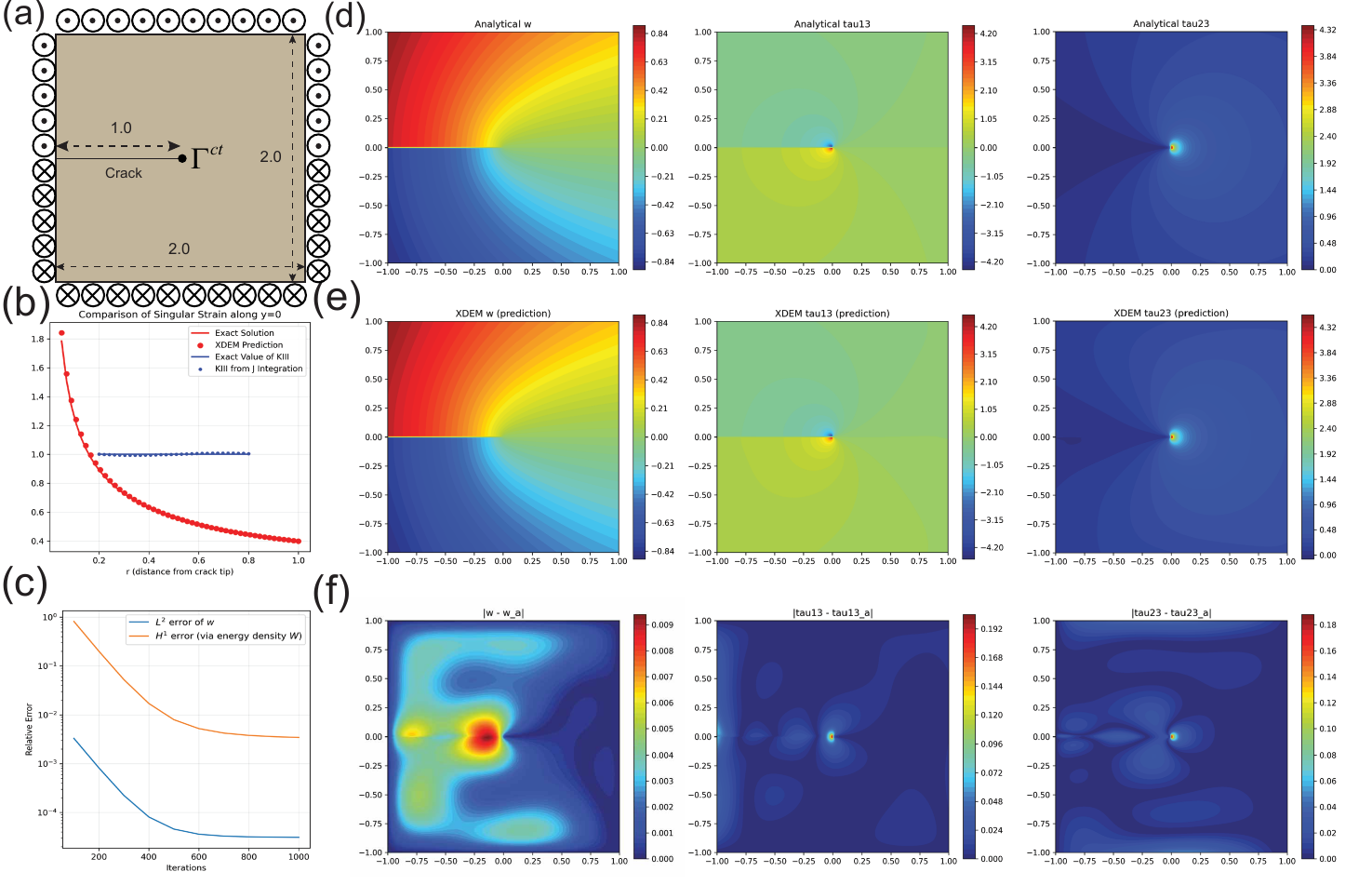}
		\par\end{centering}
	\caption{Performance of XDEM for the Mode III crack: (a) schematic of the mode III. This is an anti-plane shear problem where the displacement is perpendicular to the $x$-$y$ plane. Symbols with a cross ($\otimes$) indicate displacements directed into the plane, while dots ($\odot$) indicate displacements directed out of the plane. $K_{3}=1~\mathrm{MPa}\sqrt{\mathrm{mm}}$, $G=1~\mathrm{MPa}$, specimen dimensions of $2~\mathrm{mm}\times2~\mathrm{mm}$, and crack length of $1~\mathrm{mm}$; (b) comparison between the analytical and XDEM-predicted singular strain ($x>0$, $y=0$) and the stress intensity factors computed from the $J$-integral at various contour radii; (c) evolution of $\mathcal{L}_{2}$ and $\mathcal{H}_{1}$ errors during iterations; (d-f) analytical solution, XDEM prediction, and absolute error contour, respectively.}
	\label{fig:XDEM_mode3}
\end{figure}

\Cref{fig:XDEM_mode3}(b) shows the stress intensity factors and singular strain predictions obtained by XDEM at different radii. The analytical expression of the singular strain is given by
\begin{equation}
	\varepsilon_{z\theta}\big|_{\mathrm{interface}}=
	\frac{K_{3}}{G}\sqrt{\frac{2}{\pi}}\frac{1}{r}\frac{\partial w}{\partial\theta}
	=\frac{1}{2\sqrt{r}}\cos\!\left(\frac{\theta}{2}\right)\bigg|_{\theta=0}
	=\frac{K_{3}}{G}\frac{1}{\sqrt{2\pi r}}.
\end{equation}
It can be observed that XDEM accurately captures the singular strain near the crack tip. Furthermore, we evaluate the accuracy of the stress intensity factors computed from the $J$-integral at different contour radii.  
\Cref{fig:XDEM_mode3}(c) presents the evolution of $\mathcal{L}_{2}$ and $\mathcal{H}_{1}$ errors during training, where $\mathcal{L}_{2}$ measures the $L^{2}$ norm of displacement errors, and $\mathcal{H}_{1}$ represents the $H^{1}$ norm of the strain energy density error. Rapid convergence of both metrics confirms the efficiency of XDEM.  
\Cref{fig:XDEM_mode3}(d-f) illustrate the analytical solution, XDEM prediction, and absolute error contour, respectively.

\subsubsection{Mixed-mode crack}

The displacement fields in XDEM are expressed as:
\begin{equation}
	\begin{aligned}
		u_{1}(\boldsymbol{x},\varrho;\boldsymbol{\theta}_{\boldsymbol{u}}) &= \frac{b+x}{2b}\cdot\frac{b-x}{2b}\Big[NN_{x}(\boldsymbol{x},\varrho;\boldsymbol{\theta}_{\boldsymbol{u}}) + \sum_{i=2}T(\boldsymbol{x};\Gamma^{c(i)}) \cdot X_{1}(\boldsymbol{x};\Gamma^{c(i)})\Big], \\
		u_{2}(\boldsymbol{x},\varrho;\boldsymbol{\theta}_{\boldsymbol{u}}) &= \frac{h+y}{2h}\Big[NN_{y}(\boldsymbol{x},\varrho;\boldsymbol{\theta}_{\boldsymbol{u}}) + \sum_{i=2}T(\boldsymbol{x};\Gamma^{c(i)}) \cdot X_{2}(\boldsymbol{x};\Gamma^{c(i)})\Big].
	\end{aligned}
	\label{eq:crack_mixed_dis}
\end{equation}

\subsection{Crack propagation\label{subsec:SI_Crack-propagation}}

In \Cref{subsec:SI_Stress-intensity-factor}, we demonstrated that XDEM can accurately predict stress intensity factors (SIFs), laying the groundwork for its application to crack growth problems. In this section, we further validate XDEM for simulating crack propagation, covering three classical scenarios: straight crack growth, kinking, inclusion, and crack initiation.

\subsubsection{Straight crack propagation}

\begin{figure}
	\begin{centering}
		\includegraphics[scale=0.35]{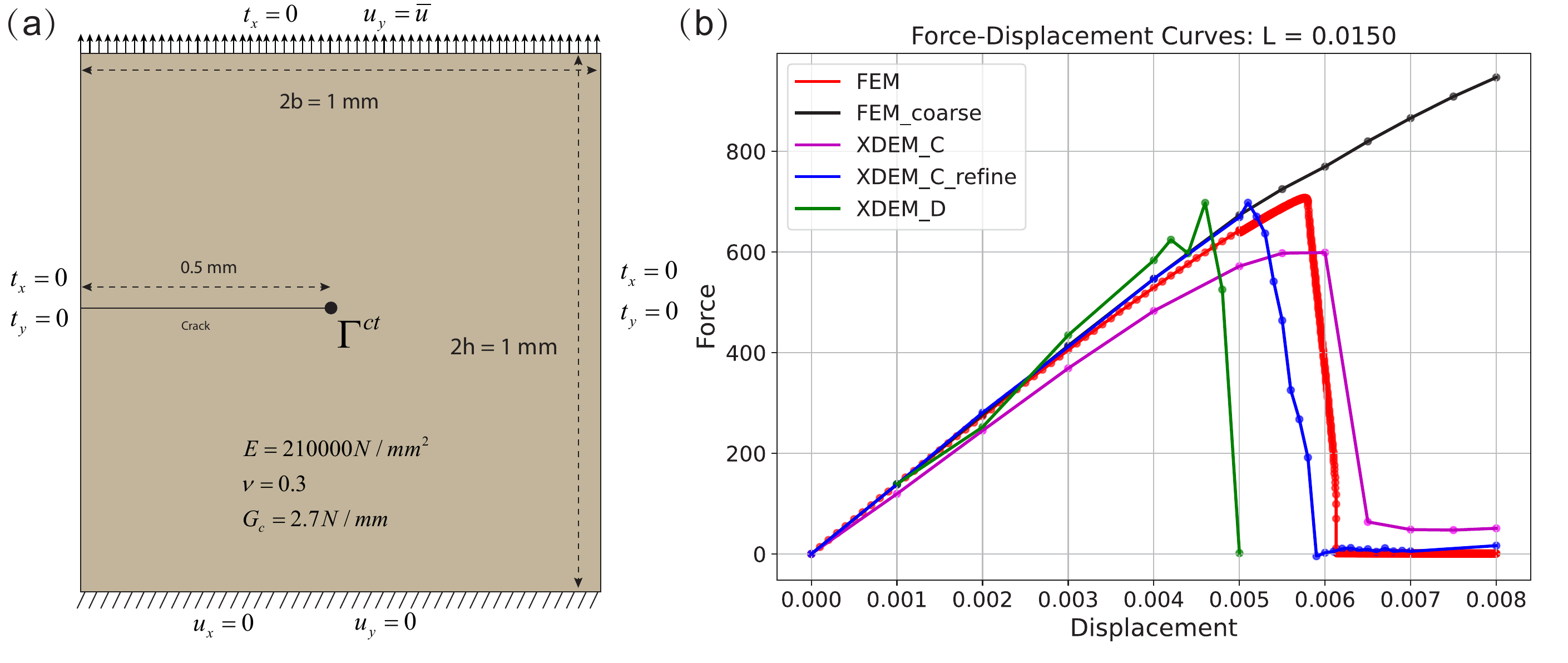}
		\par\end{centering}
	\caption{Single-edge notched specimen under tensile loading: (a) Mode-I (opening) crack geometry; (b) load--displacement curves predicted by XDEM. \label{fig:Single-edge-notched_tension}}
\end{figure}

The single-edge notched tension (SENT) specimen is a classical benchmark in fracture mechanics and is widely used for method validation. \Cref{fig:Single-edge-notched_tension}a shows the geometry and material settings for the mode-I crack, where the crack propagates in a straight path. Both the discrete and continuous variants of XDEM are capable of simulating crack growth for this problem. The displacement field used in XDEM is:
\begin{equation}
	\begin{aligned}
		u_{1}(\boldsymbol{x},\varrho;\boldsymbol{\theta}_{\boldsymbol{u}}) &= \frac{h+y}{2h}\Big[NN_{x}(\boldsymbol{x},\varrho;\boldsymbol{\theta}_{\boldsymbol{u}})+T(\boldsymbol{x};\Gamma^{ct})\,X_{1}(\boldsymbol{x};\Gamma^{ct})\Big], \\
		u_{2}(\boldsymbol{x},\varrho;\boldsymbol{\theta}_{\boldsymbol{u}}) &= \frac{h+y}{2h}\cdot\frac{h-y}{2h}\Big[NN_{y}(\boldsymbol{x},\varrho;\boldsymbol{\theta}_{\boldsymbol{u}})+T(\boldsymbol{x};\Gamma^{ct})\,X_{2}(\boldsymbol{x};\Gamma^{ct})\Big] + \frac{h+y}{2h}\,\bar{u},
	\end{aligned}
	\label{eq:crack_propagation_dis}
\end{equation}
where $\Gamma^{ct}$ denotes the crack tip and $\bar{u}$ is the prescribed displacement load.

For the discrete model (XDEM-D), the loading is divided into increments. At each increment, we compute the $J$-integral at the current crack tip. If $J>G_{c}$, we evaluate $K_{I}$ and $K_{II}$ and determine the direction $\theta_{c}$ of maximum circumferential (hoop) stress in the local polar coordinate system (centered at the crack tip with $\theta=0$ aligned with the crack tangent); see \Cref{subsec:Calculation-of-sif} for details. The crack is then advanced by a small prescribed length $\delta a=0.05$ along $\theta_{c}$. After extension by $\delta a$, $K_{I}$ and $K_{II}$ are recomputed and the condition $J\lessgtr G_{c}$ is rechecked, repeating the process until $J\le G_{c}$, at which point the algorithm proceeds to the next load step. The network architecture and training settings follow those in \Cref{subsec:SI_Mode-I-crack}, with $5000$ training iterations per load increment.  

For the continuous model (XDEM-C), the loading is likewise partitioned into increments and the variational problem \Cref{eq:variational_principle_DEM} is minimized at each step. The phase-field irreversibility is enforced using a history-field approach. The displacement field is approximated by a KAN with architecture $[2,5,5,5,2]$, while the phase field is represented by an RBF network with architecture $[2,1000,1]$. We adopt a monolithic optimization strategy: the first load step is trained with $3000$ Adam iterations, and subsequent steps with $1000$ Adam iterations. In XDEM-C we use the AT2 model, $w(\phi)=(1-\phi)^2$, and the history-field enforcement of irreversibility.

\Cref{fig:Single-edge-notched_tension}b shows the load--displacement curves predicted by XDEM. As a reference, we employ a phase-field FEM solution with length-scale $l_{0}=0.015$. Specifically, a user-defined element (UEL) implementation in \textsc{abaqus} is used with mesh refinement near the crack tip; see \Cref{subsec:FEM_UEL_detail} for details. XDEM accurately reproduces the load--displacement response using only $100\times 100$ uniformly distributed collocation points, without the tip-focused mesh refinement required by FEM. Moreover, XDEM permits larger load increments. When FEM uses the same coarse increment as XDEM (see the ``FEM\_coarse'' and ``XDEM\_C'' curves in \Cref{fig:Single-edge-notched_tension}b), FEM fails to capture the correct response.  

\begin{figure}
	\begin{centering}
		\includegraphics[scale=0.27]{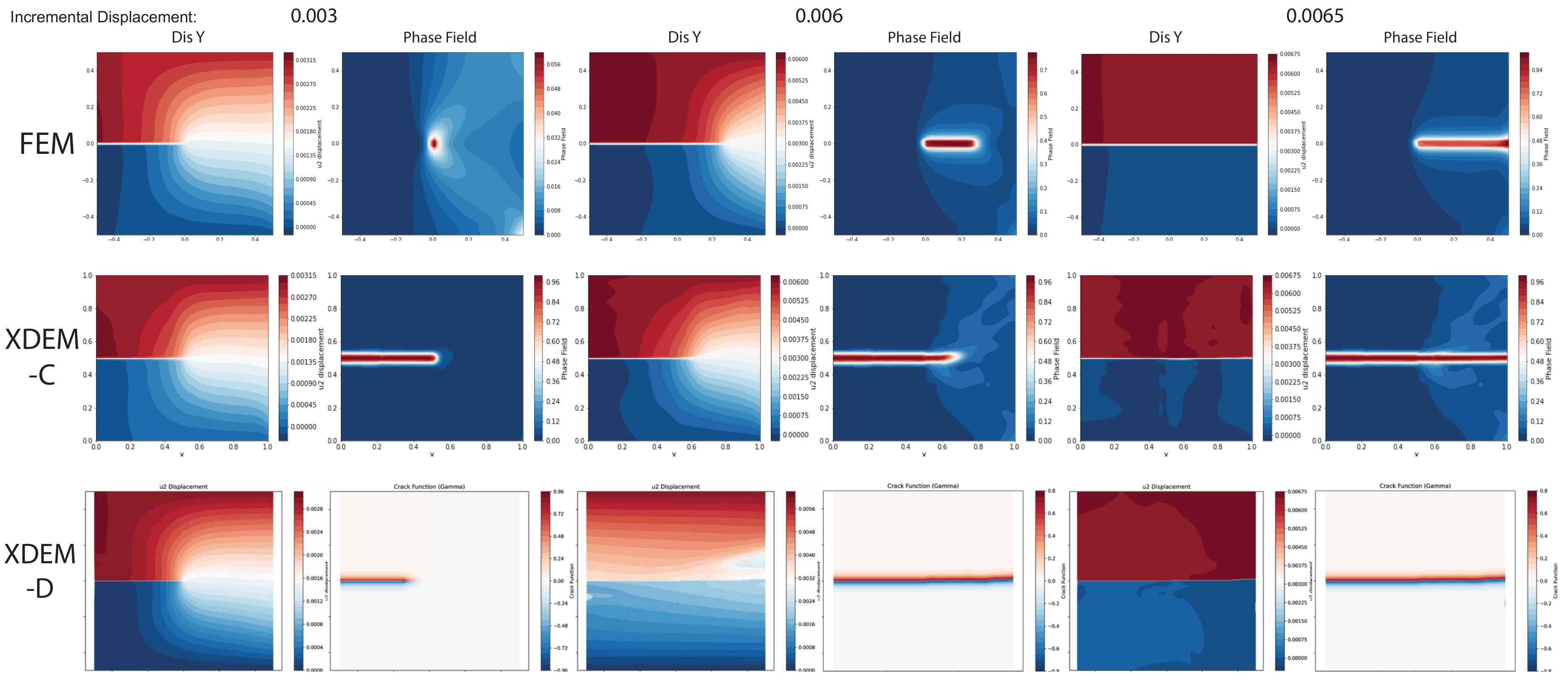}
		\par\end{centering}
	\caption{Displacement and phase-field contours for the SENT test: the first and second columns show $u_y$ and the phase field at $\bar{u}=0.003\,\mathrm{mm}$; the third and fourth columns at $\bar{u}=0.006\,\mathrm{mm}$; and the fifth and sixth columns at $\bar{u}=0.0065\,\mathrm{mm}$. Rows (top to bottom): FEM, XDEM-C, and XDEM-D (where the crack function replaces the phase field for crack visualization). \label{fig:single_crack_tension_phase_contourf}}
\end{figure}

\Cref{fig:single_crack_tension_phase_contourf} presents contour plots of the displacement and phase fields predicted by XDEM, demonstrating accurate field reconstruction. Note that, in XDEM-D, the crack function serves as an indicator of the crack location: discontinuities in the crack function correspond to the crack surface.  

\begin{figure}
	\begin{centering}
		\includegraphics[scale=0.7]{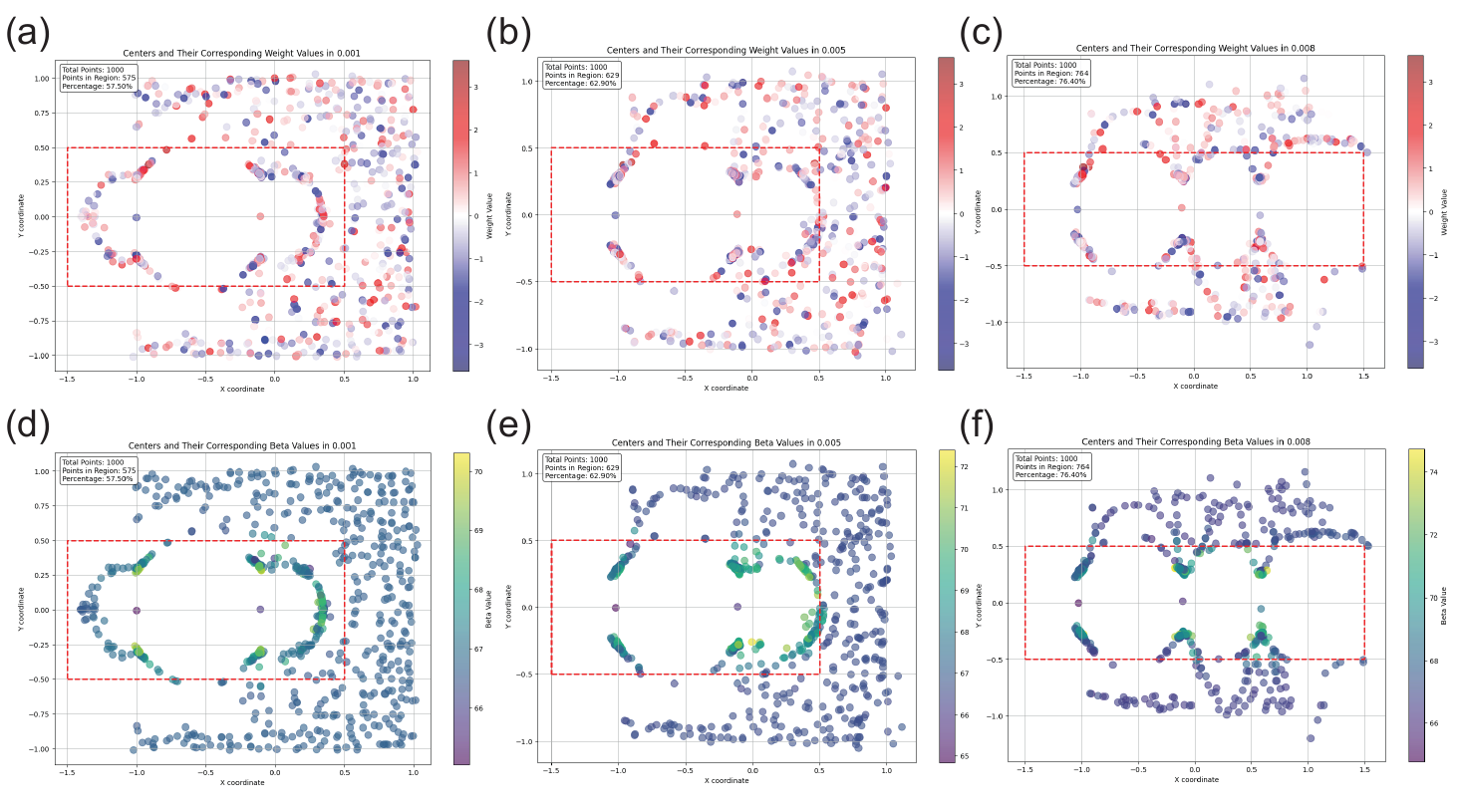}
		\par\end{centering}
	\caption{Locations of RBF centers in XDEM-C at three load steps ($\bar{u}=0.001$, $0.005$, and $0.008$). The first and second rows show the distributions of $w_{i}$ and $\beta_{i}$, respectively. \label{fig:XDEM-C_single_RBF_centers}}
\end{figure}

Since XDEM-C employs an RBF network for the phase field, the centers $\boldsymbol{c}_{i}$ adaptively concentrate near the crack, as shown in \Cref{fig:XDEM-C_single_RBF_centers}. More than half of the centers move toward the vicinity of the crack where gradients are steep. The associated weights $w_{i}$ exhibit alternating signs, and most $\beta_{i}$ values cluster around $1/l_{0}$, with higher magnitudes near the crack. We emphasize that the centers $\boldsymbol{c}_{i}$ do not lie directly on the crack, but rather in its neighborhood. This is expected because the phase field is nearly saturated (smooth) both inside the crack and far away from it (approaching $1$ and $0$, respectively). In contrast, the transition zone features sharp variations, and the RBF centers naturally concentrate there to allocate more basis functions where they are most needed.

\begin{figure}
	\begin{centering}
		\includegraphics[scale=0.5]{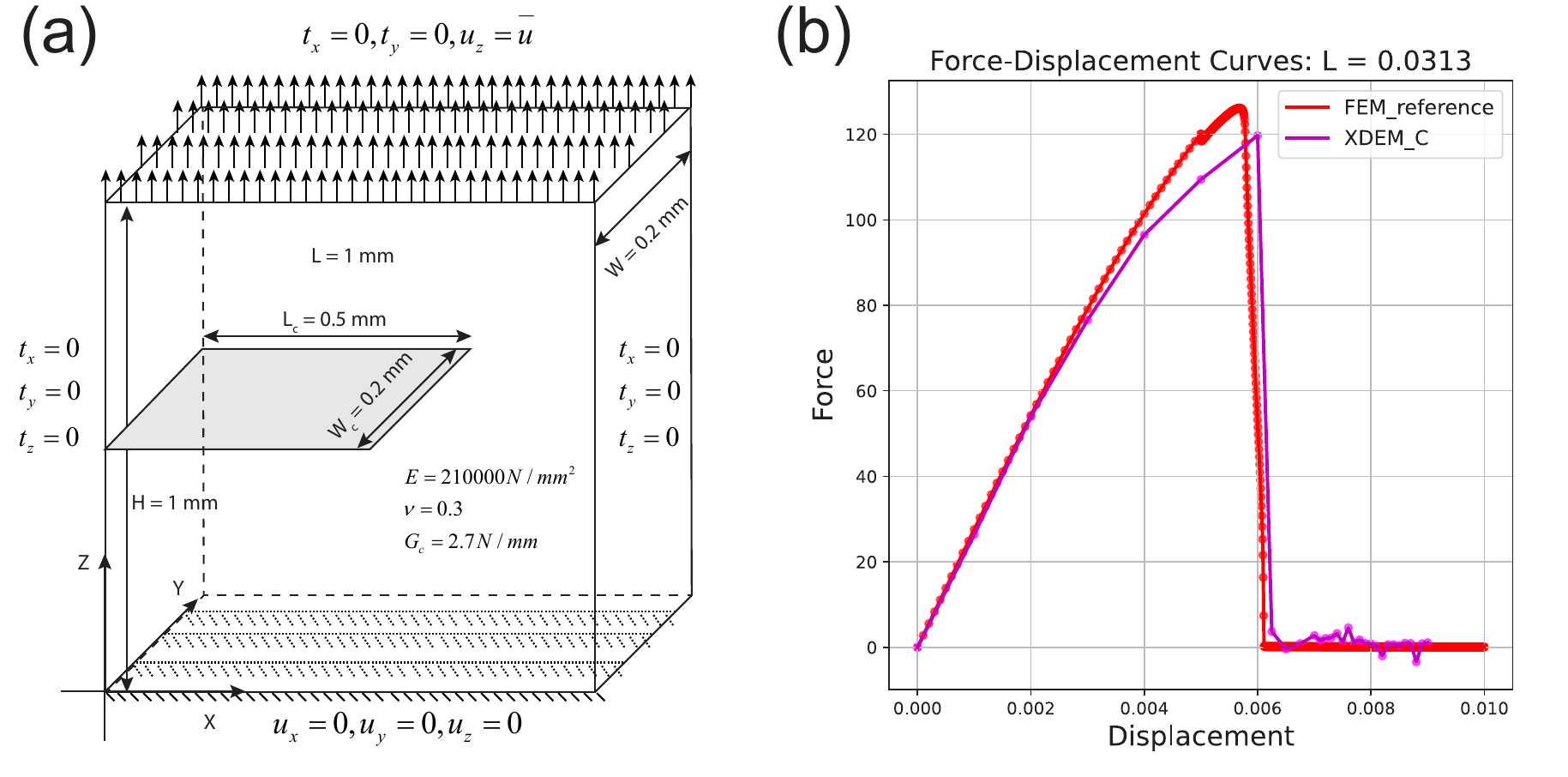}
		\par\end{centering}
	\caption{(a) Geometry of the 3D crack problem. (b) Load--displacement response predicted by XDEM. \label{fig:3D_crack}}
\end{figure}

XDEM in the discrete formulation faces challenges in tracking crack surfaces in three dimensions. Therefore, it is more natural and effective to employ the continuous formulation of XDEM for 3D crack problems. Here, we demonstrate the application of XDEM-C to the 3D crack configuration shown in \Cref{fig:3D_crack}a. The phase-field length scale is chosen as $l=0.0313$. Since the problem setup is essentially analogous to the 2D SENT test in \Cref{fig:Single-edge-notched_tension}, we use the 2D FEM phase-field solution as a reference. The load--displacement response is shown in \Cref{fig:3D_crack}b. It is worth noting that conventional DEM can only solve this problem if collocation points are heavily concentrated around the crack; with uniformly distributed points, DEM fails to converge. In contrast, XDEM achieves accurate results using far fewer, uniformly distributed collocation points.
Although the accuracy of XDEM-C still shows some discrepancy compared to traditional FEM, to the best of our knowledge, the 3D load-displacement curve is the best result available in the current literature.
We believe that XDEM-C still has room for improvement in the future, especially regarding the issue of overcoming the energy multiminima barrier in phase-field fracture, which is a key challenge for the future breakthroughs of XDEM-C.

In XDEM, collocation points are distributed uniformly: $70$ points along the $x$- and $z$-directions, and $8$ points along the $y$-direction. The displacement field is represented by a KAN network with architecture $[3,5,5,5,3]$, while the phase field is approximated by an RBF network with architecture $[3,2000,1]$. For the phase-field distribution, collocation points are placed uniformly with $20$ in $x$, $20$ in $z$, and $5$ in $y$, giving a total of $20\times20\times5=2000$ points. A monolithic optimization strategy is adopted: the first load step is trained for $3000$ Adam iterations, and subsequent steps for $1000$ iterations each. Transfer learning is employed using LoRA with rank $r=1$.

Since the energy principle requires that trial displacement fields satisfy essential boundary conditions a priori, the displacement field is constructed as
\begin{equation}
	\begin{aligned}
		u_{x}(x,y,z;\boldsymbol{\theta}_{\boldsymbol{u}}) &= NN_{x}(x,y,z;\boldsymbol{\theta}_{\boldsymbol{u}})\,z, \\
		u_{y}(x,y,z;\boldsymbol{\theta}_{\boldsymbol{u}}) &= NN_{y}(x,y,z;\boldsymbol{\theta}_{\boldsymbol{u}})\,z, \\
		u_{z}(x,y,z;\boldsymbol{\theta}_{\boldsymbol{u}}) &= NN_{z}(x,y,z;\boldsymbol{\theta}_{\boldsymbol{u}})\,z(1-z) + z\,\bar{u}.
	\end{aligned}
\end{equation}

\begin{figure}
	\begin{centering}
		\includegraphics[scale=0.27]{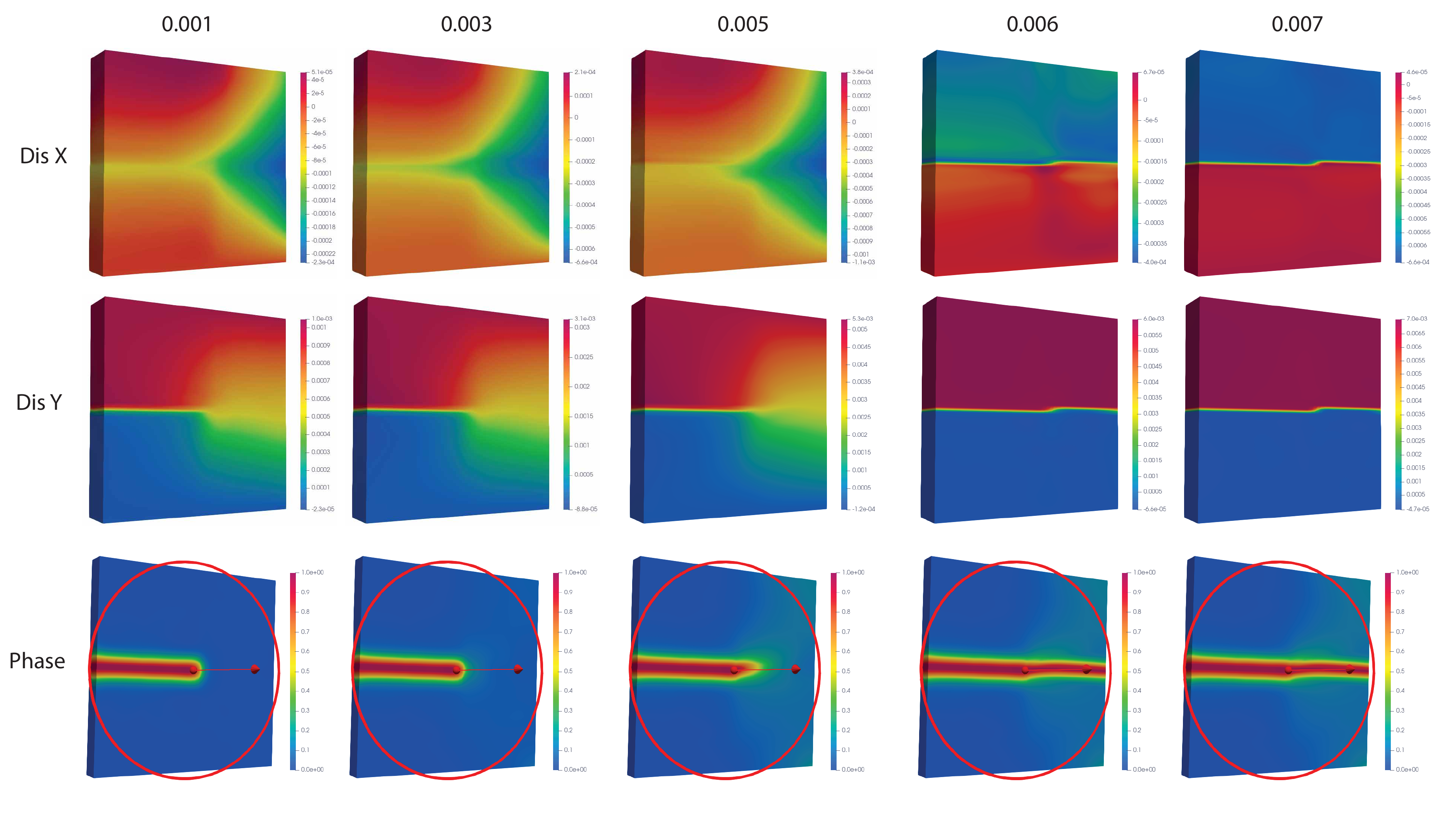}
		\par\end{centering}
	\caption{XDEM predictions of displacement and phase fields in the 3D crack problem. Columns 1--6 correspond to applied displacements $\bar{u}=0.001$, $0.003$, $0.005$, $0.006$, and $0.007$~mm, respectively. Rows (top to bottom) show $u_x$, $u_y$, and the phase field. \label{fig:3D_contourf}}
\end{figure}

\Cref{fig:3D_contourf} presents the predicted displacement and phase-field contours by XDEM. The results closely resemble those of the 2D case, further confirming the accuracy and robustness of the proposed framework in 3D fracture simulations.

\subsubsection{Crack kinking}

Next, we consider the crack kinking problem using XDEM. The benchmark
setup is the single-edge notched specimen under shear loading, where
the crack path deflects downward at approximately $70^{\circ}$, as
shown in \Cref{fig:Single-edge-notched_tension}a. The displacement
field is formulated as
\begin{equation}
	\begin{aligned}
		u_{1}(\boldsymbol{x},\varrho;\boldsymbol{\theta}_{\boldsymbol{u}}) & = \Big(\tfrac{h+y}{2h}\Big)\Big(\tfrac{h-y}{2h}\Big)\Big[NN_{x}(\boldsymbol{x},\varrho;\boldsymbol{\theta}_{\boldsymbol{u}}) 
		+ T(\boldsymbol{x};\Gamma^{ct})X_{1}(\boldsymbol{x};\Gamma^{ct})\Big] + \Big(\tfrac{h+y}{2h}\Big)\bar{u}, \\
		u_{2}(\boldsymbol{x},\varrho;\boldsymbol{\theta}_{\boldsymbol{u}}) & = \Big(\tfrac{h+y}{2h}\Big)\Big(\tfrac{h-y}{2h}\Big)\Big(\tfrac{b+x}{2b}\Big)\Big(\tfrac{b-x}{2b}\Big)\Big[NN_{y}(\boldsymbol{x},\varrho;\boldsymbol{\theta}_{\boldsymbol{u}}) 
		+ T(\boldsymbol{x};\Gamma^{ct})X_{2}(\boldsymbol{x};\Gamma^{ct})\Big],
	\end{aligned}
	\label{eq:crack_kinking_dis}
\end{equation}
where $\Gamma^{ct}$ denotes the crack tip and $\bar{u}$ is the applied displacement.

\begin{figure}
	\begin{centering}
		\includegraphics[scale=0.35]{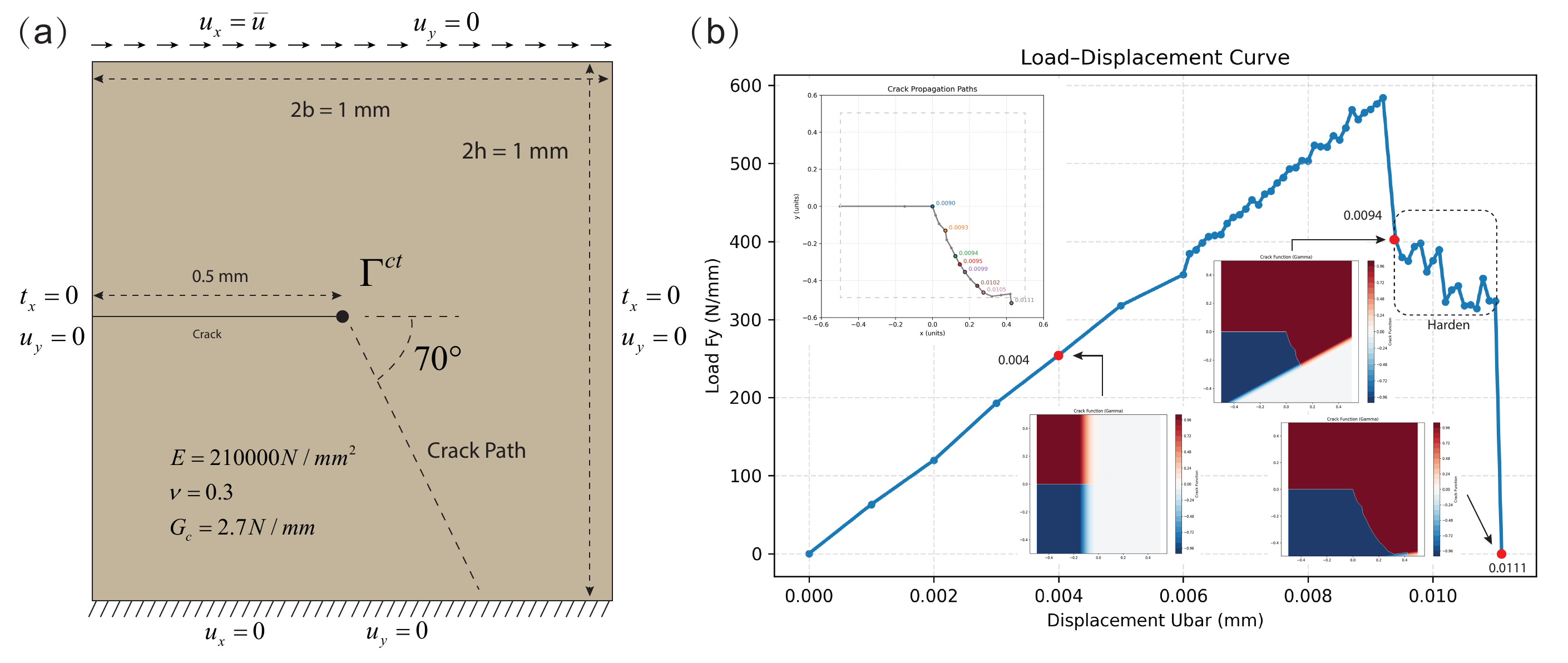}
		\par\end{centering}
	\caption{Single-edge notched specimen under shear loading: (a) Mode II crack in kinking configuration; (b) Load-displacement curve predicted by XDEM.\label{fig:Single-edge-notched_shear}}
\end{figure}

\begin{figure}
	\begin{centering}
		\includegraphics[scale=0.33]{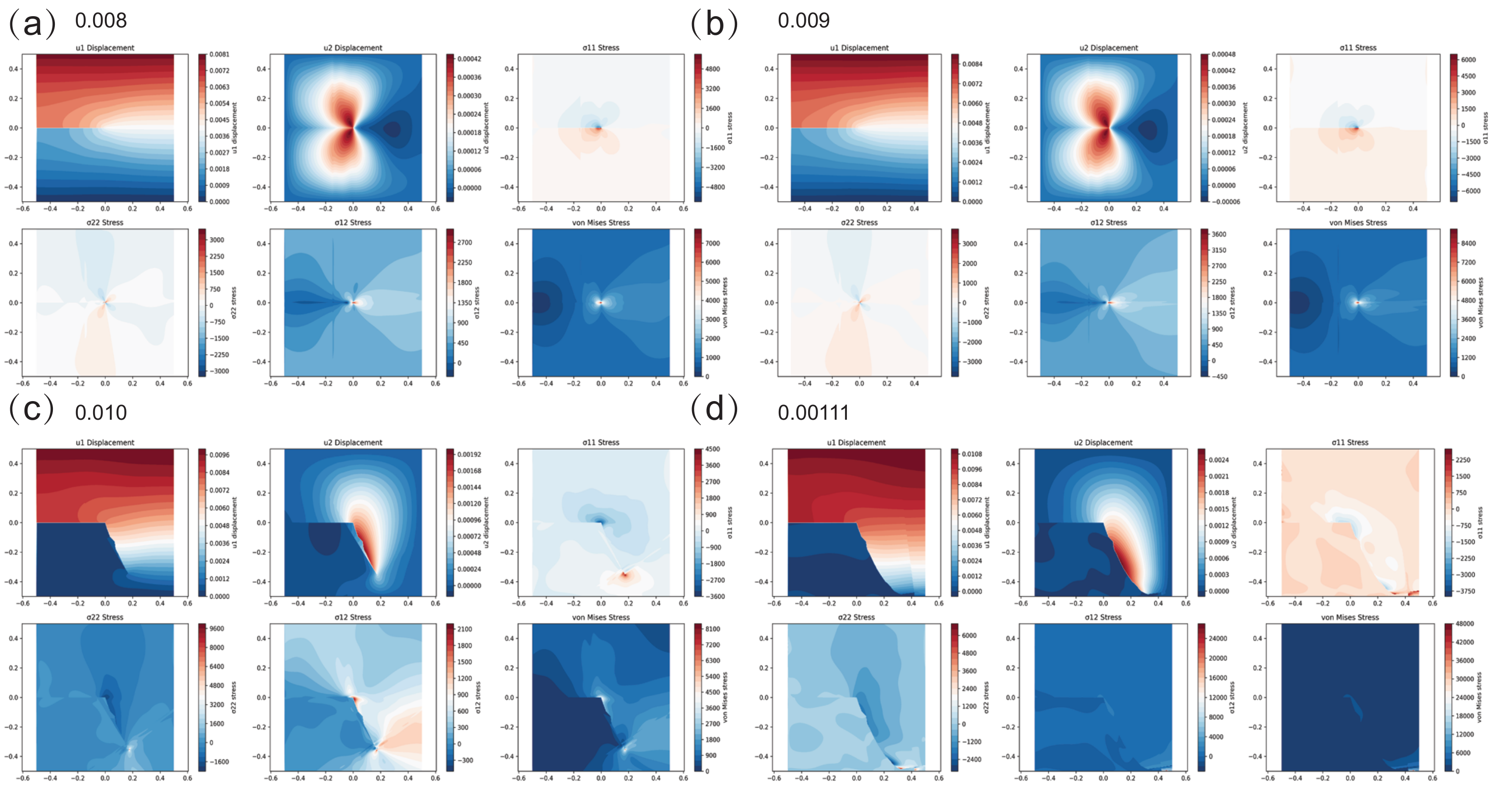}
		\par\end{centering}
	\caption{Displacement and stress contours predicted by XDEM for the single-edge notched specimen under shear loading at different displacement levels. The applied displacement $\bar{u}=0.00111$ corresponds to complete failure.\label{fig:shear_contourf}}
\end{figure}

In this case, collocation points are uniformly distributed with a resolution of $100\times 100$. The loading increments are $\Delta \bar{u}=0.001$ for the first six steps, followed by finer increments of $\Delta \bar{u}=0.0001$. The load-displacement curve predicted by XDEM is shown in \Cref{fig:Single-edge-notched_shear}b. We clearly observe a hardening stage, consistent with the reference results of \citet{goswami2020adaptiveCMAME}. \Cref{fig:Single-edge-notched_shear}b also presents the predicted crack propagation path and the corresponding crack function under different loading steps. To demonstrate that the crack function is not unique, we also tested an alternative embedding function that satisfies \Cref{eq:embedding_function}, incorporating an additional degradation near the crack front to better capture the interaction of multiple crack segments. The displacement and stress contours in \Cref{fig:shear_contourf} show distinct discontinuities across the crack and stress concentration at the crack tip.

The displacement field formulation in XDEM for the Bittencourt problem is given by
\begin{equation}
	\begin{aligned}
		u_{1}(\boldsymbol{x},\varrho;\boldsymbol{\theta}_{\boldsymbol{u}}) &= \frac{w_{2}}{w_{2}+1}\Big[NN_{x}(\boldsymbol{x},\varrho;\boldsymbol{\theta}_{\boldsymbol{u}}) 
		+ T(\boldsymbol{x};\Gamma^{ct})X_{1}(\boldsymbol{x};\Gamma^{ct})\Big], \\
		u_{2}(\boldsymbol{x},\varrho;\boldsymbol{\theta}_{\boldsymbol{u}}) &= \frac{w_{1}w_{2}w_{3}}{(w_{1}+1)(w_{2}+1)}\Big[NN_{y}(\boldsymbol{x},\varrho;\boldsymbol{\theta}_{\boldsymbol{u}}) 
		+ T(\boldsymbol{x};\Gamma^{ct})X_{2}(\boldsymbol{x};\Gamma^{ct})\Big] + \frac{y+4}{8}\bar{u}, \\
		w_{1} &= (9+x)^{2}+(4+y)^{2}, \\
		w_{2} &= (-9+x)^{2}+(4+y)^{2}, \\
		w_{3} &= x^{2}+(4-y)^{2},
	\end{aligned}
	\label{eq:crack_bittencourt_dis}
\end{equation}
where $\Gamma^{ct}$ denotes the crack tip and $\bar{u}$ is the applied displacement.

\subsubsection{Crack inclusion}

The displacement field is assumed as

\begin{equation}
	\begin{aligned}
		u_{1}(\boldsymbol{x},\varrho;\boldsymbol{\theta}_{\boldsymbol{u}}) & = \left(\frac{4+y}{8}\right) \left[ NN_{x}(\boldsymbol{x},\varrho;\boldsymbol{\theta}_{\boldsymbol{u}}) + T(\boldsymbol{x};\Gamma^{ct}) X_{1}(\boldsymbol{x};\Gamma^{ct}) \right], \\
		u_{2}(\boldsymbol{x},\varrho;\boldsymbol{\theta}_{\boldsymbol{u}}) & = \left(\frac{4+y}{8}\right) \left(\frac{4-y}{8}\right) \left[ NN_{y}(\boldsymbol{x},\varrho;\boldsymbol{\theta}_{\boldsymbol{u}}) + T(\boldsymbol{x};\Gamma^{ct}) X_{2}(\boldsymbol{x};\Gamma^{ct}) \right] + \left(\frac{4+y}{8}\right)\bar{u}.
	\end{aligned}
	\label{eq:crack_dis_inclusion}
\end{equation}

\subsubsection{Crack initiation}

\begin{figure}
	\begin{centering}
		\includegraphics[scale=0.35]{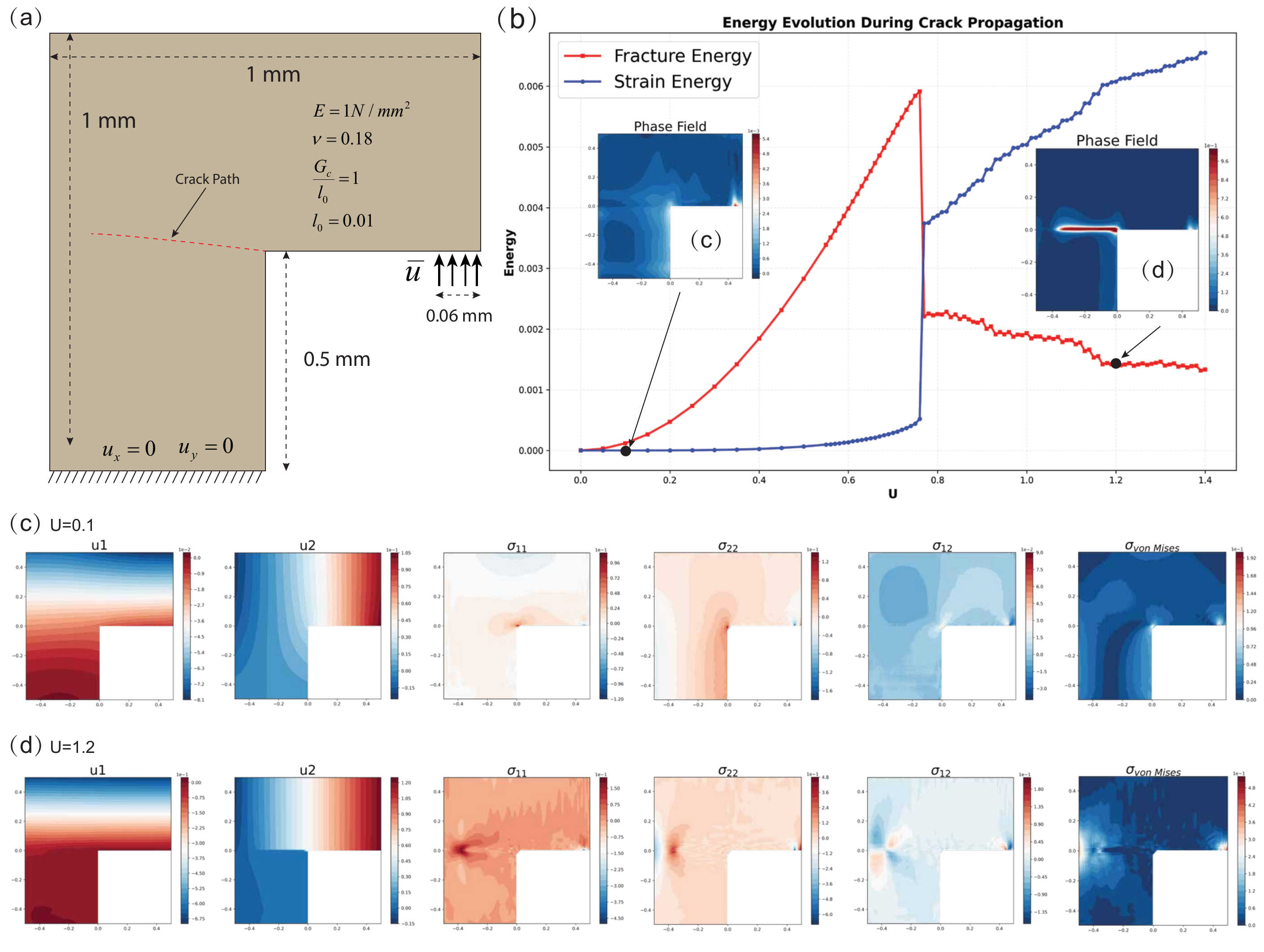}
		\par\end{centering}
	\caption{Performance of XDEM in the crack initiation problem: (a) schematic
		of the setup; (b) evolution of the energy functional and phase-field
		distribution; (c) displacement and stress fields at $\bar{u}=0.1$;
		(d) displacement and stress fields at $\bar{u}=1.2$.\label{fig:XDEM_crack_initiation}}
\end{figure}

Crack initiation problems are often difficult to handle using discrete
crack models. Therefore, we employ the continuous XDEM formulation
to address this case, as illustrated in \Cref{fig:XDEM_crack_initiation}a.
The displacement field setup follows \cite{manav2024phase}. We employ a penalization term to enforce the irreversibility of the phase field.

\Cref{fig:XDEM_crack_initiation}b presents the evolution of
the total energy functional and the phase-field contours at different
loading levels. The results predicted by XDEM show very good agreement
with the reference solution of \cite{manav2024phase}. One notable
advantage of XDEM is its flexibility in network selection. In this
example, we approximate both the displacement and phase-field using
Kolmogorov Arnold Networks (KANs) with architecture $[2,15,15,15,3]$.
For differentiation, we employed finite element shape function derivatives
rather than automatic differentiation (AD) algorithms. It is worth noting that XDEM only used $7700$ uniformly distributed
collocation points, which is significantly fewer than the $34149$
points used by \cite{manav2024phase}, who further required local
mesh refinement in the crack initiation region.

\Cref{fig:XDEM_crack_initiation}c and \Cref{fig:XDEM_crack_initiation}d
show the predicted displacement and stress fields at displacement
loads $\bar{u}=0.1$ and $\bar{u}=1.2$, respectively. These results
demonstrate that XDEM is able to capture the nucleation and subsequent
evolution of cracks with high accuracy and efficiency.

\section{Extended discussion on XDEM}\label{sec:SIMethod}
\subsection{Extended Deep Energy Method in discrete crack models}\label{sec:XDEM-D}

The key challenge for XDEM in discrete crack models is to accurately represent the displacement discontinuities across the crack surface and the highly localized fields near the crack tip. To this end, we incorporate the ideas of the heaviside step function in XFEM \citet{moes1999finite} and the asymptotic crack-tip solution into the Deep Energy Method. Specifically, the step function is used to capture displacement discontinuities, while the crack-tip enrichment function is employed to represent the singular behavior near the crack tip. In what follows, we describe how XDEM introduces these two components.

\subsubsection{Crack function}

The treatment of displacement discontinuities in XDEM is achieved by introducing a \emph{crack function}, which builds upon the concept of embedding functions proposed by Zhao et al. \citet{zhao2025denns}.  

In XDEM, the crack function $\varrho(\boldsymbol{x})$ is required to satisfy:
\begin{equation}
	\begin{cases}
		\lim_{\boldsymbol{x}\rightarrow\boldsymbol{x}_{c}^{-}}\varrho(\boldsymbol{x})\neq\lim_{\boldsymbol{x}\rightarrow\boldsymbol{x}_{c}^{+}}\varrho(\boldsymbol{x}), & \boldsymbol{x}_{c}\in\Gamma^{c}, \\[6pt]
		\lim_{\boldsymbol{x}\rightarrow\boldsymbol{x}_{0}}\varrho(\boldsymbol{x})=\varrho(\boldsymbol{x}_{0}), & \boldsymbol{x}_{0}\in\Omega\setminus\Gamma^{c}, \\[6pt]
		\lim_{\boldsymbol{x}\rightarrow\boldsymbol{x}_{0}}\nabla\varrho(\boldsymbol{x})=\nabla\varrho(\boldsymbol{x}_{0}), & \boldsymbol{x}_{0}\in\Omega\setminus\Gamma^{c},
	\end{cases}
	\label{eq:embedding_function}
\end{equation}
where $\Gamma^{c}$ denotes the crack surface.

\begin{figure}
	\begin{centering}
		\includegraphics[scale=0.9]{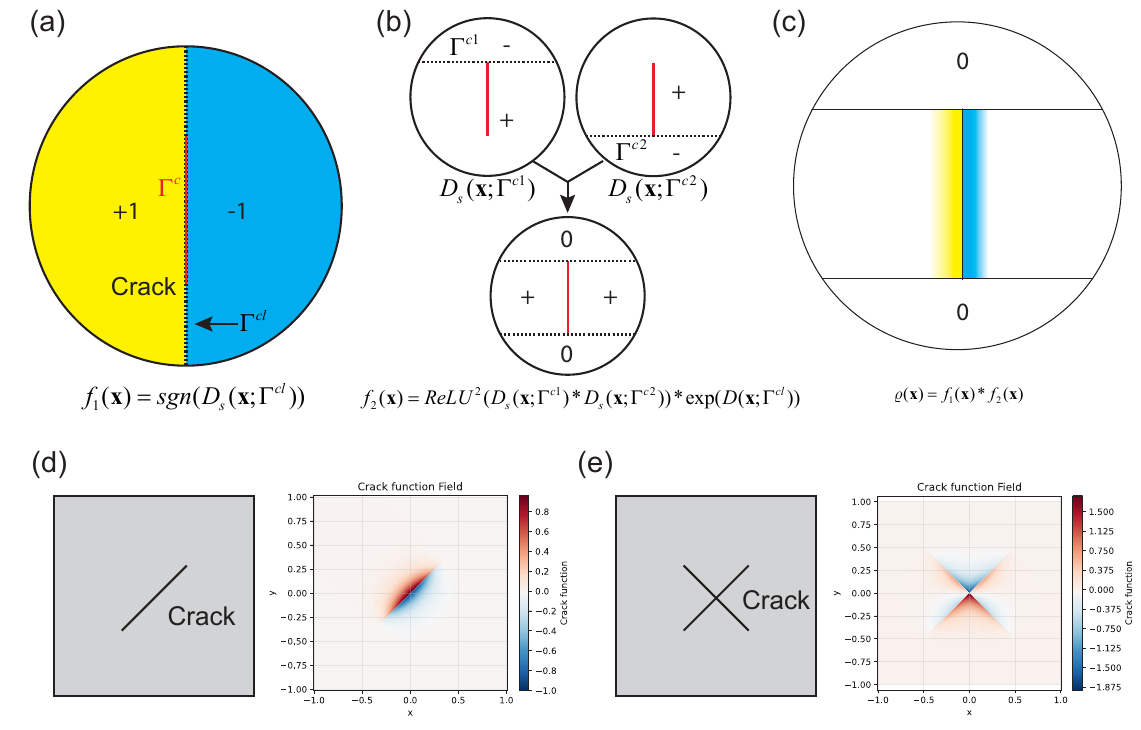}
		\par\end{centering}
	\caption{Illustration of the crack function: (a) signed distance function calculation for the crack followed by the $\mathrm{sgn}$ operator; (b) signed distance functions calculated for the two endpoints $\Gamma^{c1}$ and $\Gamma^{c2}$, processed by $\mathrm{ReLU}^{2}$ and multiplied by a decay function; (c) schematic of the crack function; (d) crack function for an inclined crack; (e) crack function for intersecting cracks. \label{fig:Embedding-function}}
\end{figure}

The crack function takes the following form:
\begin{equation}
	\begin{aligned}
		\varrho(\boldsymbol{x}) &= f_{1}(\boldsymbol{x}) \cdot f_{2}(\boldsymbol{x}), \\
		f_{1}(\boldsymbol{x}) &= \mathrm{sgn}\!\left(D_{s}(\boldsymbol{x};\Gamma^{cl})\right), \\
		f_{2}(\boldsymbol{x}) &= \mathrm{ReLU}^{2}\!\left(D_{s}(\boldsymbol{x};\Gamma^{c1}) \cdot D_{s}(\boldsymbol{x};\Gamma^{c2})\right)\,
		\exp\!\left(D(\boldsymbol{x};\Gamma^{cl})\right),
	\end{aligned}
\end{equation}
where $\mathrm{sgn}$, $D_{s}$, and $D$ are the sign function, the signed distance function, and the distance function, respectively, defined as:
\begin{align}
	\mathrm{sgn}(x) &= 
	\begin{cases}
		-1, & x\leq 0, \\
		1, & x>0,
	\end{cases} \\
	D_{s}(\boldsymbol{x};\Gamma^{cl}) &= \mathrm{sgn}\!\big(\boldsymbol{n}\cdot(\boldsymbol{x}-\hat{\boldsymbol{x}})\big)\,\|\boldsymbol{x}-\hat{\boldsymbol{x}}\|, \\
	D(\boldsymbol{x};\Gamma^{cl}) &= \|\boldsymbol{x}-\hat{\boldsymbol{x}}\|,
\end{align}
with $\hat{\boldsymbol{x}}$ denoting the closest point on the crack extension line $\Gamma^{cl}$ to $\boldsymbol{x}$, and $\boldsymbol{n}$ being the outward normal vector at $\hat{\boldsymbol{x}}$. $\Gamma^{c1}$ and $\Gamma^{c2}$ represent the perpendicular extensions of the two endpoints of $\Gamma^{c}$. \Cref{fig:Embedding-function} illustrates the construction of the crack function $\varrho(\boldsymbol{x})$ and its components.

The primary role of the crack function is to embed discontinuity information of cracks into the neural network. There are two possible strategies for incorporating the crack function:  
(1) include it as an additional input to the neural network, i.e., $NN(\boldsymbol{x},\varrho;\boldsymbol{\theta}_{\boldsymbol{u}})$, which is simple but increases the input dimension by one; or  
(2) incorporate it into the output of the neural network, which avoids increasing the input width but requires a more complex implementation. In this work, we focus on the first approach for its simplicity.

\subsubsection{Extended function}

To enable the neural network to better capture the near-tip singular fields, XDEM introduces the Williams series expansion \citet{williams1957stress}:
\begin{equation}
	\left[\begin{array}{c}
		u_{x}\\
		u_{y}
	\end{array}\right]
	= \sum_{n=0}^{+\infty} A_{n}^{I}\,\frac{r^{\tfrac{n}{2}}}{2G}
	\left[\begin{array}{c}
		f_{xn}^{I}\\
		f_{yn}^{I}
	\end{array}\right]
	+ \sum_{n=0}^{+\infty} A_{n}^{II}\,\Big(-\frac{r^{\tfrac{n}{2}}}{2G}\Big)
	\left[\begin{array}{c}
		f_{xn}^{II}\\
		f_{yn}^{II}
	\end{array}\right],
	\label{eq:williams_series}
\end{equation}
where $A_{n}^{I}$ and $A_{n}^{II}$ are real constants corresponding to mode I and mode II cracks, $G$ is the shear modulus, $f_{xn}^{I}$ and $f_{yn}^{I}$ are angular distribution functions for mode I, and $f_{xn}^{II}$ and $f_{yn}^{II}$ are the corresponding functions for mode II:  
\begin{equation}
	\begin{aligned}
		f_{xn}^{I} &= \kappa\cos\!\Big(\tfrac{n}{2}\theta\Big) - \tfrac{n}{2}\cos\!\Big[\Big(\tfrac{n}{2}-2\Big)\theta\Big] 
		+ \Big[\tfrac{n}{2}+(-1)^{n}\Big]\cos\!\Big(\tfrac{n}{2}\theta\Big), \\
		f_{yn}^{I} &= \kappa\sin\!\Big(\tfrac{n}{2}\theta\Big) + \tfrac{n}{2}\sin\!\Big[\Big(\tfrac{n}{2}-2\Big)\theta\Big] 
		- \Big[\tfrac{n}{2}+(-1)^{n}\Big]\sin\!\Big(\tfrac{n}{2}\theta\Big), \\
		f_{xn}^{II} &= \kappa\sin\!\Big(\tfrac{n}{2}\theta\Big) - \tfrac{n}{2}\sin\!\Big[\Big(\tfrac{n}{2}-2\Big)\theta\Big] 
		+ \Big[\tfrac{n}{2}-(-1)^{n}\Big]\sin\!\Big(\tfrac{n}{2}\theta\Big), \\
		f_{yn}^{II} &= -\kappa\cos\!\Big(\tfrac{n}{2}\theta\Big) - \tfrac{n}{2}\cos\!\Big[\Big(\tfrac{n}{2}-2\Big)\theta\Big] 
		+ \Big[\tfrac{n}{2}-(-1)^{n}\Big]\cos\!\Big(\tfrac{n}{2}\theta\Big),
	\end{aligned}
\end{equation}
where $(r,\theta)$ are local polar coordinates centered at the crack tip, with $\theta=0$ aligned with the crack tangent. The parameter $\kappa$ depends on the problem type: for plane strain, $\kappa=3-4\nu$; for plane stress, $\kappa=(3-\nu)/(1+\nu)$, with $\nu$ being Poisson's ratio.

\begin{figure}
	\begin{centering}
		\includegraphics[scale=0.7]{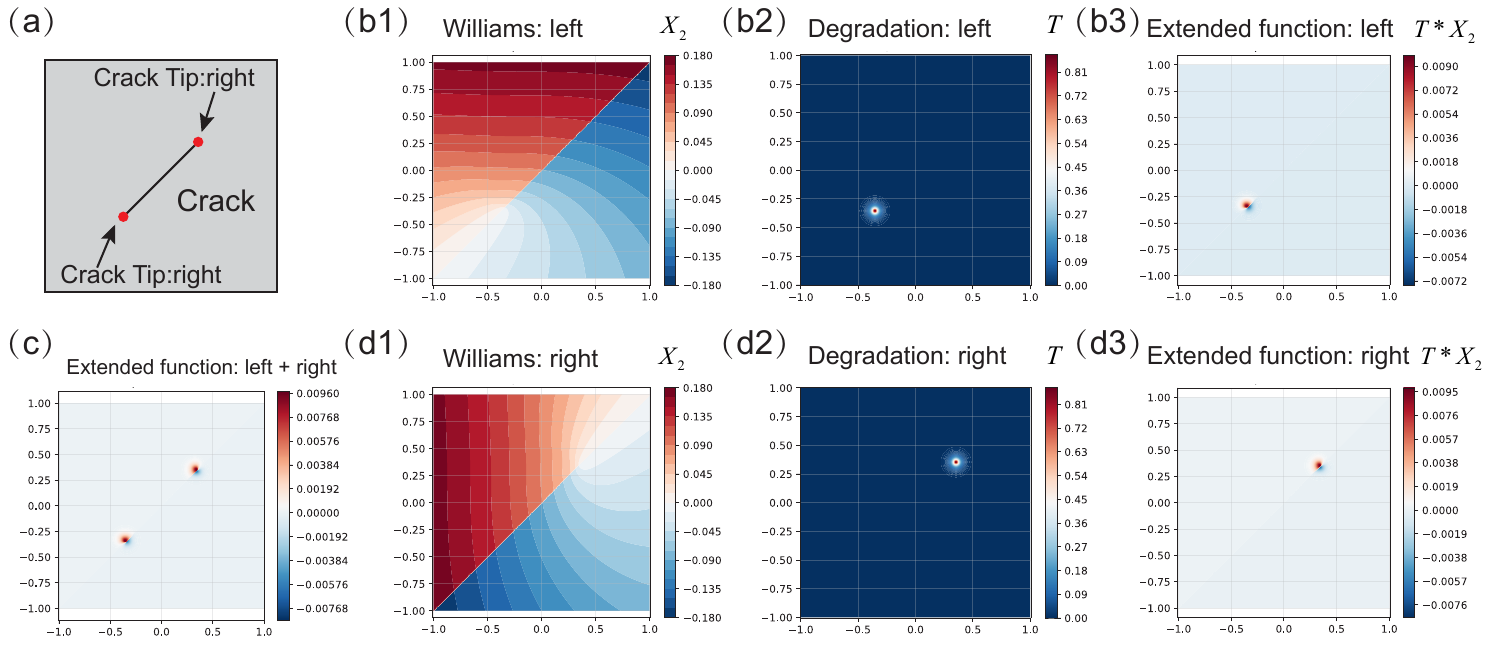}
		\par\end{centering}
	\caption{Illustration of the extended function: (a) both crack tips of an inclined crack require extended functions; (b1--3) Williams series solution $X_{2}$, decay function $T$, and extended function $T\cdot X_{2}$ at the left crack tip; (c) combined extended functions at both crack tips; (d1--3) Williams series solution $X_{2}$, decay function $T$, and extended function $T\cdot X_{2}$ at the right crack tip. \label{fig:Extended-function}}
\end{figure}

By combining the Williams series expansion in \Cref{eq:williams_series} with the embedding function, the displacement neural network is extended as follows:
\begin{equation}
	\begin{aligned}
		u_{1}(\boldsymbol{x},\varrho;\boldsymbol{\theta}_{\boldsymbol{u}}) &= D(\boldsymbol{x};\Gamma_{\boldsymbol{u}})\,\Big[NN_{x}(\boldsymbol{x},\varrho;\boldsymbol{\theta}_{\boldsymbol{u}})
		+ \sum_{i=1}^{N_{tip}}T(\boldsymbol{x};\Gamma^{c(i)})\,X_{1}(\boldsymbol{x};\Gamma^{c(i)})\Big] + \bar{u}_{x}(\boldsymbol{x}), \\
		u_{2}(\boldsymbol{x},\varrho;\boldsymbol{\theta}_{\boldsymbol{u}}) &= D(\boldsymbol{x};\Gamma_{\boldsymbol{u}})\,\Big[NN_{y}(\boldsymbol{x},\varrho;\boldsymbol{\theta}_{\boldsymbol{u}})
		+ \sum_{i=1}^{N_{tip}}T(\boldsymbol{x};\Gamma^{c(i)})\,X_{2}(\boldsymbol{x};\Gamma^{c(i)})\Big] + \bar{u}_{y}(\boldsymbol{x}), \\
		X_{1}(\boldsymbol{x};\Gamma^{c(i)}) &= \sum_{n=0}^{+\infty}\alpha_{n}^{I}\,\frac{r^{\tfrac{n}{2}}}{2G}f_{xn}^{I} 
		+ \sum_{n=0}^{+\infty}\alpha_{n}^{II}\,\Big(-\frac{r^{\tfrac{n}{2}}}{2G}\Big)f_{xn}^{II}, \\
		X_{2}(\boldsymbol{x};\Gamma^{c(i)}) &= \sum_{n=0}^{+\infty}\beta_{n}^{I}\,\frac{r^{\tfrac{n}{2}}}{2G}f_{yn}^{I} 
		+ \sum_{n=0}^{+\infty}\beta_{n}^{II}\,\Big(-\frac{r^{\tfrac{n}{2}}}{2G}\Big)f_{yn}^{II}, \\
		T(\boldsymbol{x};\Gamma^{c(i)}) &= \exp(-mr^{q}),
	\end{aligned}
	\label{eq:DEM_extended}
\end{equation}
where $\alpha_{n}^{I}$, $\alpha_{n}^{II}$, $\beta_{n}^{I}$, and $\beta_{n}^{II}$ are learnable parameters, typically initialized according to the crack type. For example, for mode I cracks, we recommend initializing $\alpha_{1}^{I}=\beta_{1}^{I}=K_{I}/\sqrt{2\pi}=\sigma\sqrt{a}/\sqrt{2}$, where $a$ is the crack length and $\sigma$ the applied tensile stress. The remaining parameters can be initialized to zero. Since the asymptotic solution is more accurate closer to the crack tip, we introduce a decay function $T(\boldsymbol{x};\Gamma^{c(i)})=\exp(-mr^{q})$ to reduce its influence away from the crack tip, where $m$ and $q$ are hyperparameters (typically $m=b/a$, $q=1$), with $b$ being a structural length scale and $a$ the crack length. $\bar{u}_{x}(\boldsymbol{x})$ and $\bar{u}_{y}(\boldsymbol{x})$ are prescribed displacement boundary conditions, while $u_{1}$ and $u_{2}$ represent the displacement components in the $x$ and $y$ directions, respectively. $D(\boldsymbol{x};\Gamma_{\boldsymbol{u}})$ denotes the distance from $\boldsymbol{x}$ to the Dirichlet boundary $\Gamma_{\boldsymbol{u}}$. The term $T\cdot X$ constitutes the \emph{extended function} in XDEM. \Cref{fig:Extended-function} illustrates the extended function for an inclined crack.  

Substituting \Cref{eq:DEM_extended} into \Cref{eq:dis_DEM}, the optimization problem can be solved. It is important to note that the loading process must be divided into multiple increments, since fracture is path-dependent: even if the final load state is identical, different loading paths may lead to entirely different displacement fields and crack trajectories. After optimization, crack propagation follows a predefined fracture criterion. For instance, under the maximum circumferential stress criterion, the crack extends in the direction $\boldsymbol{n}_{c}$ of the maximum hoop stress, with an incremental length $\alpha_{c}\boldsymbol{n}_{c}$, where $\alpha_{c}$ is the prescribed step size.  

Since the enriched displacement space better matches the true fracture fields, XDEM requires significantly fewer collocation points to achieve high accuracy. Unless otherwise stated, we employ uniformly distributed collocation points. We recommend using the discrete XDEM formulation when dealing with relatively simple crack problems.

Unlike conventional XFEM, which suffers from blending and topological enrichment issues, our global approximation framework in XDEM completely eliminates these difficulties.
\subsection{Extended Deep Energy Method in continuous damage models}\label{sec:XDEM-C}

\begin{figure}
	\begin{centering}
		\includegraphics[scale=0.55]{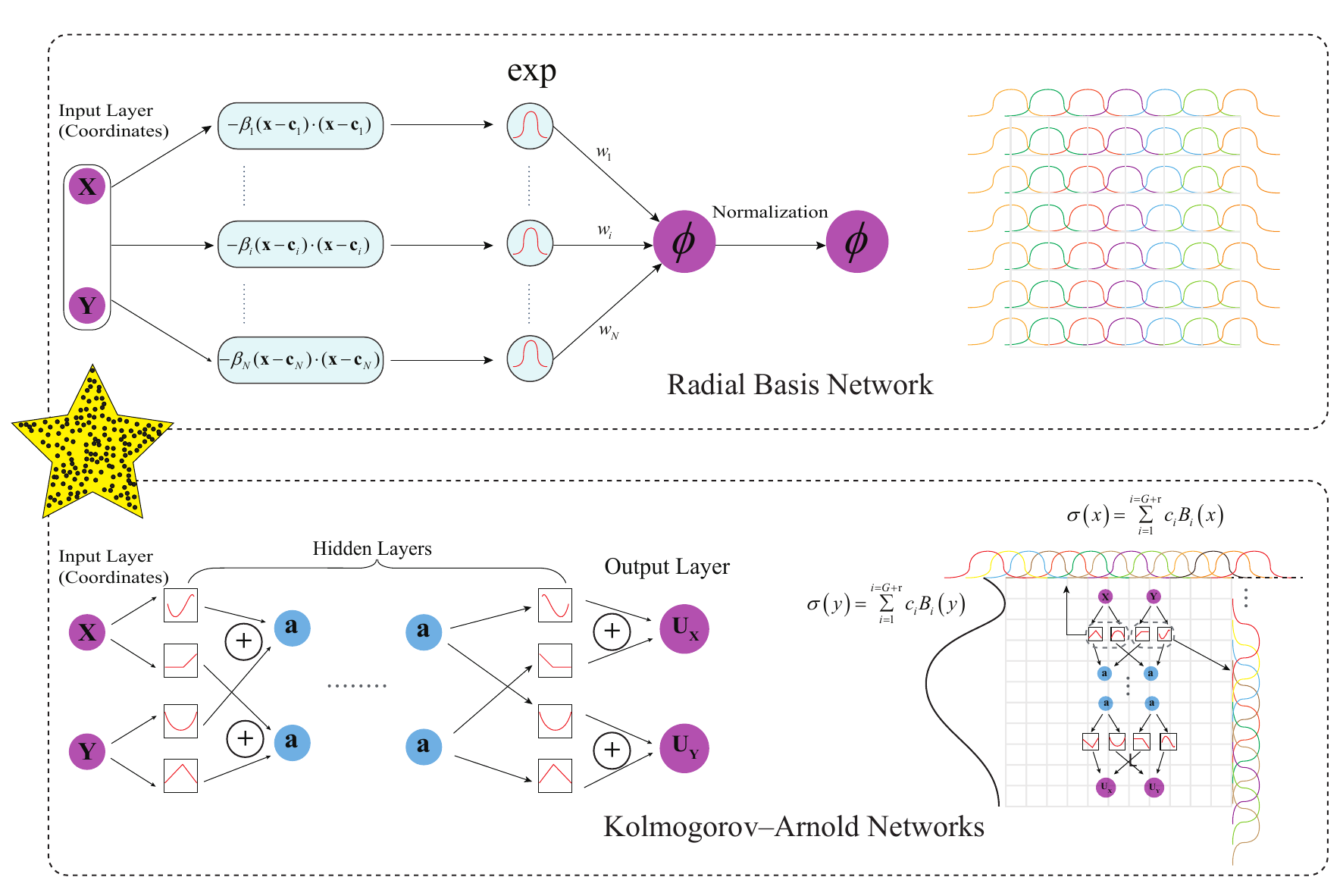}
		\par\end{centering}
	\caption{Network architectures of the Radial Basis Function Network and the Kolmogorov--Arnold Network. \label{fig:The-arch_RBF_KAN}}
\end{figure}

In the continuous damage formulation, XDEM introduces a phase-field variable to represent cracks. The displacement field and the phase field are approximated by separate neural networks. Since the phase field is more compatible with the functional space of radial basis functions (RBFs), we recommend approximating the phase field using an RBF network, with a detailed justification provided in \Cref{subsec:The-phase-field}. On the other hand, as the displacement field exhibits strong variations near crack tips, and Kolmogorov--Arnold Networks (KANs) have been shown to perform better than MLPs for problems with sharp variations \citet{wang2025kolmogorov}, we recommend approximating the displacement field with a KAN. The architectures of the RBF and KAN networks are illustrated in \Cref{fig:The-arch_RBF_KAN}, and further details are given in \Cref{subsec:RBF_KAN}. It should be noted that there is no universally optimal architecture; the choice depends on the specific problem, analogous to the use of different element types in the finite element method.  

For optimization, both monolithic and staggered strategies can be employed. The staggered scheme renders the energy functional landscape more convex, making it more robust than the monolithic scheme. However, the monolithic scheme often converges faster in practice. A detailed discussion is provided in \Cref{subsec:stagger_mono}.  

The irreversibility condition $\phi^{n+1}\geq\phi^{n}$ for the phase field is typically enforced either by a history field approach \citet{miehe2010phase} or by penalization techniques \citet{gerasimov2019penalization}; see \Cref{subsec:irreversibility_phase_field} for details. We recommend employing the continuous phase-field formulation of XDEM when dealing with complex crack problems.

\subsection{Transfer learning in Extended Deep Energy Method}\label{sec:LoRA}

Regardless of whether XDEM is applied in its discrete or continuous formulation, crack propagation simulations require discretizing the loading process into multiple increments. At each load step, the network must be retrained, which leads to significant computational cost \citet{goswami2020transfer}. However, since the displacement and phase fields in neighboring load steps are strongly correlated and share similar patterns, transfer learning can be leveraged to accelerate phase-field fracture simulations.  

\begin{figure}[!t]
	\begin{centering}
		\includegraphics[scale=0.60]{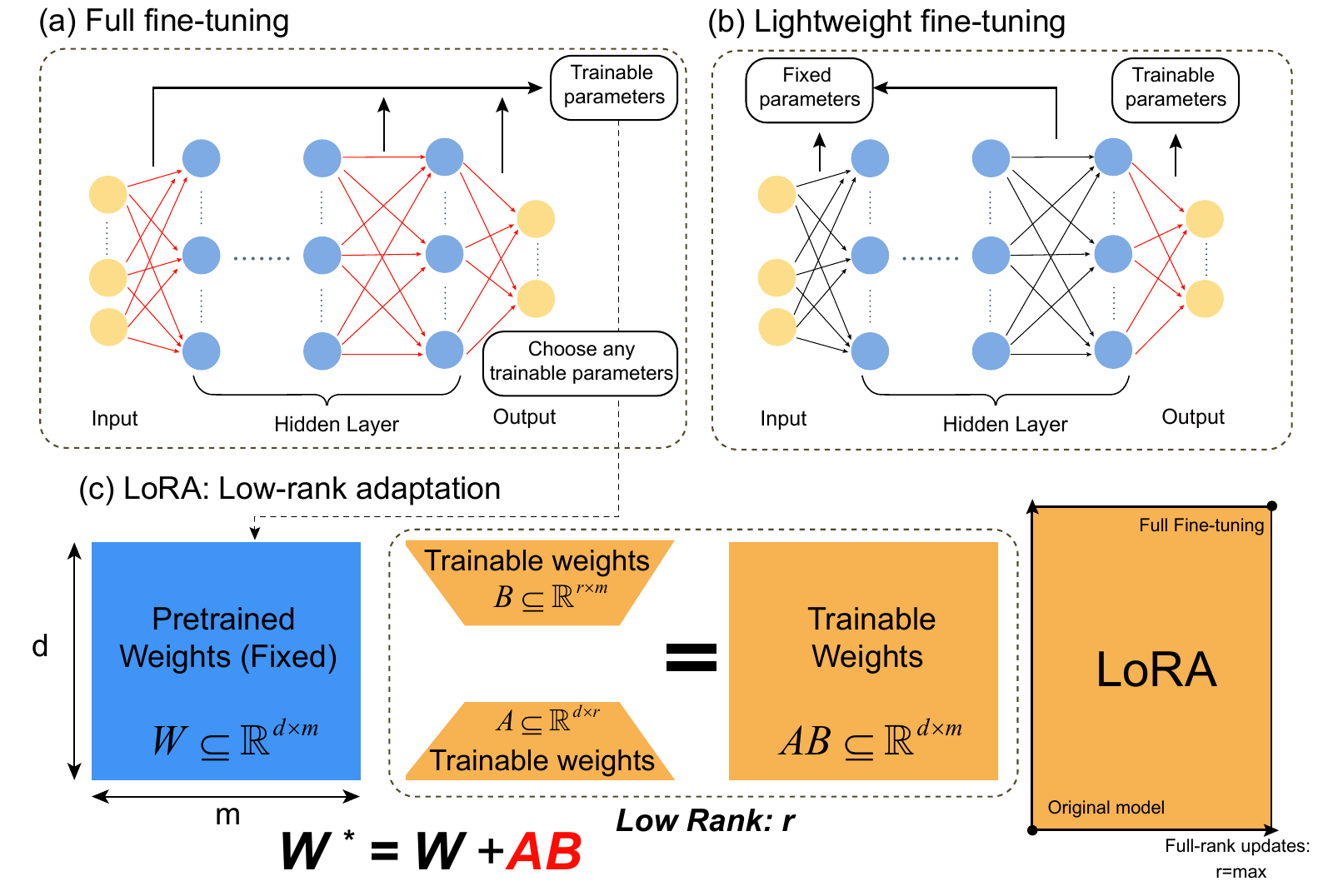}
		\par\end{centering}
	\caption{Three common parameter-based transfer learning strategies:  
		(a) Full fine-tuning: all parameters are updated (red arrows indicate trainable parameters).  
		(b) Lightweight fine-tuning: only a subset of parameters is updated (red arrows).  
		(c) LoRA: the blue matrix $\boldsymbol{W}$ denotes fixed pretrained weights, while the yellow matrices $\boldsymbol{A}$ and $\boldsymbol{B}$ represent trainable low-rank factors. During inference, the adapted weights are given by $\boldsymbol{W}^{*}=\boldsymbol{W}+\boldsymbol{A}\boldsymbol{B}$. \label{fig:Parameter-based-transfer-learn}}
\end{figure}

In this work, we adopt the Low-Rank Adaptation (LoRA) method \citet{hu2021lora} for efficient fine-tuning, which substantially reduces computational cost. LoRA provides a more general and flexible parameter-efficient transfer learning strategy compared with both full and lightweight fine-tuning \citet{wang2025transfer}, as illustrated in \Cref{fig:Parameter-based-transfer-learn}.  

The idea of LoRA is to approximate weight updates through the product of two low-rank matrices $\boldsymbol{A}\boldsymbol{B}$, where $\boldsymbol{A}\in\mathbb{R}^{d\times r}$, $\boldsymbol{B}\in\mathbb{R}^{r\times m}$, and $r\ll\min(d,m)$. These low-rank updates are added to the pretrained weight matrix $\boldsymbol{W}\in\mathbb{R}^{d\times m}$, as shown in \Cref{fig:Parameter-based-transfer-learn}c:
\begin{equation}
	\boldsymbol{W}^{*}=\boldsymbol{W}+\alpha\boldsymbol{A}\boldsymbol{B},
	\label{eq:lora_weight_change}
\end{equation}
where $r$ is the rank of LoRA, typically set to $1$ in our XDEM implementation. It is important to note that $\boldsymbol{W}$ represents the fixed pretrained weights from the previous task and is not updated. Only $\boldsymbol{A}$ and $\boldsymbol{B}$ are trainable, and their number of parameters is $r(d+m)$, which is significantly smaller than the $d\cdot m$ parameters required in full fine-tuning. The scalar $\alpha$ is a scaling factor that balances the contribution of pretrained weights $\boldsymbol{W}$ and the LoRA updates $\boldsymbol{A}\boldsymbol{B}$, and is set to $1$ by default. $\boldsymbol{A}$ and $\boldsymbol{B}$ are initialized with Gaussian-distributed entries (mean $0$, standard deviation $0.02$).  

LoRA can be interpreted as a generalized form of lightweight fine-tuning, while full fine-tuning can be seen as its limiting case when $r=\min(d,m)$. In essence, LoRA reduces the number of trainable parameters by decomposing the weight updates into low-rank factors.  

In XDEM, we apply LoRA specifically to the displacement network, as LoRA generalizes both full and lightweight fine-tuning. Concretely, we adapt the parameters $c_{m}^{(i,j)}$ and $W_{ij}$ in the KAN network according to \Cref{eq:lora_weight_change}, considering KAN network:
\begin{equation}
	\begin{aligned}
		c_{m}^{(ij)*} &= c_{m}^{(ij)} + \boldsymbol{A}_{c}\boldsymbol{B}_{c}, \\
		W_{ij}^{*} &= W_{ij} + \boldsymbol{A}_{w}\boldsymbol{B}_{w},
	\end{aligned}
\end{equation}
where $\boldsymbol{A}_{c}\in\mathbb{R}^{l_{o}\times r}$, $\boldsymbol{B}_{c}\in\mathbb{R}^{r\times[l_{i}(G+r)]}$, $\boldsymbol{A}_{w}\in\mathbb{R}^{l_{o}\times r}$, and $\boldsymbol{B}_{w}\in\mathbb{R}^{r\times l_{i}}$.  

It should be noted that transfer can be performed incrementally between consecutive load steps. However, in the later stages of displacement-controlled loading, non-physical fracture patterns may appear (see \Cref{subsec:DEM_zero_energy}). This issue arises due to the delicate balance between the expressive power of neural networks and the accuracy of numerical integration. 
Although it is fundamentally unrelated to transfer learning algorithms like LoRA, this issue primarily stems from the numerical integration precision. In the future, however, a transfer learning algorithm that improves the integration accuracy in subsequent load steps could potentially be developed for solving fracture mechanics problems using PINNs.
Therefore, in practice, XDEM often uses the parameters obtained from the first load step as the baseline for transfer learning.  
For the fracture propagation results in the manuscript, we use the parameters obtained from the first load step as the baseline for transfer learning.

\section{Discussion} \label{sec:Discussion}

\subsection{Zero-energy modes in DEM: Non-physical (virtual) fracture\label{subsec:DEM_zero_energy}}

\begin{figure}
	\begin{centering}
		\includegraphics[scale=0.6]{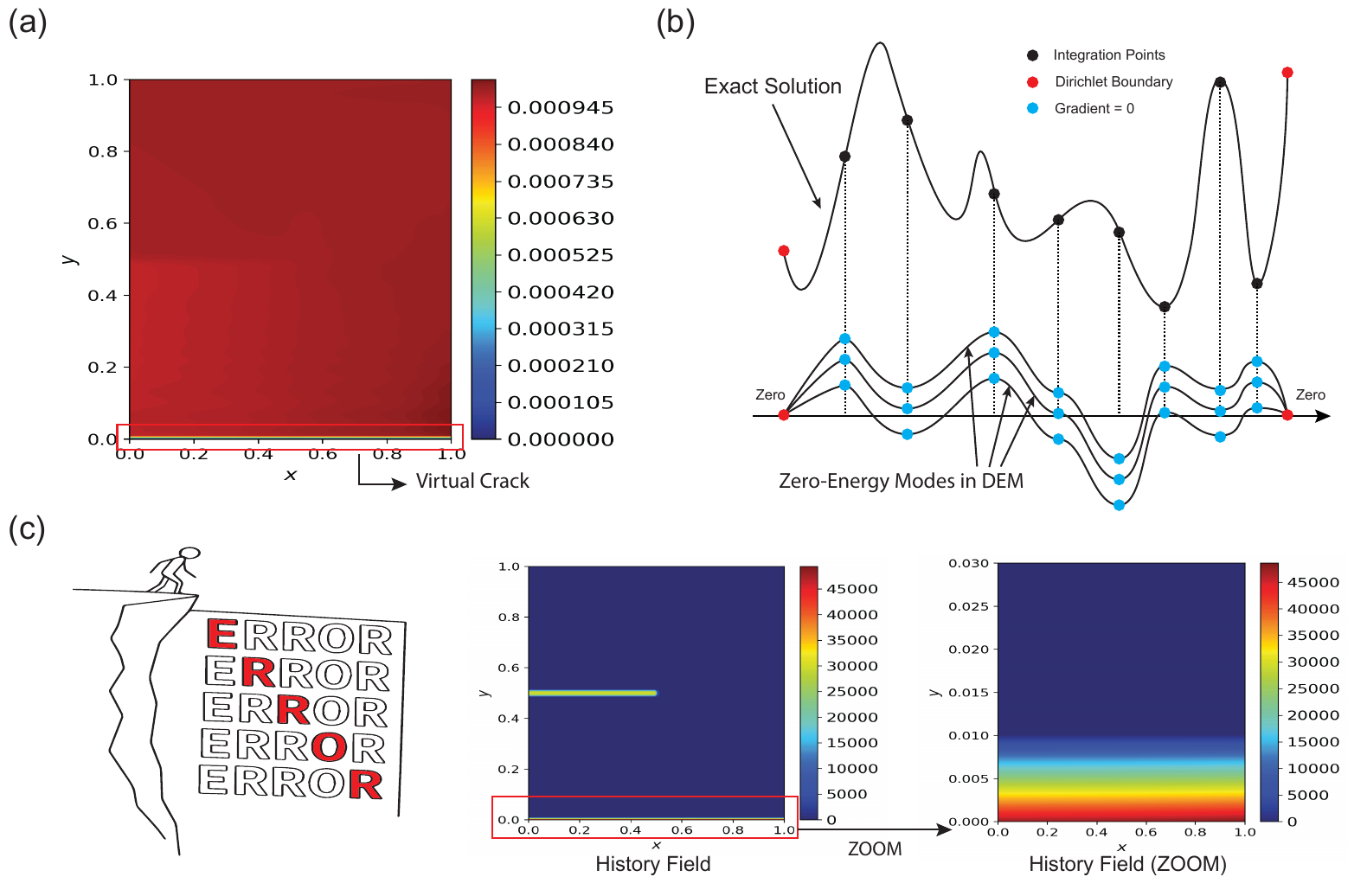}
		\par\end{centering}
	\caption{Non-physical fracture phenomenon: (a) DEM failure with insufficient
		collocation points, showing artificial discontinuities at the boundary;
		(b) illustration of zero-energy modes in DEM; (c) once zero-energy
		modes occur in XDEM, they affect all subsequent load steps.\label{fig:virtual_crack}}
\end{figure}

During our experiments, we observed that when the number of collocation
points in DEM is insufficient, spurious non-physical fracture patterns
may appear. An example is shown in \Cref{fig:virtual_crack}a,
where the $y$-displacement field under mode I loading exhibits an
artificial discontinuity. In the following, we explain the cause of
this numerical artifact.

Such virtual fracture is a common failure mode of DEM training. To
illustrate the mechanism, let us first consider a simple linear elastic
problem. Under external loading (i.e., Neumann boundary conditions),
if there is a mismatch between the strain energy and the external
work in the total potential, the strain energy is often underestimated.
As a result, the neural network tends to focus on minimizing the external
potential, whose gradient dominates and overwhelms the contribution
from the internal energy. Consequently, displacements on the Neumann
boundary may grow excessively in order to reduce the overall energy.
In principle, the internal strain energy should also increase, but
due to inaccurate integration between boundary collocation points
and interior points near the boundary, the strain energy contribution
is underestimated. This creates a loophole where the neural network
artificially reduces the energy by introducing displacement jumps
at the boundary.  

In the case of pure displacement loading (Dirichlet boundary conditions),
the external work vanishes, and the network focuses solely on minimizing
strain energy. When integration accuracy is insufficient, the network
tends to approximate the displacement field as uniformly as possible
in the domain, which again may lead to spurious fracture-like discontinuities,
as illustrated in \Cref{fig:virtual_crack}a.

Fundamentally, this phenomenon can be interpreted as a manifestation
of zero-energy modes, as illustrated in \Cref{fig:virtual_crack}b:
\begin{equation}
	\Pi(u^{\text{exact}})=\Pi(u^{\text{exact}}+u^{\text{zero}}),
\end{equation}
where $u^{\text{zero}}$ denotes a zero-energy mode displacement field.
In this mode, the strain (derivatives) vanishes at collocation points,
yet may vary arbitrarily in between. At essential boundary points,
the displacements remain zero (here we only consider displacement
fields). Such $u^{\text{zero}}$ can be added to the exact solution
$u^{\text{exact}}$ without affecting the energy functional $\Pi$,
and hence cannot be eliminated by the optimizer. Unlike finite element
methods, where spurious zero-energy modes are finite-dimensional,
the zero-energy modes in DEM are essentially infinite-dimensional
and thus more difficult to control.  

In fracture simulations, the impact of such spurious modes becomes
magnified. Once a non-physical crack emerges, it contaminates all
subsequent load steps, as illustrated in \Cref{fig:virtual_crack}c,
leading to cascading failure. A practical remedy is to reinitialize
the optimization once such virtual fracture is detected.  

In summary, the root cause of virtual fracture is insufficient integration
accuracy, which typically occurs when the number of collocation points
is too small relative to the number of training iterations. With too
few points, numerical integration becomes inaccurate, and excessive
training steps make the zero-energy mode more likely to appear. The
recommended solution is to maintain a reasonable proportionality between
the number of iterations and the number of collocation points.

\subsection{Loss functions in DEM and XDEM}

\begin{figure}
	\begin{centering}
		\includegraphics[scale=0.63]{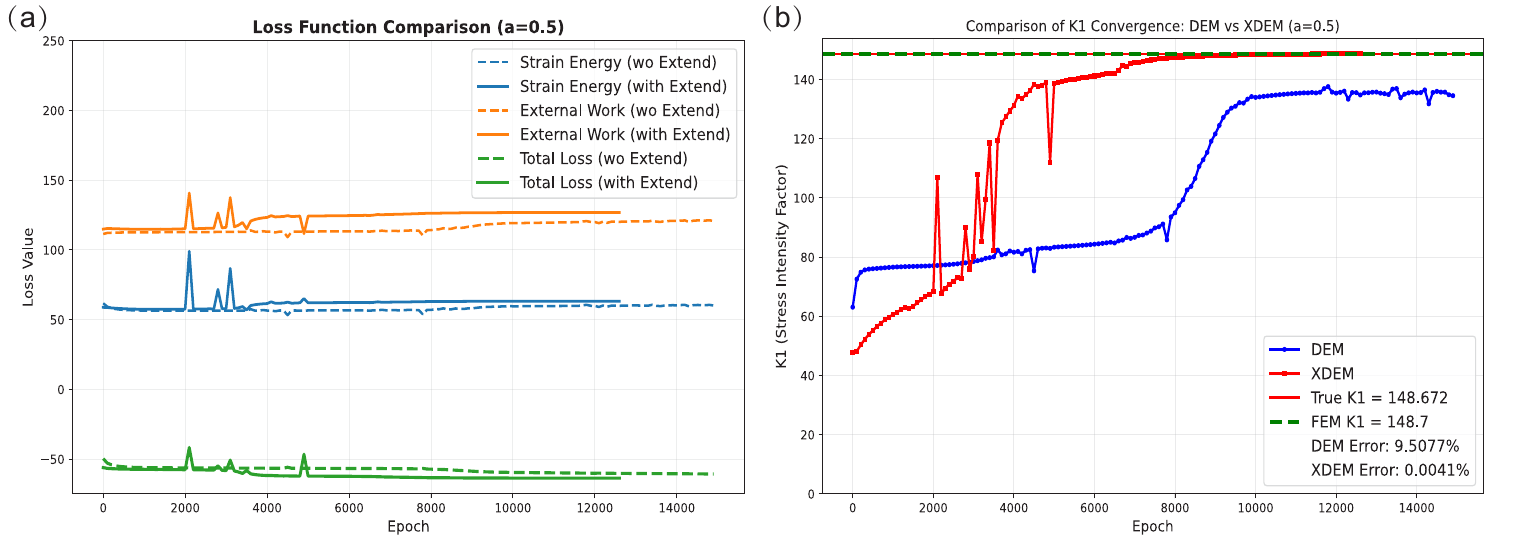}
		\par\end{centering}
	\caption{Evolution of the loss function and stress intensity factors in
		DEM and XDEM.\label{fig:loss_landscape_DEM_XDEM}}
\end{figure}

In the experiments presented in \Cref{subsec:SI_Mode-I-crack},
we observed that XDEM predicts stress intensity factors (SIFs) more
accurately and converges faster than DEM. A natural question is whether
the evolution of the energy function (loss landscape) differs significantly
between DEM and XDEM. To answer this, we analyzed the loss functions
from the same setting as in \Cref{fig:crack1_SIF}a. As shown
in \Cref{fig:loss_landscape_DEM_XDEM}, the overall difference
in energy loss between DEM and XDEM is not substantial. However, their
abilities to capture the SIF differ markedly. This discrepancy arises
because the SIF characterizes a local physical quantity at the crack
tip, which has limited impact on the global energy. This again confirms
the role of the extended function in XDEM: it enhances the representation
of crack-tip fields and allows XDEM to more effectively capture local
features relevant to fracture.

\subsection{Message passing}

The inclusion of extended functions in XDEM significantly improves
its accuracy and efficiency. In principle, neural networks are universal
approximators with strong representation capabilities~\cite{hornik1989multilayer}.
Thus, even without extended functions, the network could eventually
achieve comparable results if trained with a sufficient number of
iterations. In practice, however, PINNs often fail to reach this potential,
not because of inadequate expressive power, but due to optimization
difficulties.

\begin{figure}
	\begin{centering}
		\includegraphics[scale=0.70]{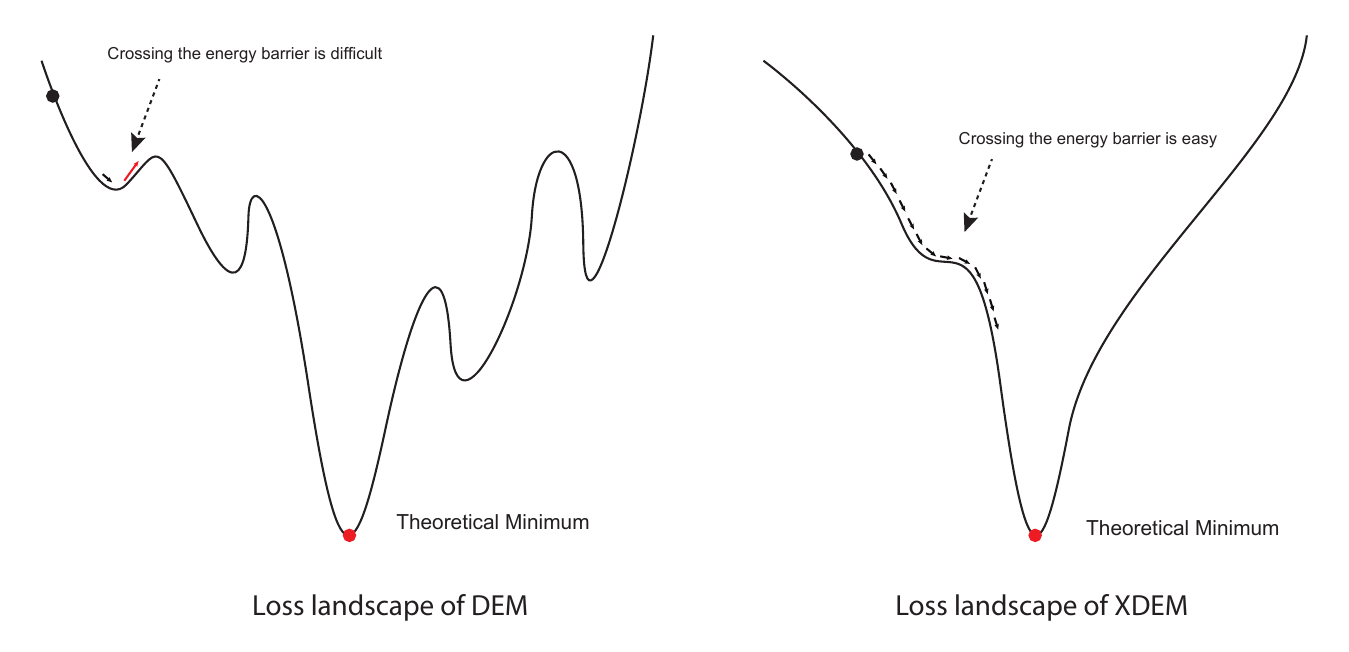}
		\par\end{centering}
	\caption{Comparison of the loss landscapes between DEM and XDEM. Note that this is only a schematic. \label{fig:Message_passing}}
\end{figure}

In XDEM, we alleviate this issue by introducing the crack function
to approximate displacement discontinuities, and the extended function
to represent crack-tip fields. These additions effectively reduce
the non-convexity of the loss landscape, thereby easing the optimization
process. We refer to this phenomenon as \emph{message passing}, as
illustrated in \Cref{fig:Message_passing}.

\subsection{Irreversibility of the Phase Field} \label{subsec:irreversibility_phase_field}

The irreversibility of the phase field is typically enforced in two
ways: the penalization approach and the history field approach:
\begin{equation}
	\begin{aligned}
		\text{Penalization: } & U^{ir}=\int_{\Omega}\frac{1}{2}\gamma_{ir}\langle\phi-\phi^{n}\rangle_{-}^{2}\,dV,\\
		\text{History: } & U^{ir}=\int_{\Omega}w(\phi)H(\Psi^{+},t_{n})\,dV,
	\end{aligned}
\end{equation}
where $\gamma_{ir}$ is the penalty factor, which must be prescribed
a priori. Following \citet{gerasimov2019penalization}, it can be
defined as:
\begin{equation}
	\gamma_{ir}=
	\begin{cases}
		\frac{G_{c}}{l_{0}}\frac{27}{64\delta_{tol}^{2}}, & \text{AT1},\\[6pt]
		\frac{G_{c}}{l_{0}}\left(\frac{1}{\delta_{tol}^{2}}-1\right), & \text{AT2},
	\end{cases}
\end{equation}
with $\delta_{tol}$ denoting the tolerance parameter.

In the history field formulation, $H$ is given by:
\begin{equation}
	H(\Psi^{+},t)=\max_{\tau\in[0,t]}\Psi^{+}(\boldsymbol{\varepsilon}(\boldsymbol{x},\tau)),
\end{equation}
where $\Psi^{+}$ is the tensile part of the elastic energy density.
For problems with an initial crack, the history field is initialized
as:
\begin{equation}
	H(\Psi^{+},t=0)=
	\begin{cases}
		\frac{BG_{c}}{2l_{0}}\left[1-\frac{2d(\boldsymbol{x})}{l_{0}}\right], & d(\boldsymbol{x})\leq\frac{l_{0}}{2},\\[6pt]
		0, & d(\boldsymbol{x})>\frac{l_{0}}{2},
	\end{cases}
	\label{eq:history_field}
\end{equation}
where $B$ is a tunable hyperparameter (set to $1000$ in our implementation),
and $d(\boldsymbol{x})$ denotes the shortest distance from point
$\boldsymbol{x}$ to the crack surface.

\subsection{The Phase Field in Fracture} \label{subsec:The-phase-field}

\begin{figure}
	\begin{centering}
		\includegraphics[scale=0.80]{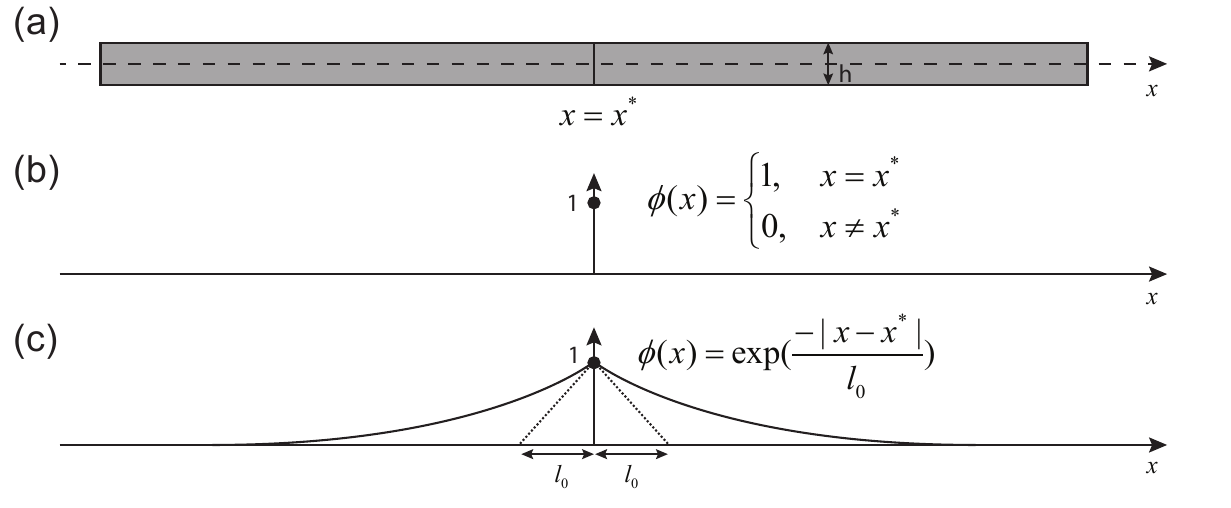}
		\par\end{centering}
	\caption{Phase-field representation of a one-dimensional infinite crack:
		(a) schematic of a one-dimensional infinite crack, where $x^{*}$ is
		the crack location; (b) discrete phase-field representation; (c) diffuse
		phase-field representation.\label{fig:fracture_1d}}
\end{figure}

We illustrate the similarity between the phase field in fracture and
the radial basis function (RBF) space using a one-dimensional infinite
crack example, as shown in \Cref{fig:fracture_1d}.  

Since the crack exists only at $x=x^{*}$, the crack can be described
by a discrete phase field:
\begin{equation}
	\phi(x)=
	\begin{cases}
		1, & x=x^{*},\\
		0, & x\neq x^{*}.
	\end{cases}
	\label{eq:discrete_crack_label}
\end{equation}
Although this description is exact, it causes difficulties in numerical
simulations. Therefore, we relax \Cref{eq:discrete_crack_label} into
a smooth formulation:
\begin{equation}
	\phi(x)=\exp\!\left(-\frac{|x-x^{*}|}{l_{0}}\right),
	\label{eq:contin_crack_label}
\end{equation}
where $l_{0}$ is the length scale parameter controlling the extent
of the diffused crack zone. The larger the $l_{0}$, the wider the
diffuse region of the crack.

It is straightforward to verify that \Cref{eq:contin_crack_label}
is the solution of the following ordinary differential equation (ODE):
\begin{equation}
	\begin{aligned}
		\text{ODE: } & \phi(x)-l_{0}^{2}\phi''(x)=0,\\
		\text{Boundary: } & \phi(\pm\infty)=0,\quad \phi(x^{*})=1,
	\end{aligned}
	\label{eq:ODEs_strong_form_phase}
\end{equation}
which corresponds to a Helmholtz-type equation with prescribed boundary
conditions.

By employing the weak form and Gaussian quadrature, the solution of
\Cref{eq:ODEs_strong_form_phase} can be reformulated as a minimization
problem of the functional $I$:
\begin{equation}
	\begin{aligned}
		\phi(x) &=\arg\min_{\phi(x)\in C_{b}} I(\phi),\\
		I(\phi) &=A_{\Gamma}\int_{-\infty}^{+\infty}\frac{1}{2}\Big(\phi^{2}(x)+l_{0}^{2}[\phi'(x)]^{2}\Big)\,dx,
	\end{aligned}
	\label{eq:functional_min_1D}
\end{equation}
where $A_{\Gamma}=h$, with $h$ denoting the height of the one-dimensional
bar (see \Cref{fig:fracture_1d}a). The admissible space is
$C_{b}=\{\phi(x)\,|\,\phi(\pm\infty)=0,\;\phi(x^{*})=1\}$, i.e., the
set of all phase-field functions satisfying the boundary conditions
in \Cref{eq:ODEs_strong_form_phase}.

It is evident that \Cref{eq:contin_crack_label} closely resembles the
RBF formulation introduced in \Cref{subsec:RBF_KAN}, which further explains
the suitability of RBF networks for approximating the phase field
in fracture.

\subsection{Radial Basis Networks and Kolmogorov-Arnold Networks \label{subsec:RBF_KAN}}

\subsubsection{Radial Basis Networks for the Phase Field}

The structure of a radial basis function (RBF) network can be expressed as
\begin{equation}
	\phi(\boldsymbol{x};\boldsymbol{\theta}_{\phi})
	=R(\boldsymbol{x};\boldsymbol{\theta}_{\phi})
	=\sum_{i=1}^{N}w_{i}\exp\!\left[-\beta_{i}(\boldsymbol{x}-\boldsymbol{c}_{i})\cdot(\boldsymbol{x}-\boldsymbol{c}_{i})\right],
	\label{eq:RBF_NN}
\end{equation}
where $w_{i}$, $\beta_{i}$, and $\boldsymbol{c}_{i}$ are trainable network parameters.  

In practice, we adopt a slightly modified form by introducing normalization at the output:
\begin{equation}
	\phi(\boldsymbol{x};\boldsymbol{\theta}_{\phi})
	=R(\boldsymbol{x};\boldsymbol{\theta}_{\phi})
	=\frac{\sum_{i=1}^{N}w_{i}\exp\!\left[-\beta_{i}(\boldsymbol{x}-\boldsymbol{c}_{i})\cdot(\boldsymbol{x}-\boldsymbol{c}_{i})\right]}
	{\sum_{i=1}^{N}\exp\!\left[-\beta_{i}(\boldsymbol{x}-\boldsymbol{c}_{i})\cdot(\boldsymbol{x}-\boldsymbol{c}_{i})\right]}.
	\label{eq:RBF_normal}
\end{equation}
The normalized form \Cref{eq:RBF_normal} has the advantage of being able to represent constant functions simply by setting all $w_{i}$ equal, which is not possible in \Cref{eq:RBF_NN}. This property is particularly important for fracture phase-field modeling, where the field is zero over most of the domain. Hence, the normalized RBF form is more consistent with the intrinsic distribution of the phase field.

\subsubsection{Kolmogorov-Arnold Networks for the Displacement Field}

In Kolmogorov-Arnold Networks (KANs) \cite{liu2024kan}, consider a layer with $l_{i}$ input neurons and $l_{o}$ output neurons. The activation functions of this layer are denoted by $\phi_{ij}$, where $i\in\{1,2,\cdots,l_{o}\}$ and $j\in\{1,2,\cdots,l_{i}\}$. Each activation $\phi_{ij}$ is constructed from B-spline basis functions according to the number of grid points $G$ and the spline order $r$:
\begin{equation}
	\phi_{ij}(\boldsymbol{X})=
	\left[
	\begin{array}{cccc}
		\sum_{m=1}^{G_{1}+r_{1}}c_{m}^{(1,1)}B_{m}(x_{1}) &
		\sum_{m=1}^{G_{2}+r_{2}}c_{m}^{(1,2)}B_{m}(x_{2}) & \cdots &
		\sum_{m=1}^{G_{l_{i}}+r_{l_{i}}}c_{m}^{(1,l_{i})}B_{m}(x_{l_{i}})\\
		\sum_{m=1}^{G_{1}+r_{1}}c_{m}^{(2,1)}B_{m}(x_{1}) &
		\sum_{m=1}^{G_{2}+r_{2}}c_{m}^{(2,2)}B_{m}(x_{2}) & \cdots &
		\sum_{m=1}^{G_{l_{i}}+r_{l_{i}}}c_{m}^{(2,l_{i})}B_{m}(x_{l_{i}})\\
		\vdots & \vdots & \ddots & \vdots\\
		\sum_{m=1}^{G_{1}+r_{1}}c_{m}^{(l_{o},1)}B_{m}(x_{1}) &
		\sum_{m=1}^{G_{2}+r_{2}}c_{m}^{(l_{o},2)}B_{m}(x_{2}) & \cdots &
		\sum_{m=1}^{G_{l_{i}}+r_{l_{i}}}c_{m}^{(l_{o},l_{i})}B_{m}(x_{l_{i}})
	\end{array}
	\right],
\end{equation}
where $G_{j}$ is the number of grid points and $r_{j}$ is the order
of the B-spline in the $j$-th input direction. Each coefficient
$c_{m}^{(i,j)}$ corresponds to a B-spline weight, with $(G_{j}+r_{j})$
parameters in the $j$-th direction. Note that both the grid division
and spline order can be chosen independently for each input direction.

To enhance the representational power of the activation functions, we introduce scaling matrices $S_{ij}$ with the same shape as $\phi_{ij}$:
\begin{equation}
	S_{ij}=
	\begin{bmatrix}
		s_{11} & s_{12} & \cdots & s_{1l_{i}}\\
		s_{21} & s_{22} & \cdots & s_{2l_{i}}\\
		\vdots & \vdots & \ddots & \vdots\\
		s_{l_{o}1} & s_{l_{o}2} & \cdots & s_{l_{o}l_{i}}
	\end{bmatrix}.
\end{equation}
The scaling operation is applied element-wise as $\boldsymbol{\phi}=\boldsymbol{\phi}\odot\boldsymbol{S}$,
where $\odot$ denotes the Hadamard product.

The final output of the layer is given by
\begin{equation}
	\boldsymbol{Y}=\tanh\Bigg\{\sum_{\text{columns}}[\boldsymbol{\phi}(\boldsymbol{X})\odot\boldsymbol{S}]
	+\boldsymbol{W}\cdot\sigma(\boldsymbol{X})\Bigg\},
\end{equation}
where $\boldsymbol{W}$ is a linear transformation matrix and $\sigma$ is a nonlinear activation function. The residual term $\boldsymbol{W}\cdot\sigma(\boldsymbol{X})$ plays a role analogous to ResNet \cite{he2016deep}, while the scaling factors $\boldsymbol{S}$ and $\boldsymbol{W}$ act similarly to normalization layers. The additional nonlinear mapping $\sigma$ improves smoothness, preventing the B-spline approximation from producing rough outputs.

If the grid and spline order are the same for all input directions, the number of trainable parameters $c_{m}^{(i,j)}$ is $(G+r)$ for each direction. The total number of trainable parameters in KAN is summarized in \Cref{tab:trainable_para_KAN}. Further details can be found in the original KAN paper \cite{liu2024kan} and its applications to PINNs, such as KINN \cite{wang2025kolmogorov,shukla2024comprehensive}.

\begin{table}
	\caption{Trainable parameters in KAN.\label{tab:trainable_para_KAN}}
	\centering{}%
	\begin{tabular}{|c|c|c|c|}
		\hline 
		Type of parameters & Variable & Number & Description \tabularnewline
		\hline 
		$c_{m}^{(i,j)}$ & spline\_weight & $l_{o}\times l_{i}\times(G+r)$ & Coefficients of B-spline in activation function $\boldsymbol{\phi}$ \tabularnewline
		\hline 
		$W_{ij}$ & base\_weight & $l_{i}\times l_{o}$ & Linear transformation after nonlinear activation $\sigma$ \tabularnewline
		\hline 
		$S_{ij}$ & spline\_scaler & $l_{i}\times l_{o}$ & Scaling factors for activation functions \tabularnewline
		\hline 
	\end{tabular}
\end{table}

\subsection{Monolithic and Staggered Schemes} \label{subsec:stagger_mono}

We derive the first variation of the AT2 model in \Cref{eq:con_PFM} as
\begin{equation}
	\begin{aligned}
		\delta\Pi &=\int_{\Omega}\Big\{w^{'}(\phi)\varPsi^{+}(\boldsymbol{\varepsilon})
		+\frac{G_{c}}{l_{0}}(\phi-l_{0}^{2}\phi_{,ii})
		+w^{'}(\phi)H(\varPsi^{+})\Big\}\delta\phi \, dV \\
		&-\int_{\Omega}\Big\{[w(\phi)\sigma_{ij}^{+}+\sigma_{ij}^{-}]_{,j}+f_{i}\Big\}\delta u_{i}\, dV
		+\int_{\Gamma^{\boldsymbol{t}}}\Big\{[w(\phi)\sigma_{ij}^{+}+\sigma_{ij}^{-}]n_{j}-\bar{t}_{i}\Big\}\delta u_{i}\, dS \\
		&+\int_{\Gamma}l_{0}^{2}\phi_{,i}n_{i}\delta\phi \, dS ,
	\end{aligned}
	\label{eq:first_variational_hist}
\end{equation}
where $\delta H(\varPsi^{+})=0$.

The second variation of \Cref{eq:first_variational_hist} is
\begin{equation}
	\begin{aligned}
		\delta^{2}\Pi &= \int_{\Omega}\Big[ 
		w^{''}(\phi)\delta\phi^{2}\varPsi^{+}(\boldsymbol{\varepsilon})
		+2w^{'}(\phi)\delta\phi\sigma_{ij}^{+}(\boldsymbol{\varepsilon})\delta\varepsilon_{ij}
		+w(\phi)\delta\varepsilon_{kl}\frac{\partial^{2}\varPsi^{+}}{\partial\varepsilon_{ij}\partial\varepsilon_{kl}}\delta\varepsilon_{ij}\\
		&\quad +\delta\varepsilon_{kl}\frac{\partial^{2}\varPsi^{-}}{\partial\varepsilon_{ij}\partial\varepsilon_{kl}}\delta\varepsilon_{ij}
		\Big]\, dV \\
		&+\int_{\Omega}\Big\{
		\frac{G_{c}}{l_{0}}\big[\delta\phi^{2}+l_{0}^{2}(\nabla\delta\phi)\cdot(\nabla\delta\phi)\big]
		+w^{''}(\phi)\delta\phi^{2}H(\varPsi^{+})
		\Big\}\, dV .
	\end{aligned}
	\label{eq:second_variational_hist}
\end{equation}

We assume $w(\phi)=(1-\phi)^{2}$. It can be shown that for any nonzero $\delta\boldsymbol{\varepsilon}$,
\begin{equation}
	\begin{aligned}
		&w(\phi)\delta\varepsilon_{kl}\frac{\partial^{2}\varPsi^{+}}{\partial\varepsilon_{ij}\partial\varepsilon_{kl}}\delta\varepsilon_{ij}
		+\delta\varepsilon_{kl}\frac{\partial^{2}\varPsi^{-}}{\partial\varepsilon_{ij}\partial\varepsilon_{kl}}\delta\varepsilon_{ij} \\
		&= w(\phi)\big[\lambda\langle\delta\varepsilon_{ii}\rangle_{+}^{2}+2\mu\langle\delta\varepsilon_{ij}\rangle_{+}^{2}\big]
		+\big[\lambda\langle\delta\varepsilon_{ii}\rangle_{-}^{2}+2\mu\langle\delta\varepsilon_{ij}\rangle_{-}^{2}\big] \geq 0 .
	\end{aligned}
	\label{eq:second_strain_energy}
\end{equation}

Thus, the only term affecting the positive definiteness of $\delta^{2}\Pi$ in \Cref{eq:second_variational_hist} is
\[
\int_{\Omega} 2w^{'}(\phi)\delta\phi\sigma_{ij}^{+}(\boldsymbol{\varepsilon})\delta\varepsilon_{ij}.
\]
Rewriting \Cref{eq:second_variational_hist} in matrix form yields
\begin{equation}
	\delta^{2}\Pi =\int_{\Omega}
	\begin{bmatrix}
		\delta\boldsymbol{\varepsilon}\\
		\delta\phi
	\end{bmatrix}^{T}
	\begin{bmatrix}
		\boldsymbol{A} & \boldsymbol{B}\\
		\boldsymbol{B}^{T} & C
	\end{bmatrix}
	\begin{bmatrix}
		\delta\boldsymbol{\varepsilon}\\
		\delta\phi
	\end{bmatrix}\, dV ,
\end{equation}
with
\begin{equation}
	\begin{aligned}
		\boldsymbol{A} &= w(\phi)\frac{\partial^{2}\varPsi^{+}}{\partial\boldsymbol{\varepsilon}\partial\boldsymbol{\varepsilon}}
		+\frac{\partial^{2}\varPsi^{-}}{\partial\boldsymbol{\varepsilon}\partial\boldsymbol{\varepsilon}}, \\
		\boldsymbol{B} &= \boldsymbol{\sigma}^{+}(\boldsymbol{\varepsilon})w^{'}(\phi), \\
		C &= w^{''}(\phi)[\varPsi^{+}(\boldsymbol{\varepsilon})+H(\varPsi^{+})]
		+\frac{G_{c}}{l_{0}}\big[1+l_{0}^{2}(\nabla)\cdot(\nabla)\big].
	\end{aligned}
\end{equation}

For $\delta\boldsymbol{\varepsilon}=A^{-1}B\delta\phi$, the minimization becomes
\begin{equation}
	\min_{\delta\boldsymbol{\varepsilon}}(\delta^{2}\Pi) 
	= \delta\phi^{T}M\delta\phi, \quad
	M = C-\boldsymbol{B}^{T}\boldsymbol{A}^{-1}\boldsymbol{B}.
\end{equation}
Hence, a necessary and sufficient condition for $\delta^{2}\Pi\geq0$ is $M>0$, i.e.,
\begin{equation}
	w^{''}(\phi)\big[\varPsi^{+}(\boldsymbol{\varepsilon})+H(\varPsi^{+})\big]
	+\frac{G_{c}}{l_{0}}\big[1+l_{0}^{2}(\nabla)\cdot(\nabla)\big]
	>
	[w^{'}(\phi)]^{2}\boldsymbol{\sigma}^{+}(\boldsymbol{\varepsilon})
	\Big[w(\phi)\frac{\partial^{2}\varPsi^{+}}{\partial\boldsymbol{\varepsilon}\partial\boldsymbol{\varepsilon}}
	+\frac{\partial^{2}\varPsi^{-}}{\partial\boldsymbol{\varepsilon}\partial\boldsymbol{\varepsilon}}\Big]^{-1}
	\boldsymbol{\sigma}^{+}(\boldsymbol{\varepsilon}).
	\label{eq:requirement_second}
\end{equation}

From \Cref{eq:requirement_second}, we conclude:
\begin{itemize}
	\item At early loading stages, \Cref{eq:requirement_second} always holds since $\boldsymbol{\varepsilon}=0$ initially, making the right-hand side vanish. Physically, the system is stable with no crack initiation.
	\item At the critical load, \Cref{eq:requirement_second} reduces to an equality, corresponding to $\delta^{2}\Pi\geq0$ and the onset of crack nucleation.
	\item Under overload, \Cref{eq:requirement_second} is violated, implying instability and crack propagation.
\end{itemize}

\begin{figure}
	\begin{centering}
		\includegraphics{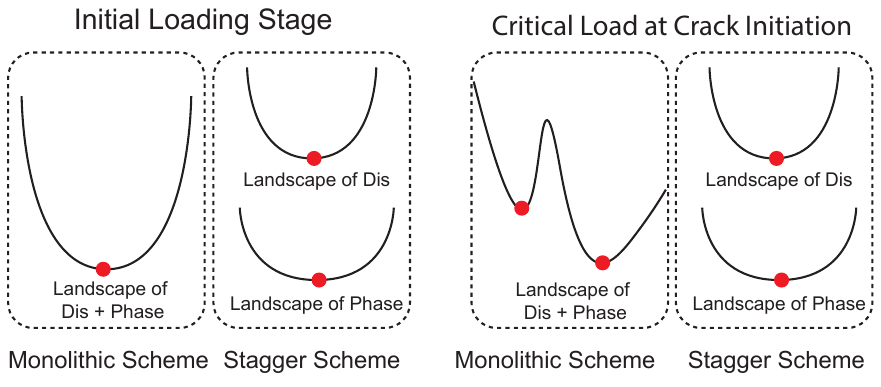}
		\par\end{centering}
	\caption{Energy landscape under different load levels. Red dots indicate minima.  
		Left: Early loading stage, where both monolithic and staggered schemes yield a convex landscape.  
		Right: Critical load stage, where the monolithic scheme results in a non-convex landscape with multiple minima, while the staggered scheme ensures convexity in either the displacement or phase-field space. \label{fig:different_phase_landscape}}
\end{figure}

Therefore, algorithmically, when the load is below the crack initiation threshold, $\phi$ and $\boldsymbol{u}$ can be optimized simultaneously using a monolithic scheme for faster convergence. Near the critical load, a staggered scheme is preferable since the cross term $\int_{\Omega}2w^{'}(\phi)\delta\phi\sigma_{ij}^{+}(\boldsymbol{\varepsilon})\delta\varepsilon_{ij}=0$, ensuring $\delta^{2}\Pi\geq0$ and thus improved robustness.   As shown in \Cref{fig:different_phase_landscape}, the energy landscape varies across different loading stages.

\subsection{Calculation of Stress Intensity Factor } \label{subsec:Calculation-of-sif}

In XDEM, two approaches are employed to evaluate the stress intensity factors (SIFs): the $J$-integral method and the interaction integral method.

First, consider the $J$-integral, defined as
\begin{equation}
	J=\oint_{\Gamma}(wn_{1}-n_{\alpha}\sigma_{\alpha\beta}u_{\beta,1})d\Gamma
\end{equation}
where $w$ is the strain energy density function, $n_{1}$ is the $x$-component of the unit normal vector to the closed contour $\Gamma$ in the local coordinate system, $\sigma_{\alpha\beta}$ is the stress tensor, and $u_{\beta,1}$ denotes the gradient of the displacement in the local $x$-direction. All physical quantities must be evaluated in the local coordinate system, with the $x$-axis aligned with the crack tangent. The contour $\Gamma$ encloses the crack tip.

In linear elastic fracture mechanics (LEFM), the $J$-integral represents the energy release rate:
\begin{equation}
	J = G_{c} = \frac{K_{1}^{2}}{E} + \frac{K_{2}^{2}}{E} + \frac{K_{3}^{2}}{2G},
\end{equation}
where $K_{1}$, $K_{2}$, and $K_{3}$ denote the mode I, mode II, and mode III SIFs, respectively. $E$ and $G$ are the elastic and shear moduli. For plane strain conditions, $E$ is replaced by $E/(1-\nu^{2})$, where $\nu$ is the Poisson's ratio.

It is evident that the $J$-integral couples $K_{1}$, $K_{2}$, and $K_{3}$. To decouple the mode I and mode II components, we introduce the phase angle $\psi$, defined as
\begin{equation}
	\psi=\arctan\frac{K_{2}}{K_{1}}=\lim_{r\to0,\theta=0}\arctan\frac{\tau_{12}}{\sigma_{22}} .
\end{equation}
Once $\psi$ is obtained, the individual SIFs are recovered as
\begin{equation}
	\begin{aligned}
		K_{1} &= \sqrt{EJ}\cos(\psi),\\
		K_{2} &= \sqrt{EJ}\sin(\psi).
	\end{aligned}
\end{equation}
However, accurate evaluation of $\psi$ requires information very close to the crack tip, where steep gradients make the computation challenging. Therefore, this method is generally more reliable for pure mode I or pure mode II cracks.

For mixed-mode cracks, the interaction integral method is more robust. We define
\begin{equation}
	\begin{aligned}
		M^{(1)} &= \oint_{\Gamma}\Big(w^{mix(1)}\delta_{1j}-\sigma_{\alpha\beta}u_{\alpha,1}^{(1)}-\sigma_{\alpha\beta}^{(1)}u_{\alpha,1}\Big)n_{j}\, d\Gamma,\\
		M^{(2)} &= \oint_{\Gamma}\Big(w^{mix(2)}\delta_{1j}-\sigma_{\alpha\beta}u_{\alpha,1}^{(2)}-\sigma_{\alpha\beta}^{(2)}u_{\alpha,1}\Big)n_{j}\, d\Gamma,\\
		w^{mix(1)} &= \tfrac{1}{2}\big(\sigma_{\alpha\beta}\varepsilon_{\alpha\beta}^{(1)}+\sigma_{\alpha\beta}^{(1)}\varepsilon_{\alpha\beta}\big),\\
		w^{mix(2)} &= \tfrac{1}{2}\big(\sigma_{\alpha\beta}\varepsilon_{\alpha\beta}^{(2)}+\sigma_{\alpha\beta}^{(2)}\varepsilon_{\alpha\beta}\big),
	\end{aligned}
\end{equation}
where the superscripts $(1)$ and $(2)$ denote auxiliary (virtual) fields, constructed from the $K$-field representation in \Cref{eq:williams_series} by taking the $n=1$ term. Once $M^{(1)}$ and $M^{(2)}$ are computed, the SIFs are obtained as
\begin{equation}
	\begin{aligned}
		K_{1} &= \frac{E}{2}M^{(1)},\\
		K_{2} &= \frac{E}{2}M^{(2)}.
	\end{aligned}
\end{equation}
The interaction integral is widely regarded as more reliable for mixed-mode fracture problems.

After obtaining $K_{1}$ and $K_{2}$, the crack propagation angle can be estimated using the maximum hoop stress criterion:
\begin{equation}
	\begin{aligned}
		\sigma_{\theta} &= \frac{1}{\sqrt{2\pi r}}\cos\frac{\theta}{2}\Big[K_{1}(1+\cos\theta)-3K_{2}\sin\theta\Big],\\
		\theta &= \arccos\frac{3K_{2}^{2}\pm\sqrt{K_{1}^{4}+8K_{1}^{2}K_{2}^{2}}}{K_{1}^{2}+9K_{2}^{2}}.
	\end{aligned}
\end{equation}
Among the two candidate angles $\theta$, the one corresponding to the maximum $\sigma_{\theta}$ is selected as the crack propagation direction.

\subsection{UEL Details in FEM} \label{subsec:FEM_UEL_detail}

Although the phase-field fracture model offers clear theoretical
advantages, standard finite element software typically does not provide
elements with an additional phase-field degree of freedom (DOF). Hence,
a user-defined element (UEL) is required in order to couple the
mechanical displacements with the phase field. In our implementation
within \textsc{ABAQUS}, the UEL carries three DOFs per node in 2D:
two displacement components $(x,y)$ and the scalar phase field. The
coupled system is solved in a staggered manner. Below we summarize
the essential FEM details.

In the finite element discretization, the displacement field and the
phase field are interpolated as
\begin{equation}
	\begin{aligned}
		\boldsymbol{u}(\boldsymbol{x})
		&=\sum_{I=1}^{N_{\text{node}}}\boldsymbol{N}^{\boldsymbol{u}}_{I}(\boldsymbol{x})\,\boldsymbol{u}_{I}
		=\sum_{e=1}^{N_{e}}\sum_{I=1}^{m}\boldsymbol{N}^{\boldsymbol{u}e}_{I}(\boldsymbol{x})\,\boldsymbol{u}^{e}_{I}
		=\sum_{e=1}^{N_{e}}\boldsymbol{N}^{\boldsymbol{u}e}(\boldsymbol{x})\,\boldsymbol{u}^{e},\\[3pt]
		\phi(\boldsymbol{x})
		&=\sum_{I=1}^{N_{\text{node}}}N^{\phi}_{I}(\boldsymbol{x})\,\phi_{I}
		=\sum_{e=1}^{N_{e}}\sum_{I=1}^{m}N^{\phi e}_{I}(\boldsymbol{x})\,\phi^{e}_{I}
		=\sum_{e=1}^{N_{e}}\boldsymbol{N}^{\phi e}(\boldsymbol{x})\,\boldsymbol{\phi}^{e},
	\end{aligned}
	\label{eq:u_fai}
\end{equation}
where $\boldsymbol{u}_{I}=\big[u^{x}_{I}\;\;u^{y}_{I}\big]^{T}$ is the nodal displacement and $\phi_{I}$ the nodal
phase field. $N_{\text{node}}$ and $N_{e}$ are the total numbers of
nodes and elements, respectively; $\boldsymbol{u}^{e}_{I}$ and $\phi^{e}_{I}$
are the element-level nodal DOFs. For 2D problems we use
\begin{equation}
	\begin{aligned}\boldsymbol{N}_{I}^{\boldsymbol{u}}(\boldsymbol{x})=\left[\begin{array}{cc}
			N_{I}(\boldsymbol{x})\\
			& N_{I}
		\end{array}\right]; & N_{I}^{\phi}(\boldsymbol{x})=N_{I}\\
		\boldsymbol{N}^{\boldsymbol{u}e}(\boldsymbol{x})=\left[\begin{array}{cccc}
			\boldsymbol{N}_{I}^{\boldsymbol{u}e}(\boldsymbol{x}) & \boldsymbol{N}_{2}^{\boldsymbol{u}e}(\boldsymbol{x}) & \cdots & \boldsymbol{N}_{m}^{\boldsymbol{u}e}(\boldsymbol{x})\end{array}\right]; & \boldsymbol{N}^{\phi e}(\boldsymbol{x})=\left[\begin{array}{cccc}
			N_{I}^{\phi e}(\boldsymbol{x}) & N_{2}^{\phi e}(\boldsymbol{x}) & \cdots & N_{m}^{\phi e}(\boldsymbol{x})\end{array}\right]\\
		\boldsymbol{u}^{e}=\left[\begin{array}{cccc}
			\boldsymbol{u}_{1}^{e} & \boldsymbol{u}_{2}^{e} & \cdots & \boldsymbol{u}_{m}^{e}\end{array}\right]^{T}; & \boldsymbol{\phi}^{e}=\left[\begin{array}{cccc}
			\phi_{1}^{e} & \phi_{2}^{e} & \cdots & \phi_{m}^{e}\end{array}\right]^{T}
	\end{aligned}
	.
\end{equation}
Here $m$ denotes the number of nodes per element; we adopt 4-node
quadrilateral elements, so $m=4$. Unless otherwise stated, vectors
are written as column vectors.

For the 4-node quadrilateral element, the shape functions in the
parent (reference) coordinates $\boldsymbol{\xi}=(\xi,\eta)\in[-1,1]\times[-1,1]$ are
\begin{equation}
	\begin{aligned}
		N^{e}_{1}(\boldsymbol{\xi})&=\tfrac{(1-\xi)(1-\eta)}{4},\qquad
		N^{e}_{2}(\boldsymbol{\xi})=\tfrac{(1+\xi)(1-\eta)}{4},\\
		N^{e}_{3}(\boldsymbol{\xi})&=\tfrac{(1+\xi)(1+\eta)}{4},\qquad
		N^{e}_{4}(\boldsymbol{\xi})=\tfrac{(1-\xi)(1+\eta)}{4}.
	\end{aligned}
	\label{eq:Shape_function}
\end{equation}

\begin{figure}
	\centering
	\includegraphics[scale=0.5]{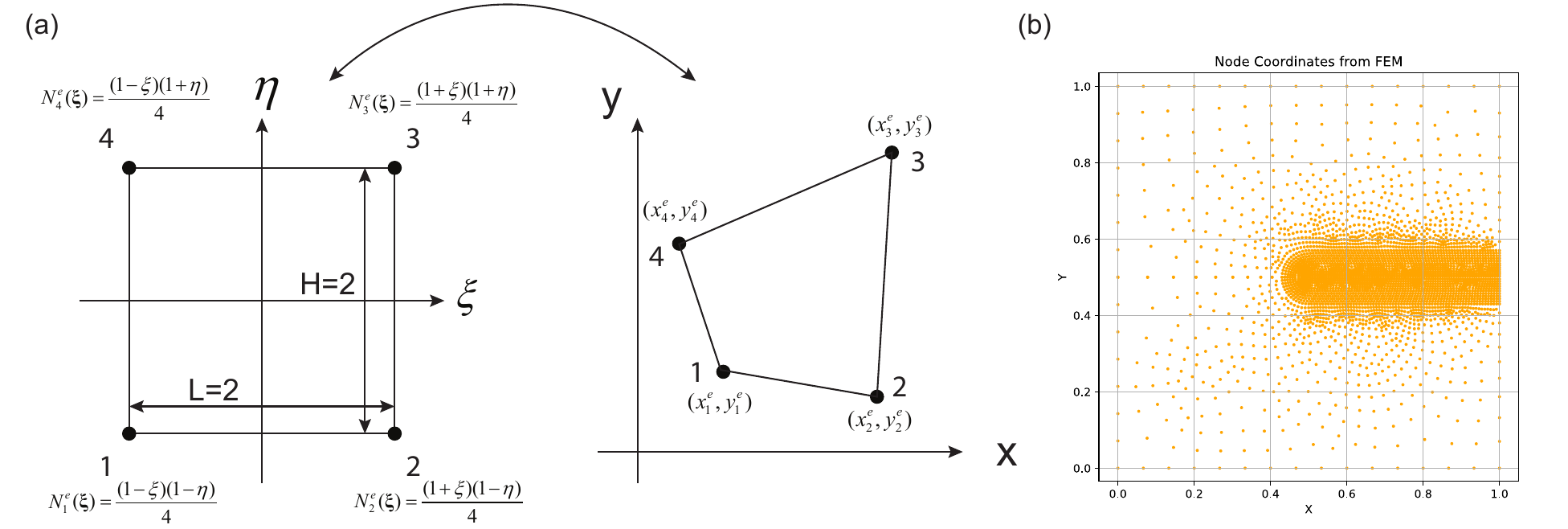}
	\caption{(a) Mapping from the parent coordinates $(\xi,\eta)$ of a standard square parent element to the physical coordinates $(x,y)$ of the actual finite element. (b) Node distribution of the mesh in FEM.\label{fig:shape_function}}
\end{figure}

The mapping from the parent element to the physical element is standard
in FEM (see \Cref{fig:shape_function}a and any FEM book
\cite{finite_element_book}).

Taking spatial derivatives for the phase field, we obtain
\begin{equation}
	\begin{aligned}
		\frac{\partial\phi(\boldsymbol{x})}{\partial\boldsymbol{x}}
		&=
		\begin{bmatrix}
			\partial\phi/\partial x\\
			\partial\phi/\partial y
		\end{bmatrix}
		=\sum_{e=1}^{N_{e}}\boldsymbol{B}^{\phi e}(\boldsymbol{x})\,\boldsymbol{\phi}^{e},\\[3pt]
		\boldsymbol{B}^{\phi e}(\boldsymbol{x})
		&=
		\begin{bmatrix}
			\partial N^{e}_{1}/\partial x & \partial N^{e}_{2}/\partial x & \partial N^{e}_{3}/\partial x & \partial N^{e}_{4}/\partial x\\
			\partial N^{e}_{1}/\partial y & \partial N^{e}_{2}/\partial y & \partial N^{e}_{3}/\partial y & \partial N^{e}_{4}/\partial y
		\end{bmatrix},\qquad
		\boldsymbol{\phi}^{e}=\big[\phi^{e}_{1}\;\phi^{e}_{2}\;\phi^{e}_{3}\;\phi^{e}_{4}\big]^{T}.
	\end{aligned}
	\label{eq:B_shape}
\end{equation}
Since $N^{e}_{I}$ is defined in the parent coordinates, chain rule gives
\begin{equation}
	\begin{aligned}
		\frac{\partial N^{e}_{I}(\boldsymbol{x})}{\partial x}
		&=\frac{\partial N^{e}_{I}(\boldsymbol{\xi})}{\partial\boldsymbol{\xi}}
		\frac{\partial\boldsymbol{\xi}}{\partial x}
		=
		\begin{bmatrix}
			\partial \xi/\partial x & \partial \eta/\partial x
		\end{bmatrix}
		\begin{bmatrix}
			\partial N^{e}_{I}/\partial \xi\\[2pt]
			\partial N^{e}_{I}/\partial \eta
		\end{bmatrix},\\[3pt]
		\frac{\partial N^{e}_{I}(\boldsymbol{x})}{\partial y}
		&=\frac{\partial N^{e}_{I}(\boldsymbol{\xi})}{\partial\boldsymbol{\xi}}
		\frac{\partial\boldsymbol{\xi}}{\partial y}
		=
		\begin{bmatrix}
			\partial \xi/\partial y & \partial \eta/\partial y
		\end{bmatrix}
		\begin{bmatrix}
			\partial N^{e}_{I}/\partial \xi\\[2pt]
			\partial N^{e}_{I}/\partial \eta
		\end{bmatrix}.
	\end{aligned}
	\label{eq:derivative_shape}
\end{equation}
Hence
\begin{align}
	\boldsymbol{B}^{e}(\boldsymbol{x})&=(\boldsymbol{J}^{e})^{-1}\,\boldsymbol{N}^{e}_{\boldsymbol{\xi}},
	\label{eq:B_matrix}\\
	(\boldsymbol{J}^{e})^{-1}&=
	\begin{bmatrix}
		\partial \xi/\partial x & \partial \eta/\partial x\\
		\partial \xi/\partial y & \partial \eta/\partial y
	\end{bmatrix},\qquad
	\boldsymbol{J}^{e}=
	\begin{bmatrix}
		\partial x/\partial \xi & \partial y/\partial \xi\\
		\partial x/\partial \eta & \partial y/\partial \eta
	\end{bmatrix},\\
	\boldsymbol{N}^{e}_{\boldsymbol{\xi}}&=
	\begin{bmatrix}
		\partial N^{e}_{1}/\partial \xi & \partial N^{e}_{2}/\partial \xi & \partial N^{e}_{3}/\partial \xi & \partial N^{e}_{4}/\partial \xi\\
		\partial N^{e}_{1}/\partial \eta & \partial N^{e}_{2}/\partial \eta & \partial N^{e}_{3}/\partial \eta & \partial N^{e}_{4}/\partial \eta
	\end{bmatrix}
	=
	\begin{bmatrix}
		-\tfrac{(1-\eta)}{4} & \tfrac{(1-\eta)}{4} & \tfrac{(1+\eta)}{4} & -\tfrac{(1+\eta)}{4}\\[3pt]
		-\tfrac{(1-\xi)}{4} & -\tfrac{(1+\xi)}{4} & \tfrac{(1+\xi)}{4} & \tfrac{(1-\xi)}{4}
	\end{bmatrix}.
\end{align}

By interpolating the physical coordinates with the same shape functions,
the Jacobian can be written as
\begin{align}
	\boldsymbol{J}^{e}
	&=
	\begin{bmatrix}
		\displaystyle \frac{\partial\big[\sum_{I=1}^{m}N_{I}^{e}(\boldsymbol{\xi})\,x^{e}_{I}\big]}{\partial \xi}
		&
		\displaystyle \frac{\partial\big[\sum_{I=1}^{m}N_{I}^{e}(\boldsymbol{\xi})\,y^{e}_{I}\big]}{\partial \xi}
		\\[6pt]
		\displaystyle \frac{\partial\big[\sum_{I=1}^{m}N_{I}^{e}(\boldsymbol{\xi})\,x^{e}_{I}\big]}{\partial \eta}
		&
		\displaystyle \frac{\partial\big[\sum_{I=1}^{m}N_{I}^{e}(\boldsymbol{\xi})\,y^{e}_{I}\big]}{\partial \eta}
	\end{bmatrix}
	=\boldsymbol{N}^{e}_{\boldsymbol{\xi}}\,\boldsymbol{X}^{e},
	\label{eq:Jacobi}\\
	\boldsymbol{X}^{e}&=
	\begin{bmatrix}
		x^{e}_{1} & x^{e}_{2} & x^{e}_{3} & x^{e}_{4}\\
		y^{e}_{1} & y^{e}_{2} & y^{e}_{3} & y^{e}_{4}
	\end{bmatrix}^{T},
\end{align}
where $(x^{e}_{I},y^{e}_{I})$ are the physical coordinates of the element
nodes.

Substituting \Cref{eq:Jacobi} into \Cref{eq:B_matrix}, and then using $\boldsymbol{B}^{\phi e}$ in \Cref{eq:B_shape}, we obtain
\begin{equation}
	\frac{\partial\phi(\boldsymbol{x})}{\partial\boldsymbol{x}}
	=\sum_{e=1}^{N_{e}}\boldsymbol{B}^{\phi e}(\boldsymbol{x})\,\boldsymbol{\phi}^{e}
	=\sum_{e=1}^{N_{e}}\big(\boldsymbol{N}^{e}_{\boldsymbol{\xi}}\boldsymbol{X}^{e}\big)^{-1}\boldsymbol{N}^{e}_{\boldsymbol{\xi}}\,\boldsymbol{\phi}^{e}.
	\label{eq:T_x}
\end{equation}
The same procedure applies to the displacement gradient.

Inserting \Cref{eq:T_x} and \Cref{eq:u_fai} into the first variation
\Cref{eq:first_variational_hist} yields
\begin{equation}
	\small
	\begin{aligned}
		\delta\Pi
		&=\int_{\Omega}\!\Big[
		w'(\phi)\,\delta\phi\,\varPsi^{+}(\boldsymbol{\varepsilon})
		+w(\phi)\,\sigma^{+}_{ij}\,\delta\varepsilon_{ij}
		+\sigma^{-}_{ij}\,\delta\varepsilon_{ij}\Big]\,dV
		+\int_{\Omega}\!\Big\{\frac{G_{c}}{l_{0}}\big[\phi\,\delta\phi+l_{0}^{2}(\nabla\phi)\!\cdot\!(\nabla\delta\phi)\big]
		+w'(\phi)\,\delta\phi\,H(\varPsi^{+})\Big\}\,dV\\
		&\quad-\int_{\Omega}\boldsymbol{f}\cdot\delta\boldsymbol{u}\,dV
		-\int_{\Gamma^{\boldsymbol{t}}}\bar{\boldsymbol{t}}\cdot\delta\boldsymbol{u}\,dS\\
		&=\sum_{e=1}^{N_{e}}\Bigg\{
		\int_{\Omega^{e}}\!\Big[
		w'(\phi)\,\boldsymbol{N}^{e}\delta\boldsymbol{\phi}^{e}\,\varPsi^{+}(\boldsymbol{\varepsilon})
		+\big(\boldsymbol{D}\langle \boldsymbol{B}^{e}\boldsymbol{u}^{e}\rangle_{+}\big)^{T}\!\boldsymbol{B}^{e}\delta\boldsymbol{u}^{e}\,w(\phi)
		+\big(\boldsymbol{D}\langle \boldsymbol{B}^{e}\boldsymbol{u}^{e}\rangle_{-}\big)^{T}\!\boldsymbol{B}^{e}\delta\boldsymbol{u}^{e}
		-\boldsymbol{f}\cdot\boldsymbol{N}^{e}\delta\boldsymbol{u}^{e}
		\Big]\,dV
		\\
		&\qquad
		+\int_{\Omega^{e}}\!\Big\{
		\frac{G_{c}}{l_{0}}\big[(\boldsymbol{N}^{e}\boldsymbol{\phi}^{e})^{T}\boldsymbol{N}^{e}\delta\boldsymbol{\phi}^{e}
		+l_{0}^{2}(\boldsymbol{B}^{e}\boldsymbol{\phi}^{e})^{T}\boldsymbol{B}^{e}\delta\boldsymbol{\phi}^{e}\big]
		+w'(\phi)\,\boldsymbol{N}^{e}\delta\boldsymbol{\phi}^{e}H(\varPsi^{+})
		\Big\}\,dV
		\Bigg\}-\int_{\Gamma^{\boldsymbol{t}e}}\!\bar{\boldsymbol{t}}\cdot\boldsymbol{N}^{e}\delta\boldsymbol{u}^{e}\,dS\\
		&=\sum_{e=1}^{N_{e}}\Bigg\{
		\int_{\Omega^{e}}\!\Big[
		w'(\phi)\,\boldsymbol{N}^{e}\,\varPsi^{+}(\boldsymbol{\varepsilon})
		+\frac{G_{c}}{l_{0}}\big(\boldsymbol{N}^{eT}\boldsymbol{N}^{e}\boldsymbol{\phi}^{e}
		+l_{0}^{2}\boldsymbol{B}^{eT}\boldsymbol{B}^{e}\boldsymbol{\phi}^{e}\big)
		+w'(\phi)\,\boldsymbol{N}^{e}H(\varPsi^{+})
		\Big]\,dV\;\delta\boldsymbol{\phi}^{e}\\
		&\qquad+\Big[
		\int_{\Omega^{e}}\big(w(\phi)\boldsymbol{D}\langle\boldsymbol{B}^{e}\boldsymbol{u}^{e}\rangle_{+}^{T}\boldsymbol{B}^{e}
		+\boldsymbol{D}\langle\boldsymbol{B}^{e}\boldsymbol{u}^{e}\rangle_{-}^{T}\boldsymbol{B}^{e}\big)\,dV
		-\int_{\Omega^{e}}\boldsymbol{f}\cdot\boldsymbol{N}^{e}\,dV
		-\int_{\Gamma^{\boldsymbol{t}e}}\bar{\boldsymbol{t}}\cdot\boldsymbol{N}^{e}\,dS
		\Big]\delta\boldsymbol{u}^{e}
		\Bigg\}.
	\end{aligned}
	\label{eq:first_variational_hist-1}
\end{equation}

The resulting nonlinear residual equations at the element level are
\begin{equation}
	\begin{aligned}
		\boldsymbol{R}_{\boldsymbol{u}^{e}}&=
		\int_{\Omega^{e}}\!\big[w(\phi)\,\boldsymbol{B}^{eT}\boldsymbol{D}\langle\boldsymbol{B}^{e}\boldsymbol{u}^{e}\rangle_{+}
		+\boldsymbol{B}^{eT}\boldsymbol{D}\langle\boldsymbol{B}^{e}\boldsymbol{u}^{e}\rangle_{-}\big]\,dV
		-\int_{\Omega^{e}}\boldsymbol{N}^{eT}\boldsymbol{f}\,dV
		-\int_{\Gamma^{\boldsymbol{t}e}}\boldsymbol{N}^{eT}\bar{\boldsymbol{t}}\,dS=\boldsymbol{0},\\[3pt]
		\boldsymbol{R}_{\phi^{e}}&=
		\int_{\Omega^{e}}\!\Big[w'(\phi)\,\boldsymbol{N}^{eT}\varPsi^{+}(\boldsymbol{\varepsilon})
		+\frac{G_{c}}{l_{0}}\big(\boldsymbol{N}^{eT}\boldsymbol{N}^{e}\boldsymbol{\phi}^{e}
		+l_{0}^{2}\boldsymbol{B}^{eT}\boldsymbol{B}^{e}\boldsymbol{\phi}^{e}\big)
		+w'(\phi)\,\boldsymbol{N}^{eT}H(\varPsi^{+})\Big]\,dV=\boldsymbol{0}.
	\end{aligned}
\end{equation}
We linearize the system via Newton's method. The element tangent matrix reads
\begin{equation}
	\boldsymbol{K}^{e}=
	\begin{bmatrix}
		\boldsymbol{K}^{e}_{\boldsymbol{u}\boldsymbol{u}} & \boldsymbol{K}^{e}_{\boldsymbol{u}\phi}\\
		\boldsymbol{K}^{e}_{\phi\boldsymbol{u}} & \boldsymbol{K}^{e}_{\phi\phi}
	\end{bmatrix}
	=
	\begin{bmatrix}
		\partial \boldsymbol{R}_{\boldsymbol{u}^{e}}/\partial \boldsymbol{u}^{e} &
		\partial \boldsymbol{R}_{\boldsymbol{u}^{e}}/\partial \boldsymbol{\phi}^{e}\\
		\partial \boldsymbol{R}_{\phi^{e}}/\partial \boldsymbol{u}^{e} &
		\partial \boldsymbol{R}_{\phi^{e}}/\partial \boldsymbol{\phi}^{e}
	\end{bmatrix},
	\qquad
	\frac{\partial \boldsymbol{R}_{\phi^{e}}}{\partial \boldsymbol{u}^{e}}
	=\bigg(\frac{\partial \boldsymbol{R}_{\boldsymbol{u}^{e}}}{\partial \boldsymbol{\phi}^{e}}\bigg)^{T}
	=\int_{\Omega^{e}}\!w'(\phi)\,\boldsymbol{B}^{eT}\boldsymbol{D}\langle\boldsymbol{B}^{e}\boldsymbol{u}^{e}\rangle_{+}\,\boldsymbol{N}^{e}\,dV,
	\label{eq:K_matrix}
\end{equation}
with
\begin{equation}
	\frac{\partial \boldsymbol{R}_{\boldsymbol{u}^{e}}}{\partial \boldsymbol{u}^{e}}
	=\frac{\partial}{\partial \boldsymbol{u}^{e}}
	\int_{\Omega^{e}}\!\big[w(\phi)\,\boldsymbol{B}^{eT}\boldsymbol{D}\langle\boldsymbol{B}^{e}\boldsymbol{u}^{e}\rangle_{+}
	+\boldsymbol{B}^{eT}\boldsymbol{D}\langle\boldsymbol{B}^{e}\boldsymbol{u}^{e}\rangle_{-}\big]\,dV,
\end{equation}
\begin{equation}
	\frac{\partial \boldsymbol{R}_{\phi^{e}}}{\partial \boldsymbol{\phi}^{e}}
	=\int_{\Omega^{e}}\!\Big[
	w''(\phi)\,\boldsymbol{N}^{eT}\boldsymbol{N}^{e}\,\varPsi^{+}(\boldsymbol{\varepsilon})
	+\frac{G_{c}}{l_{0}}\big(\boldsymbol{N}^{eT}\boldsymbol{N}^{e}
	+l_{0}^{2}\boldsymbol{B}^{eT}\boldsymbol{B}^{e}\big)
	+w''(\phi)\,\boldsymbol{N}^{eT}\boldsymbol{N}^{e}H(\varPsi^{+})
	\Big]\,dV.
\end{equation}

Due to the inherent non-convexity of the fracture energy (see the
analyses in  \Cref{subsec:stagger_mono}), the global tangent
may be indefinite, which affects numerical stability and convergence.
A common remedy is to drop the off-diagonal coupling blocks
$\boldsymbol{K}^{e}_{\boldsymbol{u}\phi}$ and $\boldsymbol{K}^{e}_{\phi\boldsymbol{u}}$,
and to employ a staggered solution: solve alternately for the displacement
and phase-field subproblems.

In our UEL implementation, we exploit the block structure and adopt
a staggered iterative scheme. To accurately resolve the diffusive
crack, the element size $h$ must be chosen sufficiently small relative
to the phase-field length scale $l_{0}$. For the single-edge notched
square specimen with a mode-I crack, a total of 3259 user elements
are used. The bottom edge is fully clamped; the crack faces are traction-free.
A uniform tensile displacement is prescribed on the top boundary.
To extract the load-displacement curve, all $y$-DOFs on the top edge
are coupled to a reference point, where the displacement is applied
and the reaction force is recorded. The mesh (with local refinement
near the crack tip) is illustrated in \Cref{fig:shape_function}b.

\renewcommand{\Crefname}{Supplementary References}

\bibliography{mybibfile}

\end{document}